\documentclass[mathleft
]{an}
\usepackage{natbib}
\usepackage{graphicx}
\usepackage[tight]{subfigure}
\usepackage{times}
\usepackage{amsmath}
\begin{document}
\Pagespan{1}{}
\Yearpublication{2012}%
\Yearsubmission{2012}%
\newcommand{\kms}{{~\rm km\; s^{-1}}}
\newcommand{\cc}{{~\rm cm^{-3}}}
\newcommand{\msyr}{{~M_{\odot}~\rm yr^{-1}}}
\newcommand{\cm}{{~\rm cm}}
\newcommand{\s}{{~\rm s}}
\newcommand{\km}{{~\rm km}}
\newcommand{\g}{{~\rm g}}
\newcommand{\K}{{~\rm K}}
\newcommand{\erg}{{~\rm erg}}
\newcommand{\yr}{{~\rm yr}}
\newcommand{\Myr}{{~\rm Myr}}
\newcommand{\Gyr}{{~\rm Gyr}}
\newcommand{\pc}{{~\rm pc}}
\newcommand{\kpc}{{~\rm kpc}}
\newcommand{\Mpc}{{~\rm Mpc}}
\newcommand{\keV}{{~\rm keV}}
\newcommand{\kev}{{~\rm keV}}
\newcommand{\AU}{{~\rm AU}}
\newcommand{\mum}{{~\rm \mu m}}
\newcommand{\days}{{~\rm days}}

\newcommand{\aap}{A\&A}

\title{The jet feedback mechanism (JFM): from supernovae to clusters of galaxies}

\author{Noam Soker\inst{1}\fnmsep\thanks{Corresponding author:\email{soker@physics.technion.ac.il}},
Muhammad Akashi\inst{1}, Avishai Gilkis\inst{1}, Shlomi Hillel\inst{1}, Oded Papish\inst{1}, Michael Refaelovich\inst{1}, and Danny Tsebrenko\inst{1}}
\titlerunning{The jet feedback mechanism}
\authorrunning{N. Soker et al. }
\institute{Department of Physics, Technion -- Israel Institute of Technology, Haifa 32000, Israel;}



\abstract{%
We study the similarities of jet-medium interactions in several quite different astrophysical systems using 2D and 3D hydrodynamical numerical simulations, and find many similarities.
The systems include cooling flow (CF) clusters of galaxies, core collapse supernovae (CCSNe), planetary nebulae (PNe),
and common envelope (CE) evolution.
The similarities include hot bubbles inflated by jets in a bipolar structure, vortices on the sides of the jets,
vortices inside the inflated bubbles, fragmentation of bubbles to two and more bubbles, and buoyancy of bubbles.
The activity in many cases is regulated by a negative feedback mechanism.
Namely, higher accretion rate leads to stronger jet activity that in turn suppresses the accretion process.
After the jets power decreases the accretion resumes, and the cycle restarts.
In the case of CF in galaxies and clusters of galaxies we also study the accretion process, which is most
likely by cold clumps, i.e., the cold feedback mechanism.
In CF clusters we find that heating of the intra-cluster medium (ICM) is
done by mixing hot shocked jet gas with the ICM, and not by shocks.
Our results strengthen the jet feedback mechanism (JFM) as a common process in many astrophysical objects.
}
\maketitle


\section{Introduction}
\label{sec:intro}

Jets are observed in a large variety of astrophysical objects.
In many of these objects jets play a significant, and even crucial, role in the evolution due to their high
energy content.
In some of these systems the effects of the jets are regulated by a negative feedback mechanism, the jet-feedback mechanism (JFM).
Such is the case in cooling flow (CF) clusters and CF galaxies, where the jets heat the intra-cluster medium (ICM), and during galaxy formation when the jets
were required to heat and expel the ISM.
A feedback mechanism might also operate in the explosion of core collapse supernova (CCSN), where it is possible that
jets launched by the newly formed neutron star (NS) or a black hole (BH) explode the star.

The aim of this report is to emphasize the connections and similarities between some of these seemingly very different kind of objects.
These similarities allow us to learn from one object on the other.
The systems studied in this paper are listed in Table \ref{table:compare}.
The report is concentrated on the similar processes of jet-medium interaction in the different objects. For that, many
works on the different objects and processes that do not consider the similarities between the different objects will not be mentioned here.
\begin{table*}
    \begin{tabular}{|l|l|l|l|l|l|}
     \hline
        Property / System    & Clusters           & galaxy formation& CCSNe            & PNe            & CE      \\ \hline
        Typical Energy (erg) &     $10^{60}$      &   $10^{59}$     & $10^{51}$        &$10^{44}$       & $10^{44}-10^{48}$        \\
        Typical Mass $(M_\odot)$& $10^{12}$       &  $10^{11}$      & $10$             &$1$             & $1$   \\
        Typical Size         & $100 \kpc$         & $10 \kpc$       & $10^9 \cm$       &$0.1 \pc$       & $10-100 R_\odot$       \\
        Typical Time         & $10^7-10^8 \yr$    & $10^7-10^8 \yr$ & $1-3 \s$         & $10-100 \yr$   & $1 \yr$        \\
        $T_{\rm bubble}(\K)$ & $10^9-10^{10} $    & $10^9-10^{10}$  &  $10^{10} $      &   $10^6 $      & $10^8-10^{10}$  \\
        $T_{\rm ambiant}(\K)$& $10^7-10^8 $       &  $10^6-10^7 $   & few$\times 10^9$ & $10^4 $        & $10^{5}$         \\
        Compact object       & BH                  & BH             & NS or BH         & MS or WD       & NS or WD            \\
        $\qquad$ mass $(M_\odot)$ & $10^8-10^{10}$& $ 10^6-10^9 $  & $1-5 $           & $1$            &$1$                  \\
        Main Effect of jets  & Heating ICM        & Expelling gas   & Exploding the star&Shaping the PN & Reducing accretion  \\
        Observations         & X-ray bubbles      & (Massive outflows)& (Axisymmetry)   & Bipolar PNs   & ~                  \\ \hline
    \end{tabular}
    \caption{Systems discussed in this paper where feedback and shaping by jets take place.
    The different listed values are typical and to an order or magnitude (or two even) accuracy only.
    Typical energy: Energy in one jet episode.
    Typical Mass, Size: of the relevant ambient gas. Typical time: the duration of the jets activity episode.
    In the last row of observations, in parenthesis are expected observations.
    Acronym: \textbf{PNe}: Planetary nebulae; \textbf{CCSNe}: core collapse supernovae; \textbf{CF}: cooling flow; \textbf{ICM}: Intra-cluster medium
    \textbf{CE}: common envelope; \textbf{BH}: black hole; \textbf{NS}: neutron star; \textbf{MS}: Main sequence star  }
    \label{table:compare}
\end{table*}
\subsection{Morphology}
\label{subsec:morphology}

The efficiency of the JFM comes from the deep potential well of the jet's launching site.
In CF clusters, CF galaxies, and during galaxy formation the jets are lunched by a super massive black hole (SMBH), while
in CCSN they are launched by the newly formed NS or BH.
In all cases the size of the compact object is several orders of magnitude smaller than that of the surrounding medium,
ensuring that few percents of the accretion gravitational energy can substantially influence the ambient medium.

However, there are two  other conditions for the JFM to work efficiently:
(1) The jets must deposit their kinetic energy in the inner region, where the feedback mechanism must work.
(2) Radiative loses, usually by photons but by neutrinos in CCSNe, must be small.
The deposition of the kinetic energy is via strong shock waves. The shocked jet's material temperature can be much higher
than the ambient medium temperature and, due to pressure balance, its density is lower. If radiative losses are small indeed, then
the post shock jet's material expands and forms a low-density bubble.
As usually there are two opposite jets, a structure of two opposite bubbles is formed.
A pair of bubbles (or several pairs) is a generic structure of the JFM when they can be observed.
This structure is termed bipolar in the case of planetary nebulae (PNe).
In young stellar objects (YSO) there are jets that interact with the dense surrounding gas. However, radiative cooling is very rapid
for YSO jets, and no bubbles are formed. Any JFM, if exist, is of very low efficiency.

\begin{figure}
\centering
\hskip -1.5 cm
\includegraphics[width =80mm]{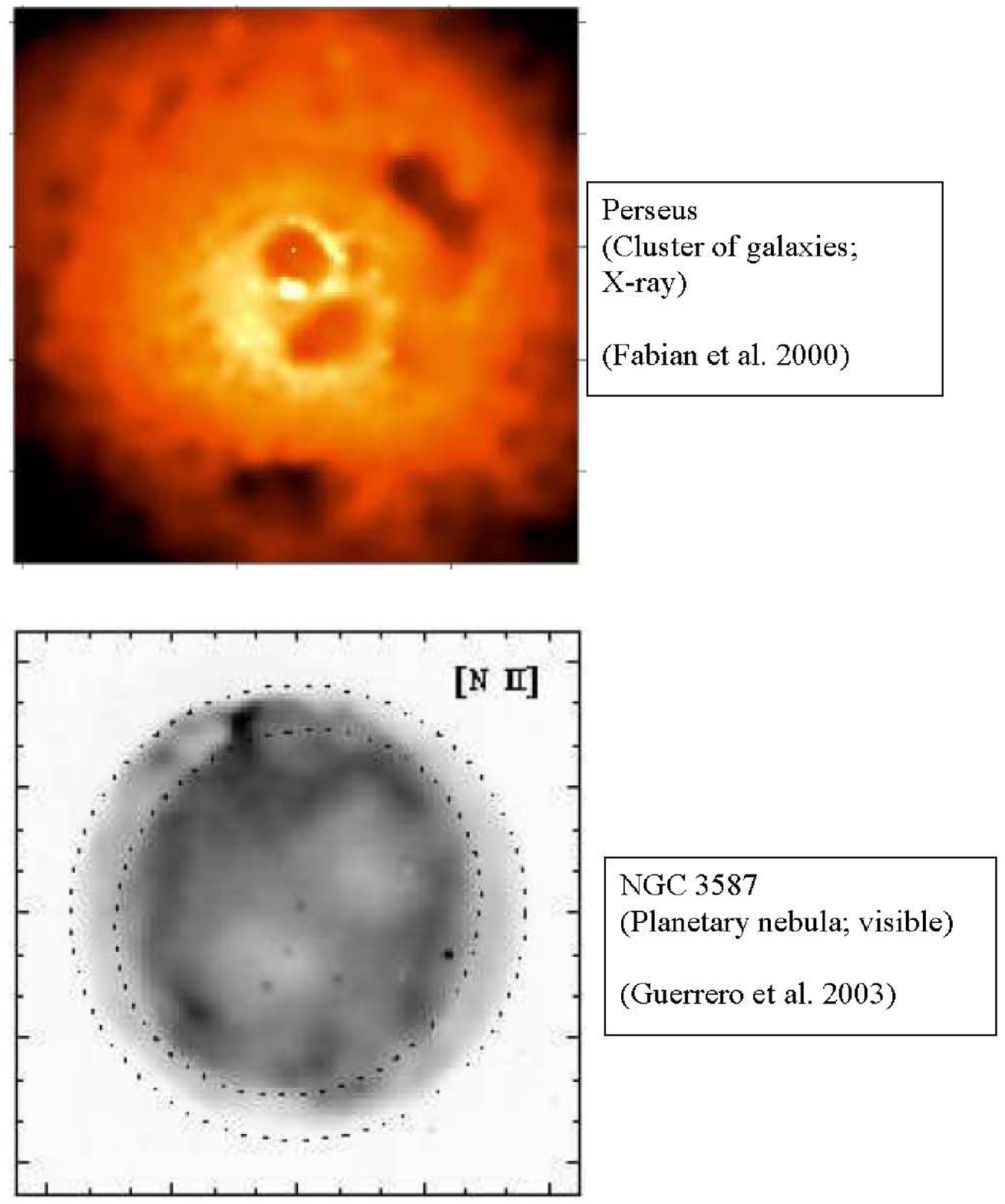} 
\hskip -0.4 cm
\includegraphics[width =80mm]{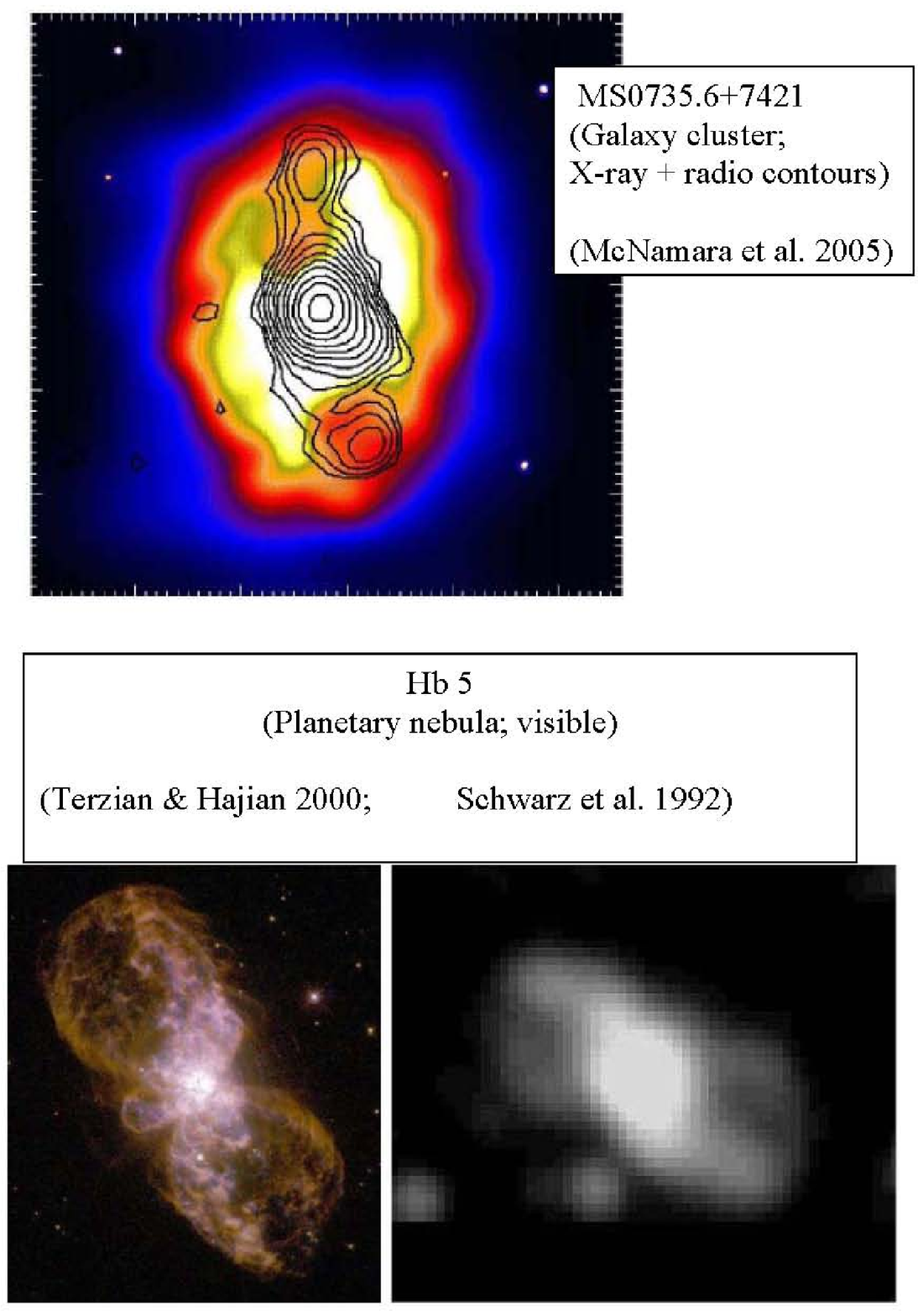} 
\caption{Comparing false color X-ray image of a galaxy cluster with a
visible image of a planetary nebula (PN). The upper panel emphasizes pairs of fat spherical bubbles near the center, while
the bottom panel emphasizes point symmetric morphology of the bubble pair, and the consequences of bad resolution (right PN image).
Images from \cite{Fabianetal2000}, \cite{Guerreroetal2003}, \cite{McNamaraetal2005}, \cite{TerzianHajian2000}, and \cite{Schwarzetal1992}. }
\label{fig:bisker12}
\end{figure}

The most prominent similarity of the bubble-pair (bipolar) structure is perhaps the very similar morphological structures found in some
PNe and some X-ray deficient bubbles in CF clusters \citep{SokerBisker2006}.
Two comparisons are presented in Figure \ref{fig:bisker12}, taken from \cite{SokerBisker2006} where more examples are given.
Despite the several orders of magnitude differences in size, energy, mass, and timescales, the similarity is
not only in the morphology, but in some basic physical processes as well. These similarities were studied in a series of papers
by one of us (\citealt{SokerBisker2006} and references therein).
Point symmetric PNe, like Hb~5 presented in the right panel of Figure \ref{fig:bisker12}, are thought to be shaped by stellar binary interactions
that caused a jet-precession.
The similarity of the morphology of Hb-5 and the pair of bubbles in the CF cluster MS~0735.6+7421 brought
\cite{Pizzolato2005} to suggest that MS~0735.6+7421 has a massive binary BHs system in its center.
Simulations of PNe jet-shaping are presented in section \ref{sec:bipolar}, while jet-inflated bubbles in CF clusters are presented in section \ref{sec:ICM}.

In PNe gravity is negligible and there is no need for a feedback mechanism.
We note though that in some PNe and other nebulae around stars, most of the kinetic energy of the expanding gas comes from the jets.
This seems to be the case in the bipolar nebula, the Homunculus, around the binary system $\eta$ Carinae \citep{Soker2004a}.
The relevance of PNe is that the bipolar structure of pairs of bubbles is well resolved, and can teach us about the interaction of jets with the ambient gas. This is demonstrated in the right panel of
Figure \ref{fig:bisker12} where a much better resolved image of Hb-5 reveals many details that cannot be resolved in MS~0735.6+7421.
In some cases the jets and bubbles cannot be observed at all. This is the case in jets that might exist in CCSNe
and during galaxy formation.
Simulations of jet-driven CCSN are presented in section \ref{sec:CCSN}.

\subsection{Negative Feedback}
\label{subsec:feedback}

There are some common processes and properties of the JFM that are common to the different astrophysical objects.
We list them here, and later implement some of them in the numerical simulations.
\begin{enumerate}
\item \emph{Accretion disk.} The jets are launched by an accretion disk around a compact object. In the case of jets launched by SMBH such disks
are inferred from observations, and the jets are directly observed. However, if CCSNe are driven by jets, they cannot be observed.
The same hold for jets during the galaxy formation epoch, as these objects are at large distances and in many cases the central region is
expected to be obscured.
In most CCSNe there is no sufficient angular momentum to explain a continuous accretion disk.
Instead, we assume that as a result of the stationary accretion shock instability (SASI; e.g. \citealt{Blondin2007},
but see \citealt{Nordhausetal2010} for suppression of this instability in 3D simulations),
or some other stochastic processes, segments of the post-shock accreted gas (inward to the stalled shock wave)
possess local angular momentum.
Accretion of dense gas from varying directions can be seen in the recent simulations by
\citep{Mueller2012b}.
Such an accretion form a \emph{jittering pair of opposite jets} \citep{papish2011}.
\item \emph{Universal jets' properties.} The properties of jets launched by the compact objects have some universal average properties.
$(i)$ The velocity of the pre-shock jets' material of $v_f \simeq v_{\rm esc}$, where $v_{\rm esc}$ is the escape velocity from the compact object.
$(ii)$ The ratio between mass lose rate in the two jets to mass accretion rate is $\eta \equiv \dot M_f/\dot M_{\rm acc} \simeq 0.1$.
\item \emph{Non-penetrating jets.} For a jet to deposit energy in the relevant inner regions it
should not penetrate through these regions; this is termed the \emph{a non-penetrating JFM}.
The condition for jet-stopping is that the time required for the jet to propagate through the surrounding
gas and break out of it must be longer than the typical time for restarting the penetration process.
The restarting can be due to transverse motion of the jets, as in the jittering-jet model or in the case of jet precession, due to the transverse relative motion
of the surrounding gas and the SMBH (hence the jets continuously encounter fresh gas), due to the orbital motion of the companion in a common envelope (CE),
and by wide jets that propagate very slowly through the ambient gas.
These cases have been explored in the past for jets launched by SMBHs.
\cite{Sternberg2007} showed that slow massive wide (SMW) jets can inflate the fat bubbles that are observed in many CFs, in clusters,
groups of galaxies, and in elliptical galaxies.
The same basic physics that prevents wide jets from penetrating through the ICM
was shown to hold for precessing jets \citep{Sternberg2008a, Falceta-Goncalves2010}, or a
relative motion of the jets to the medium \citep{Bruggen2007, Soker2009, Morsony2010, Mendygral2012}.
If the jets penetrate to a too large distance, then no bubbles are formed, while
in intermediate cases elongated and/or detached from the center bubbles are formed
(e.g., \citealt{Basson2003, Omma2004, Heinz2006, Vernaleo2006, AlouaniBibi2007, Sternberg2007, ONeill2010, Mendygral2011, Mendygral2012}).
\item \emph{Available mass.} The mass available for the inflow is very large.
Namely, the mass that is eventually accreted to the NS in CCSN or the SMBH in CFs is limited by the JFM
and not by the mass available in the surroundings.
\item \emph{Stochastic accretion.}
In the JFM considered here in CCSN and CFs (both in clusters and in galaxies), the feeding of the accretion disk is by cold clumps.
By cold we refer to a gas temperature much below the virial one.
Although the average rate of accretion is regulated by the feedback process, there might be large temporary variations
in the rates of mass and angular momentum accretion values.
The cold accretion allows both accretion and jet activity to coexist.
As well, the cold clumps falls quite rapidly, allowing fast communication between the ambient gas and the accretion disk, as required
in the JFM.
In CFs this is termed the cold feedback mechanism, and it is discussed in section \ref{subsec:accretion}.
In CCSN the stochastic accretion of clumps allows a temporary formation of accretion disk.
\item \emph{Incomplete energy deposition.} The JFM we explore here differs from other mechanisms by the requirement that its
influence on the environment is not complete.
In CF clusters and galaxies, the accretion is of cold clumps; the cold feedback mechanism.
Namely, the heating by the jets although very efficient, is not complete.
Most of the cold gas forms stars and cold filaments in the ICM, rather than feeding the SMBH.
The small fraction of the cold gas mass that does feed the SMBH is sufficient to maintain the JFM.
Hence, star formation and cold filaments in CF clusters and galaxies are generic features of the JFM considered here.
In CCSN the accretion continues for a few seconds while the jets and the bubbles they inflate expel most of the core mass.
The regulation of this continuous accretion episode is part of the feedback mechanism.
This is in contrast to most models of CCSNe explosion that start by reviving the accretion shock and terminate accretion,
and where a feedback mechanism does not exist.
\end{enumerate}

\section{Bipolar shaping of planetary nebulae}
\label{sec:bipolar}

There are a number of numerical simulations that try to reproduce the asymmetrical structure of PNe
(e.g., \citealt{Huarte-Espinosa2012} for a recent paper and references therein).
We here concentrate only on those PNs that have two pairs of bubbles similar in structure to pairs of
bubbles in CF clusters. These are most likely formed by jets.

Our simulations are performed by using version 4.0-beta of the FLASH code \citep{Fryxell2000}.
The FLASH code is an  adaptive-mesh refinement modular code used for solving
hydrodynamics or magnetohydrodynamics problems.
Here we use the unsplit PPM (piecewise-parabolic method) solver of FLASH to
simulate gas dynamics in different astrophysical environments.
The simulations are differ by adding different cooling functions and by the appropriate EOS.
In the simulations of PN shaping we include radiative cooling, but not gravity.
In the simulation of CCSN described later we do incorporate gravity.

The radiative cooling is added to the simulation at all temperatures $T>10^4\K$,
and it is carefully treated near contact discontinuities to prevent large
temperature gradients from causing unphysical results.
The cooling function for solar abundances that we use was taken from
(\citealt{SutherlandDopita1993}; their table 6).

We employ a full 3D adaptive mesh refinement (AMR, 6 levels ($2^{9}$ cells in each direction)) using
a Cartesian grid $(x,y,z)$ with outflow boundary condition at all boundary surfaces.
We define the $x-y$ (z=0) plane to be the equatorial plane of the PN and simulate the whole space (the two sides of
the equatorial plane).
The grid size is $ 2 \times 10^{17} \cm $ in the $x$ and $y$ directions, and $4 \times 10^{17} \cm $ in the $z$ direction.

At the beginning of the simulation, $t=0$, the grid is filled with a spherically-symmetric slow wind, blown by the AGB stellar progenitor,
having a uniform radial velocity of $v_{wind}=10 \kms$.
The density at $t=0$ is taken to be $\rho(t=0) = \frac {\dot M_{wind}}{4 \pi r^{2} v_{wind}}$, where we take here
a slow-wind mass loss rate of $\dot M_{wind}=  10^{-5} {M_\odot \yr^{-1}}$.
The jet is lunched from the inner $6 \times 10^{15} \cm $ region, and within a half opening angle of $\alpha = 70 \,^{\circ}$.
This is a wide jet; similar results will be obtained by rapidly precessing jets.
The jets are launched with a radial velocity of $v_{\rm jet}=600 \kms$ and the mass loss rate in each side of
$\dot M_{\rm jet} = 3 \times 10^{-8}{M_\odot \yr^{-1}}$.
For numerical reasons a weak slow wind is injected in the sector $\alpha<\theta< 90  \,^{\circ}$.
The slow wind and the ejected jet start with a temperature of $10000 \K$.
The initial jets' temperature has no influence on the results (as long it is
highly supersonic) because the jet rapidly cools due to adiabatic expansion.

Our calculations do not include the ionizing radiation and the fast wind blown
by the central star during the PN phase.
We simply aim at showing the shaping of the nebular gas to a bipolar shape by blowing wide jets.
There is no feedback heating, but there is a shaping that is very similar to the one in clusters.
The mass of the hot bubble is very small relative to the ambient mass, while the linear momentum of the jet's
material is comparable to the one of the AGB wind.

\begin{figure*}
 \subfigure[][]{\label{subfigure:arrow1}\includegraphics*[scale=0.255,clip=true,trim=25 60 407.8 0]{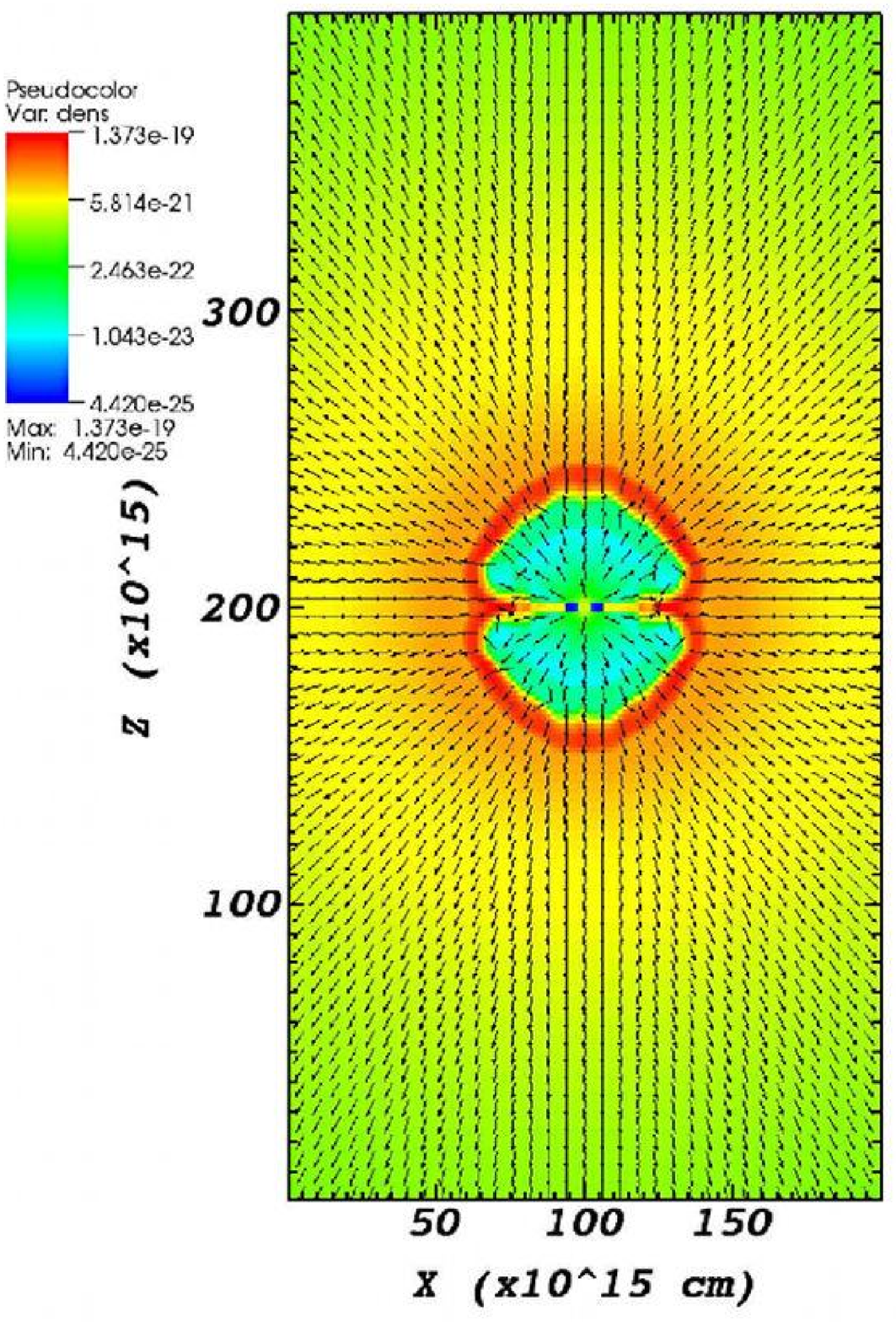}}
 \subfigure[][]{\label{subfigure:vortex1}\includegraphics[scale=0.255,clip=true,trim=25 60 382.7 0]{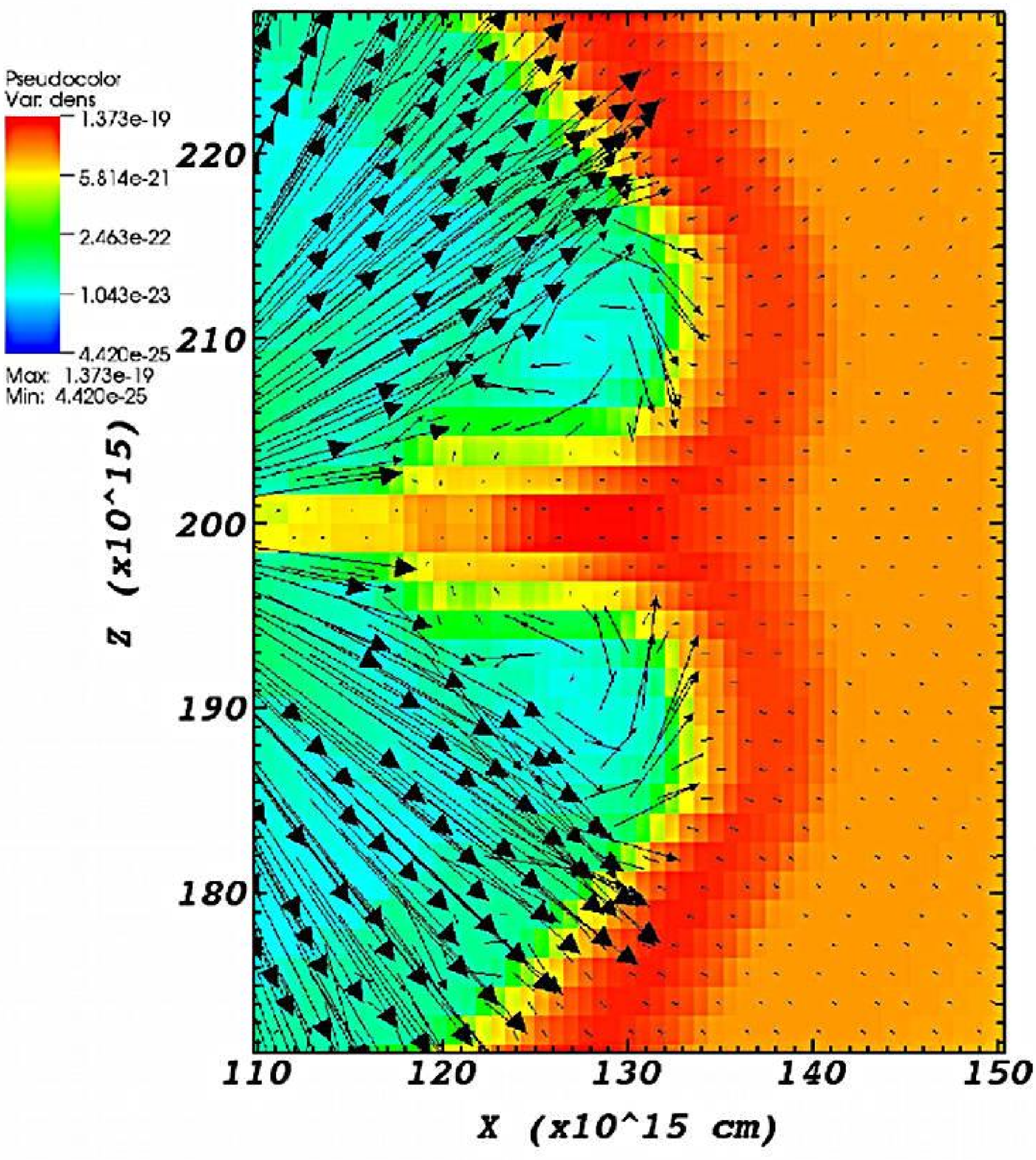}}
 \subfigure[][]{\label{subfigure:temp1}\includegraphics*[scale=0.259,clip=true,trim=0 75 350 0]{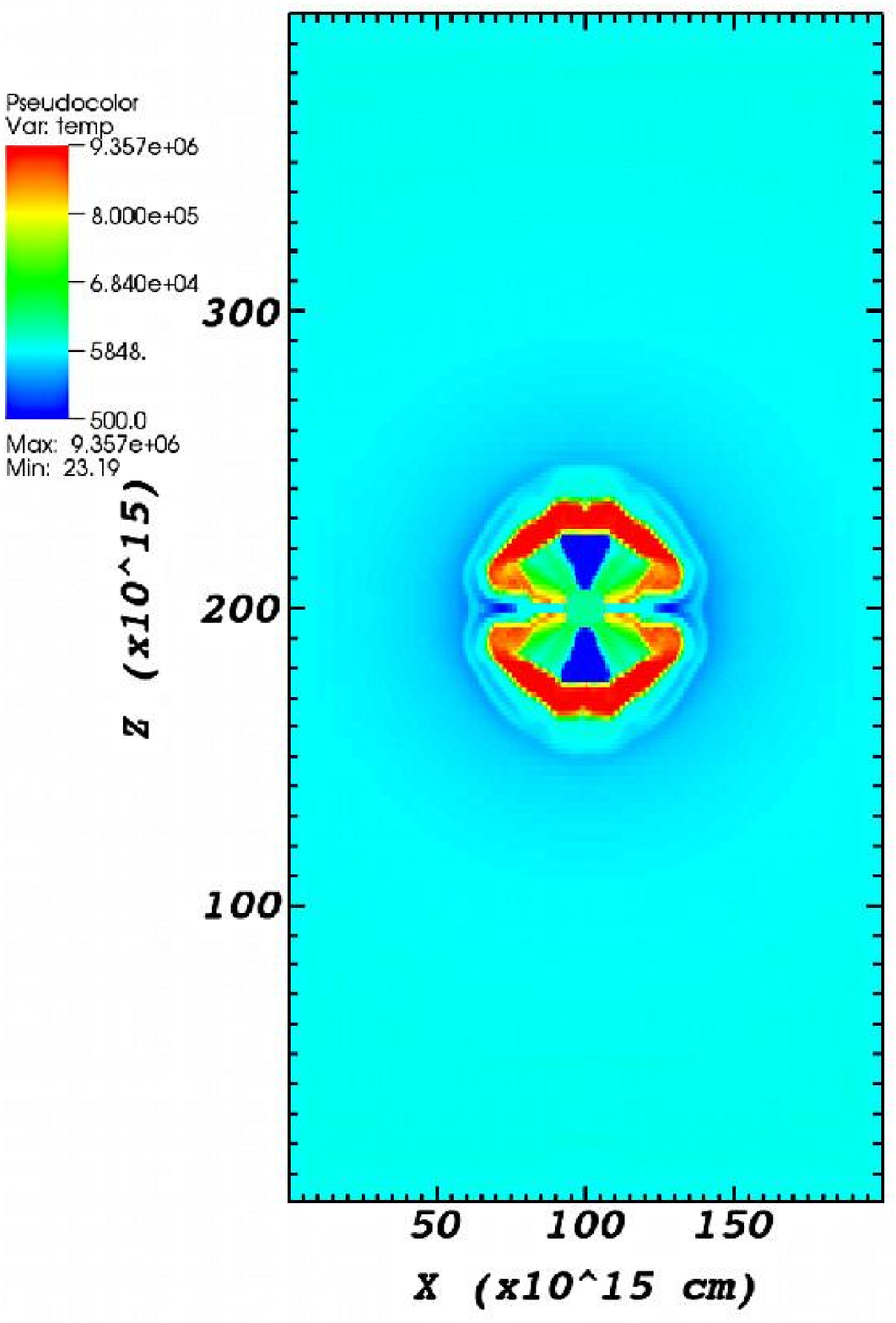}}\\
 \subfigure[][]{\label{subfigure:arrow2}\includegraphics*[scale=0.255,clip=true,trim=25 60 407.8 0]{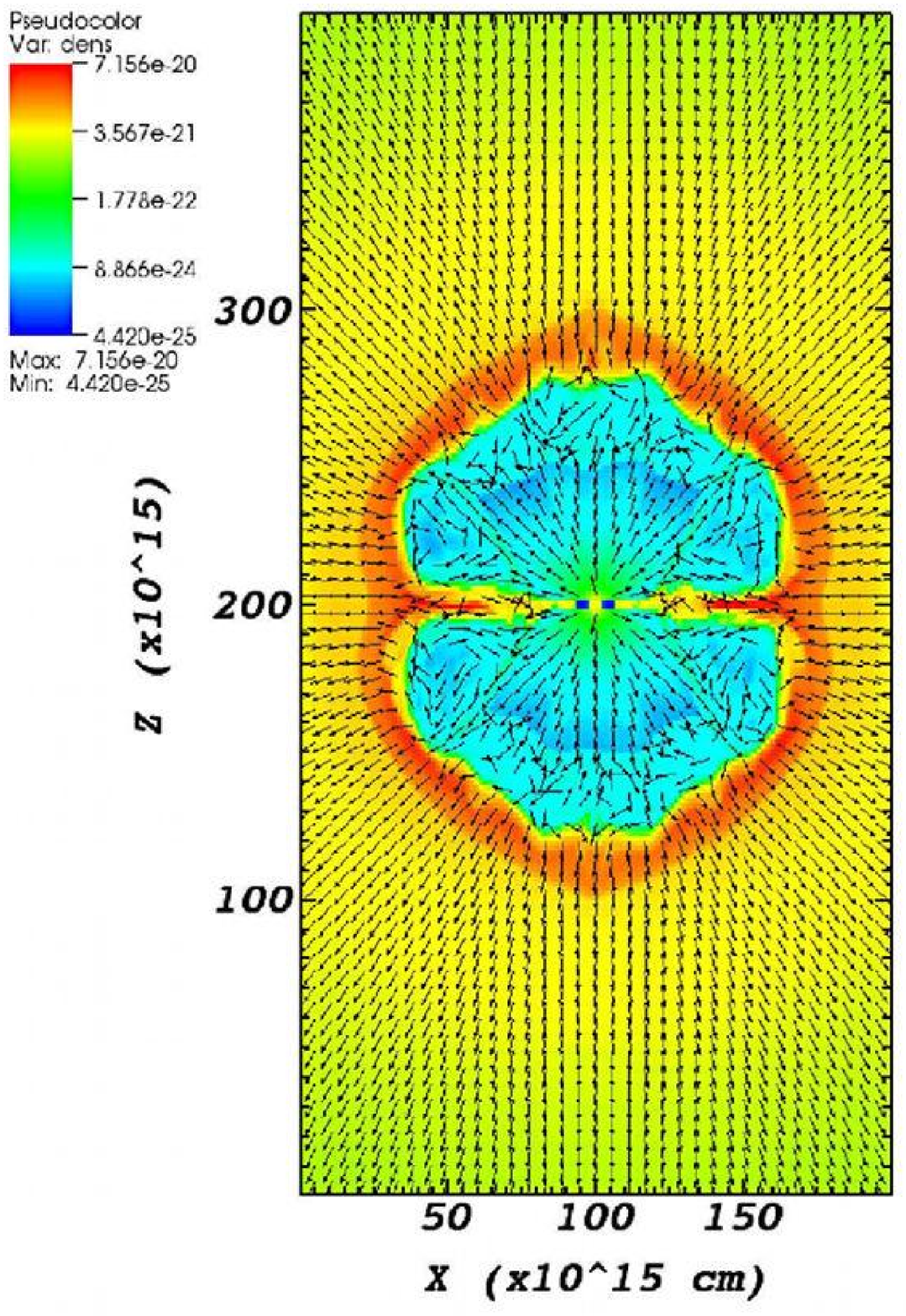}}
 \subfigure[][]{\label{subfigure:vortex2}\includegraphics*[scale=0.255,clip=true,trim=25 60 382.7 0]{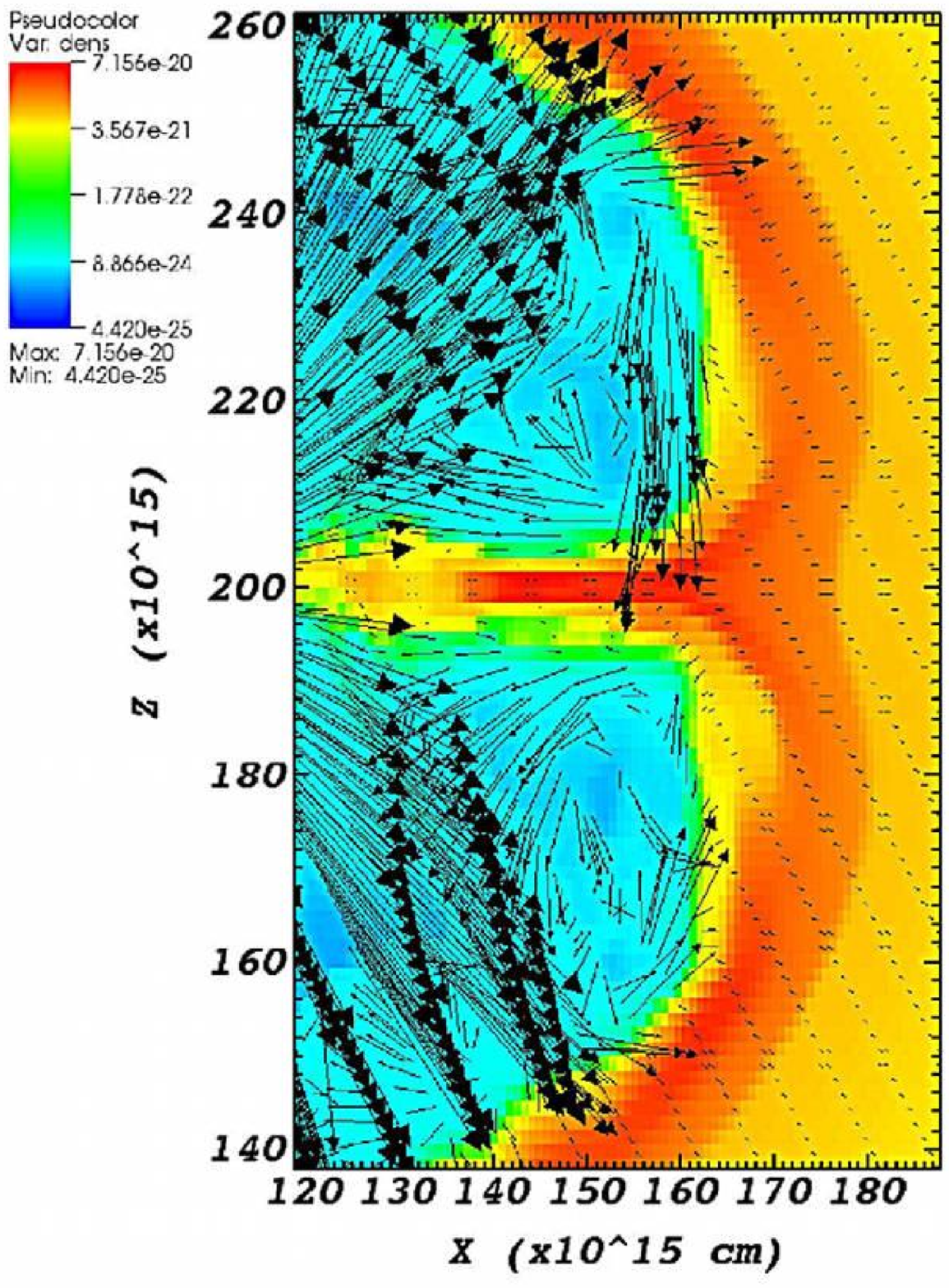}}
 \subfigure[][]{\label{subfigure:temp2}\includegraphics*[scale=0.259,clip=true,trim=0 75 350 0]{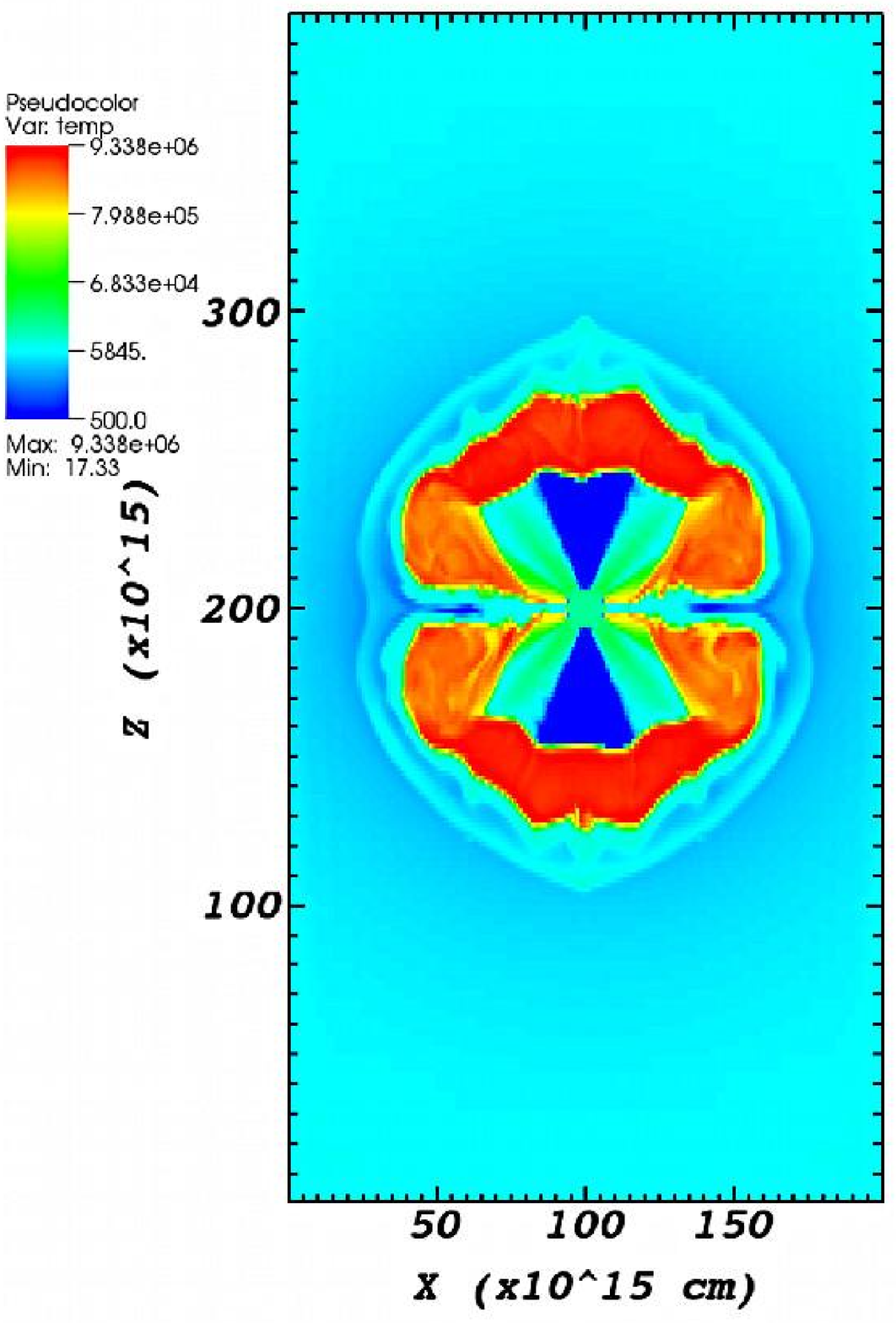}}
\caption{Color-maps of the gas density (in $\g \cm^{-3}$), density and velocity in an enlarged vortex zone, and the gas temperature (in $\K$).
Upper and lower rows are for  $t = 580 \yr$ and $t = 1140 \yr$, respectively.
The velocity arrows in panels \subref{subfigure:arrow1} and \subref{subfigure:arrow2} show only the direction,
while in the panels with enlarged vortex zones they are linear with the magnitude of the velocity. }
  \label{fig:pne}
\end{figure*}

In Figure \ref{fig:pne} we show color-maps of the the gas density, enlargement of the vortex zone, and
the gas temperature, at $t = 580 \rm yr$ and $t = 1140 \rm yr$, in the
upper row (panels \subref{subfigure:arrow1} , \subref{subfigure:vortex1}, and \subref{subfigure:temp1}),
and lower row (panels \subref{subfigure:arrow2} , \subref{subfigure:vortex2}, and \subref{subfigure:temp2}), respectively.
In all plots we show the meridional plane of the nebula. We recall that the simulation is full 3D, and include
both side (no mirror symmetry is assumed in this section).

The following features should be noticed.
\begin{enumerate}
\item The two opposite jets form a bipolar structure of two `fat bubbles'. By fat bubbles we refer to more or less spherical
bubbles attached to their origin. We will find this structure in the simulations of bubbles in clusters of galaxies in the next section.
\item A dense thin shell is formed around the bubbles. This is also seen in bubbles in clusters of galaxies.
\item A structure of two shocks and a contact discontinuity between them is formed. The forward shock runs at a very low
Mach number into the slow AGB wind. The jet is shocked in the reverse shock to very high temperatures. The result is that the bubble is
much hotter than the shocked ambient gas. The same holds in bubbles in clusters, but not in bubbles in CCSNe (see Table \ref{table:compare}).
\item Vortices are formed on the sides of the bubbles. We will encounter much prominent vortices in our 2D simulations of bubbles in clusters.
\item The vortices to the sides of the bubbles compress gas to a thin disk in the equatorial plane.
This is best seen in panels \subref{subfigure:vortex1} and  \subref{subfigure:vortex2}.
The vortices not only form the disk-like structure, but also slow it down.
This process was previously studied numerical by \cite{Akashi2008}, who also suggested that part of the
equatorial dense gas might flow back to the center and form a large (up to hundreds of AU) Keplerian disk;
Planets might be formed in such Keplerian disks \citep{Perets2010}.
\end{enumerate}

\section{Jet feedback in galaxies and clusters}
\label{sec:ICM}
\subsection{Accretion}
\label{subsec:accretion}

It is widely accepted that feedback powered by active galactic nuclei (AGN) has a key role in galaxy formation and
in cooling flows (CFs) in galaxies and in clusters of galaxies.
In galaxy formation AGN feedback heats and expels gas (e.g., \citealt{Ostriker2010} and references therein),
and by that can determine the correlation between the central SMBH mass and some
properties of the galaxy. In cooling flow clusters jets launched by the SMBH heat the gas and maintain
a small, but non zero cooling flow (see review by \citealt{McNamara2012, Fabian2012});
this is termed a moderate cooling flow in \cite{Sokeretal2001}.

There is a dispute on how the accretion onto the SMBH occurs, in particular in cooling flows.
One side argues for accretion to be of hot gas via the Bondi accretion process
(e.g., \citealp{Allen2006, Russell2010, Narayan2011}),
while the other side argues that the accretion is of dense and
cold clumps in what is termed the cold feedback mechanism \citep{Pizzolato2005}.
The cold feedback mechanism has been strengthened recently by observations
of cold gas and by more detailed studies
\citep{Revaz2008,Pope2009,Wilman2009,Wilman2011,Nesvadba2011,Cavagnolo2011,Gaspari2012a,Gaspari2012b,
McCourt2012,Sharma2012,Farage2012,Kashi2012}.

The Bondi accretion process, on the other hand, suffers from several problems \citep{McNamara2011, Cavagnolo2011, Soker2009}.
We point out yet another problematic point with it.
In a recent paper, \citet{Wong2011} resolved the region within the Bondi accretion radius of the S0 galaxy NGC~3115.
If the density and temperature profile is interpreted as resulting from a Bondi
accretion flow into the $M_{\rm BH}=2 \times 10^9 M_\odot$ central SMBH, the derived accretion rate is
$\dot M_B=2.2 \times 10^{-2} M_\odot \yr^{-1}$.
They note that for a radiation power of $0.1 \dot{M_B}\, c^2$, the expected accretion luminosity is
six orders of magnitude above the observed upper limit.
They attribute this to a process where most of the inflowing gas is blown away.

We take a different view. We argue that the Bondi accretion flow is not relevant for the
conditions in typical galaxies and clusters of galaxies.
The reason is that one cannot assume a zero pressure at the center, either because of
stellar winds or because of jets and winds blown by the AGN.

\begin{figure}
\centering
\includegraphics[width=80mm]{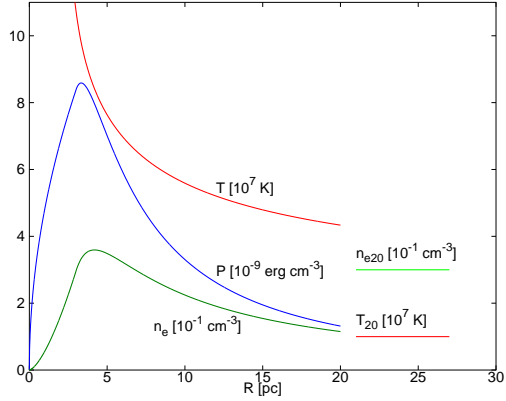}
\caption{The average density, pressure and temperature of the hot bubble as a function of radius.
There are two free parameters: the fraction of the escape velocity in which the hot gas leaves the hot bubble,
and the fraction of mass from the winds that is incorporated into the bubble.
The parameters used here are from \citep{HillelSoker2012}.
The pressure of the hot bubble at $r=20 \pc$ about equals the ISM pressure of NGC~3115.
}\label{fig:hotbubble}
\end{figure}

In application to the galaxy NGC~3115, we \citep{HillelSoker2012} have estimated the average
density and pressure of a hot bubble around the SMBH. The hot bubble is formed by stellar winds.
In Fig. \ref{fig:hotbubble}, the average density, pressure and temperature of the hot bubble are shown as a function of radius.
The pressure of the shocked stellar winds of the high-velocity circum-SMBH stars can be larger than the
ISM pressure near the center.
This accounts, we argue, for the accretion rate of NGC~3115 being much lower than the Bondi accretion rate \citep{Wong2011}.

The average density of the hot shocked stellar wind within the bubble of radius $R$ is given by
\begin{equation}
\rho_w \simeq  \left( \frac{4 \pi}{3} R^3 \right)^{-1} \frac{R}{\beta u_\ast(R)}
\int_0^R  {\left[ n_\ast(r) \eta \dot{m_\ast} \right] 4 \pi r^2 dr},
\label{eq:rho1}
\end{equation}
and the pressure is
\begin{equation}
\begin{split}
P_{e\ast} \simeq & \frac{2}{3}
\left( \frac{4 \pi}{3} R^3 \right)^{-1}
\frac{R}{\beta u_\ast (R)} \\
& \times \int_0^R{\left[ \frac{1}{2} n_\ast (r) \eta \dot{m_\ast} u_\ast^2 (r) \right] 4 \pi r^2 dr}.
\label{eq:pe2}
\end{split}
\end{equation}
Here, $u_\ast(r)$ is the velocity of a star at a distance $r$ from the SMBH, $\beta u_\ast(R)$ is the velocity at which the hot gas leaves the hot bubble, $\dot m_\ast$ is the average
mass loss rate per star, $n_\ast (r)$ is the number density of stars near the SMBH,
and $\eta$ is the fraction of mass from the stellar winds that is incorporated into the bubble.
For the stellar number density in NGC~3115 we take (based on \citealt{Kormendy1996})
\begin{equation}
n_\ast =  5\times 10^5 \pc^{-3}
\begin{cases}
1, & r \le 3 \pc \\
(r / 3 \pc)^{-3}, & r > 3 \pc.
\end{cases}
\label{eq:nast1}
\end{equation}

In conclusion, we find that in some cases this pressure exerted by the stellar winds is significant and
can substantially suppress the inflow of the ISM relative to what a simple Bondi accretion would give.
Shocked winds of circum-SMBH high-velocity stars form a bubble of hot gas whose pressure is significant.
There are some uncertainties in the model, such as the exact behavior of the stellar mass loss,
trajectories of stars around the SMBH, and the stochastic behavior of the post-shock stellar winds.
Some of these will be studied in future numerical simulations. However, the result that the stellar winds cannot be
ignored is robust.

For some values of the parameters we found that a situation might arise where the hot bubble's density is lower than the ISM density.
In this case Rayleigh-Taylor (RT) instability takes place, and a density-inversion layer is formed.
Such a structure, we claim, is similar to the density inversion found in the outer atmosphere of
red giant stars (e.g., \citealt{Harpaz1984, Freytag2008}), but not identical.
Although hot tenuous gas buoys outward and dense ISM gas moves inward, the density-inversion layer itself continues to exist.
The ISM gas is heated near the center and accumulated to the hot bubble.

Our result is more general in showing that in many cases the Bondi accretion process does not work because
one of its basic assumptions, that there is no central pressure, breaks down.
This is one of several reasons why the Bondi accretion model is not applicable in many cases.

We note that the considerations above cannot be applied straightforward
to galaxies like the Milky Way where the SMBH mass is relatively low.
The reason is that the mass of SMBH at the center of the Galaxy is much smaller than that in NGC~3115,
while the stellar velocity dispersions $\sigma$ are similar, as well as the stellar densities.
The radius of influence of the SMBH, $G M_{\rm BH}/ \sigma^2$, in NGC~3115 is large, $\sim 30 \pc$,
and contains many stars with significant total mass loss rate that can act against the ISM pressure.
In the Milky Way the radius of influence is only $\sim 0.1 \pc$. There are many stars with large mass loss rate
outside this regions. There is continuous mass loss rate by stars from large distances inward.
This is seen in the numerical simulations of \citealt{Cuadra2006}.
We note, though, that when they take a smaller inner radius in
their simulations, the accretion rate decreases. We interpret this as the building of pressure near the center.

\subsection{Bubbles inflation}
\label{subsec:bubbles}

Bubbles (cavities) devoid of X-ray emission, mostly as opposite pairs, are
observed in a large fraction of cooling flow (CF) clusters and groups of
galaxies, as well as in CF elliptical galaxies (e.g., \citealt{Dong2010}).
These bubbles are inflated by jets launched from the central active galactic nuclei (AGN),
as evident by the radio emission that fills most bubbles.
In some cases two opposite chains of bubbles that are close to each other, and even overlap, are observed,
as in Hydra A \citep{Wise2007}, and in two bubbles in the galaxy group NGC 5813 \citep{Randall2011}.
These chains were usually attributed to several episodes of jet activity.

\cite{Refaelovich2012} suggest another plausible mechanism for cavity chains creation.
Several fragmentation mechanisms act on the primary vortex that
created just behind the jet's head and split it to several smaller vortices.
These mechanisms include Kelvin-Helmholtz and Rayleigh-Taylor instabilities which the jet's cocoon flow is prone to.
Using hydrodynamical simulations with a continuous jet \cite{Refaelovich2012} show that such a fragmentation process
takes place for a wide range of jet parameters,  and causes  the fragments to appear like a chain
of cavities in the X-ray brightness map.

In the current paper we present new results which may help to discriminate observationally between the two mechanisms,
 vortex fragmentation and multiple jet-launching episodes.

To emphasize the differences between the above mentioned mechanisms we repeat the standard simulation presented by \cite{Refaelovich2012}
with the difference that the jet is switched on and off. The activity phase last $20\Myr$
followed by a pause of $50\Myr$. We simulate four such cycles spanning a period of $280\Myr$.
The total two jets power is $P_{2j} = 4 \times 10^{45} \erg \s^{-1}$, the jets Mach number is $15$ relative to the ambient medium,
their velocity is $v_j = 1.33 \times 10^{4} \km \s^{-1}$, and the half-opening angle of each jet is $\alpha=30^{\circ}$.

We use the PLUTO hydrodynamic code \citep{Mignone2007}.
The simulations are 2.5D in the sense that we use a spherical coordinate system, but impose cylindrical symmetry,
hence calculating the flow with a 2D polar grid $(r,\theta)$.

\begin{figure*}
  \centering
 \subfigure[][]{\label{subfigure:d_epi_20}\includegraphics*[scale=0.5,clip=true,trim=80 0 80 0]{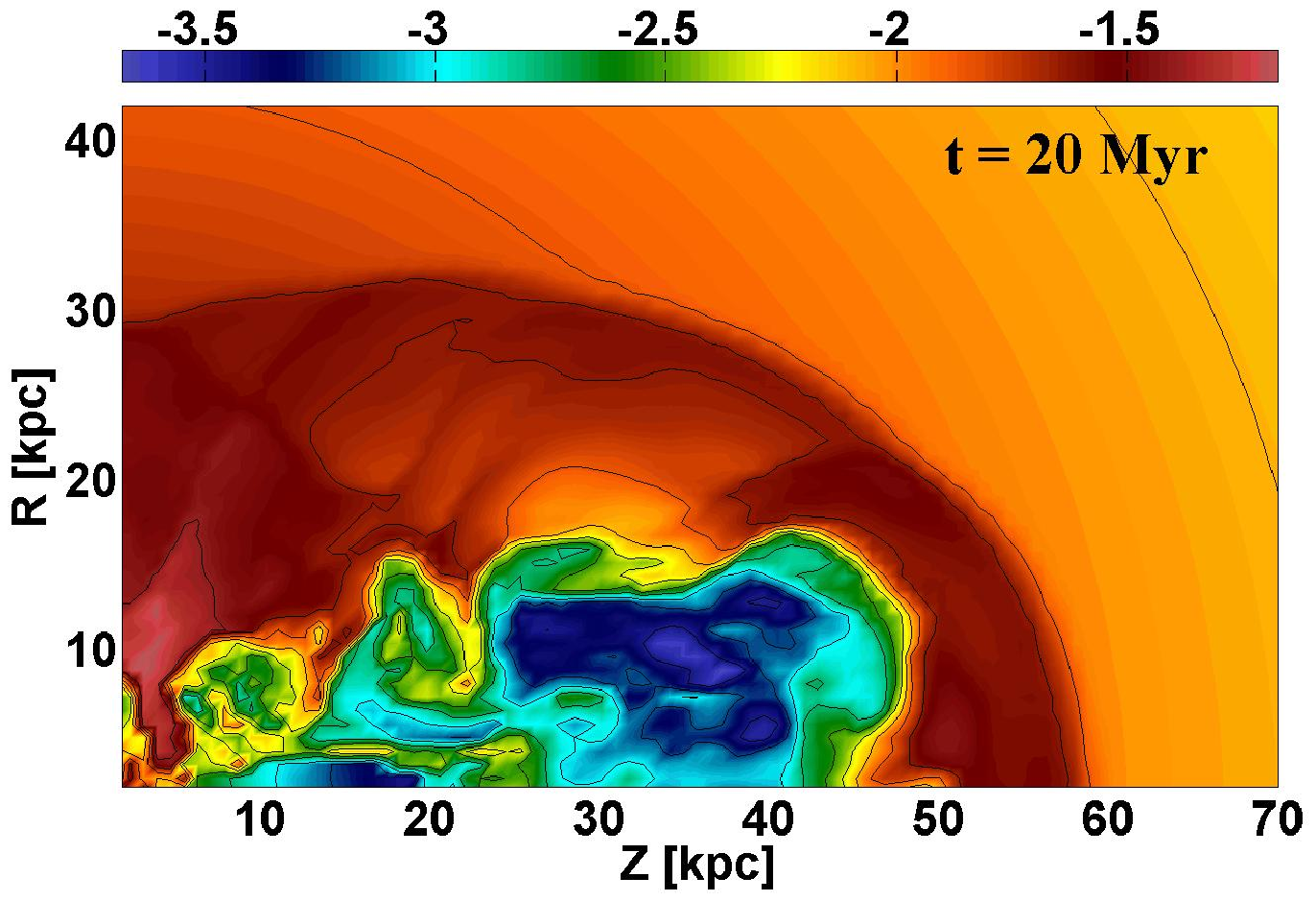}}
 \subfigure[][]{\label{subfigure:d_std_20}\includegraphics[scale=0.5,clip=true,trim=80 0 80 0]{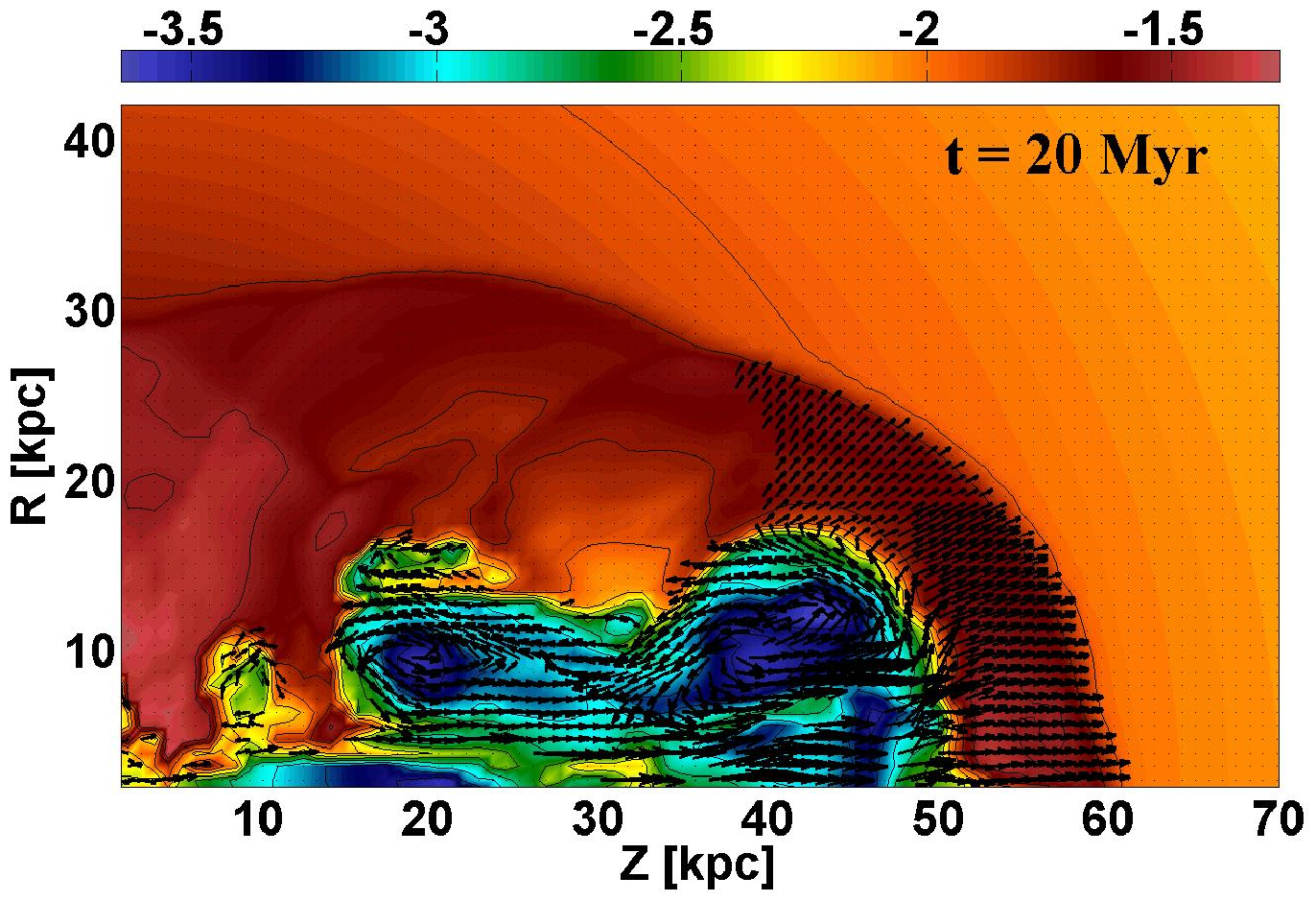}}\\
 \subfigure[][]{\label{subfigure:d_epi_69}\includegraphics*[scale=0.5,clip=true,trim=80 0 80 0]{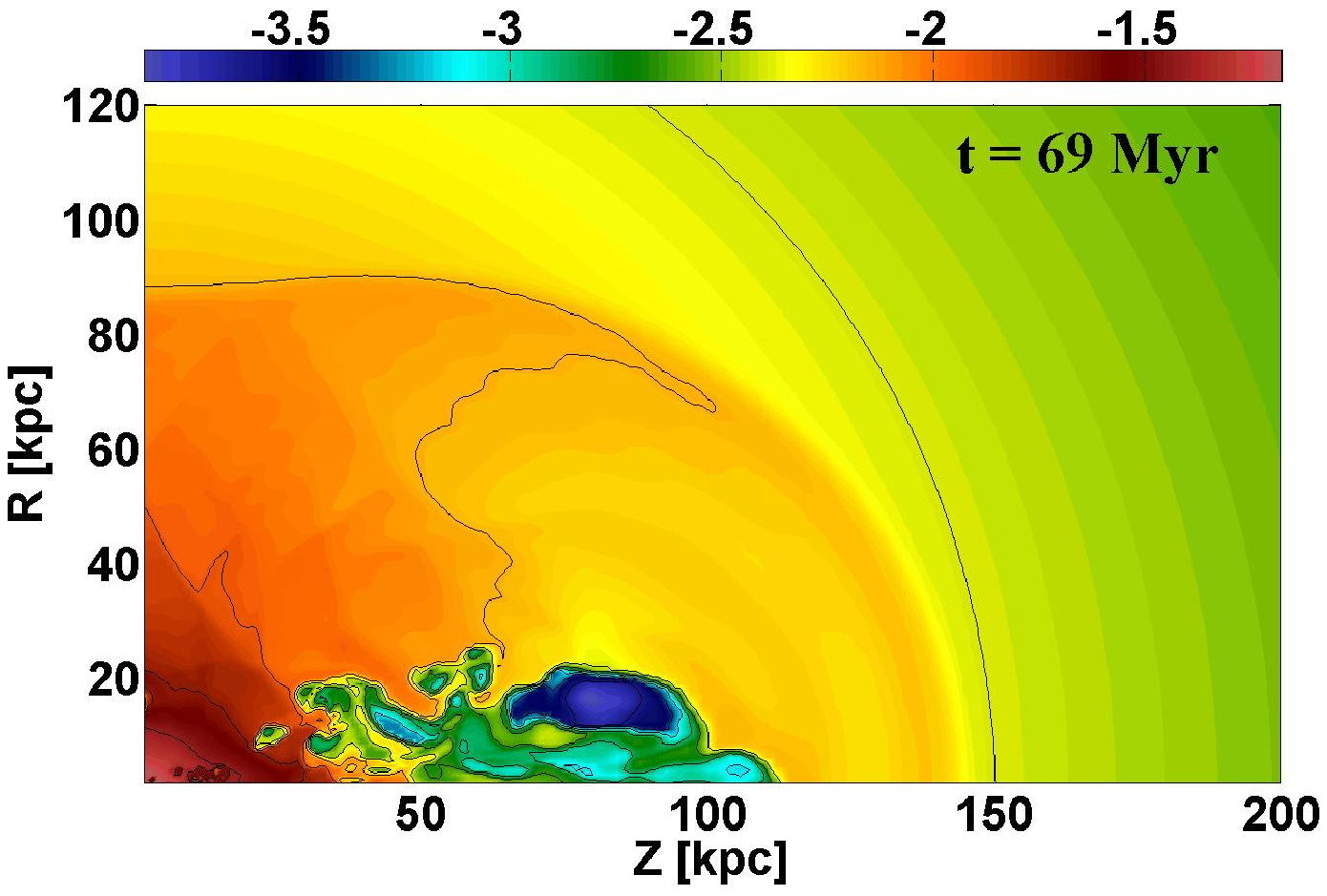}}
 \subfigure[][]{\label{subfigure:d_std_69}\includegraphics*[scale=0.5,clip=true,trim=80 0 80 0]{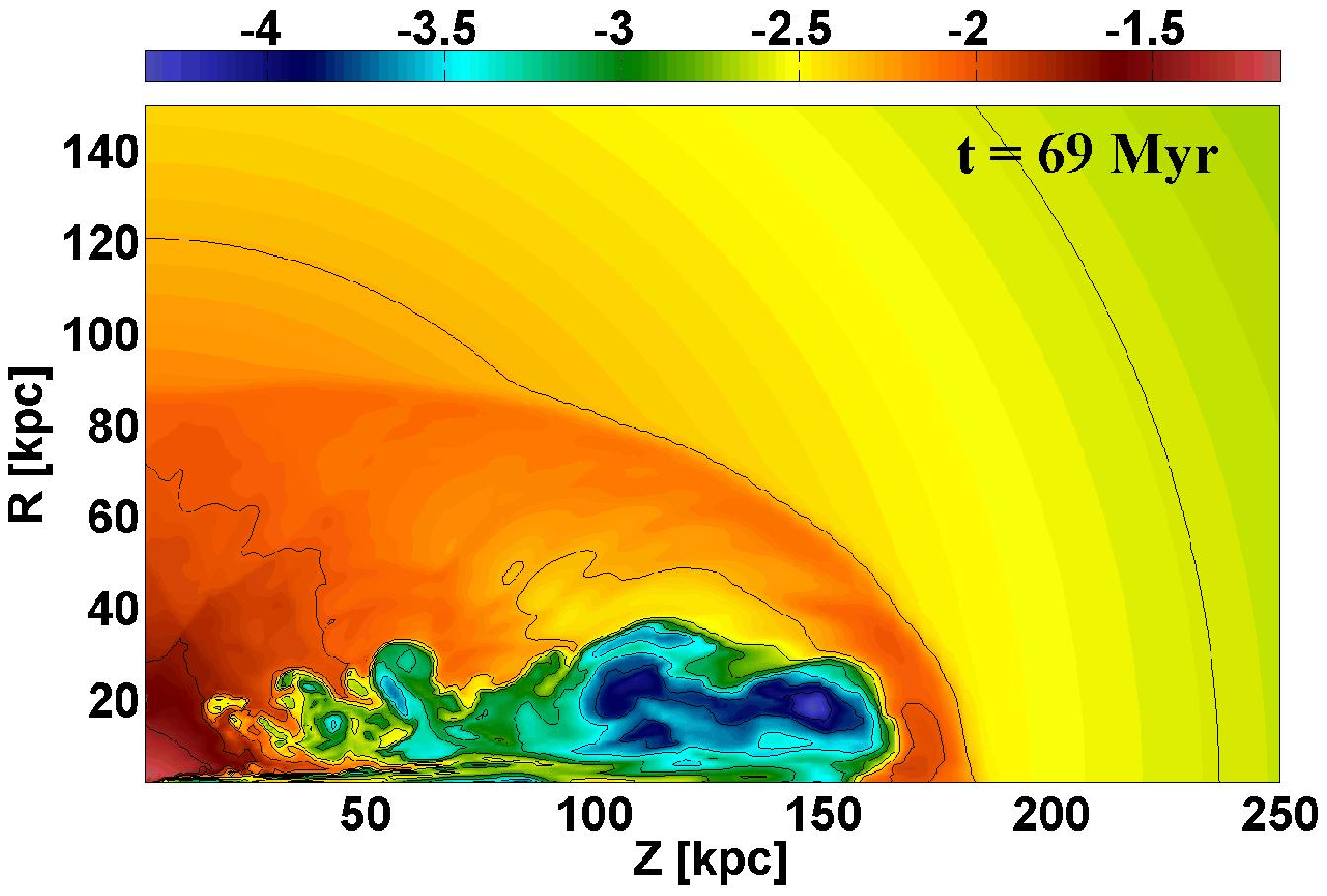}}\\
 \subfigure[][]{\label{subfigure:d_epi_180}\includegraphics*[scale=0.5,clip=true,trim=80 0 80 0]{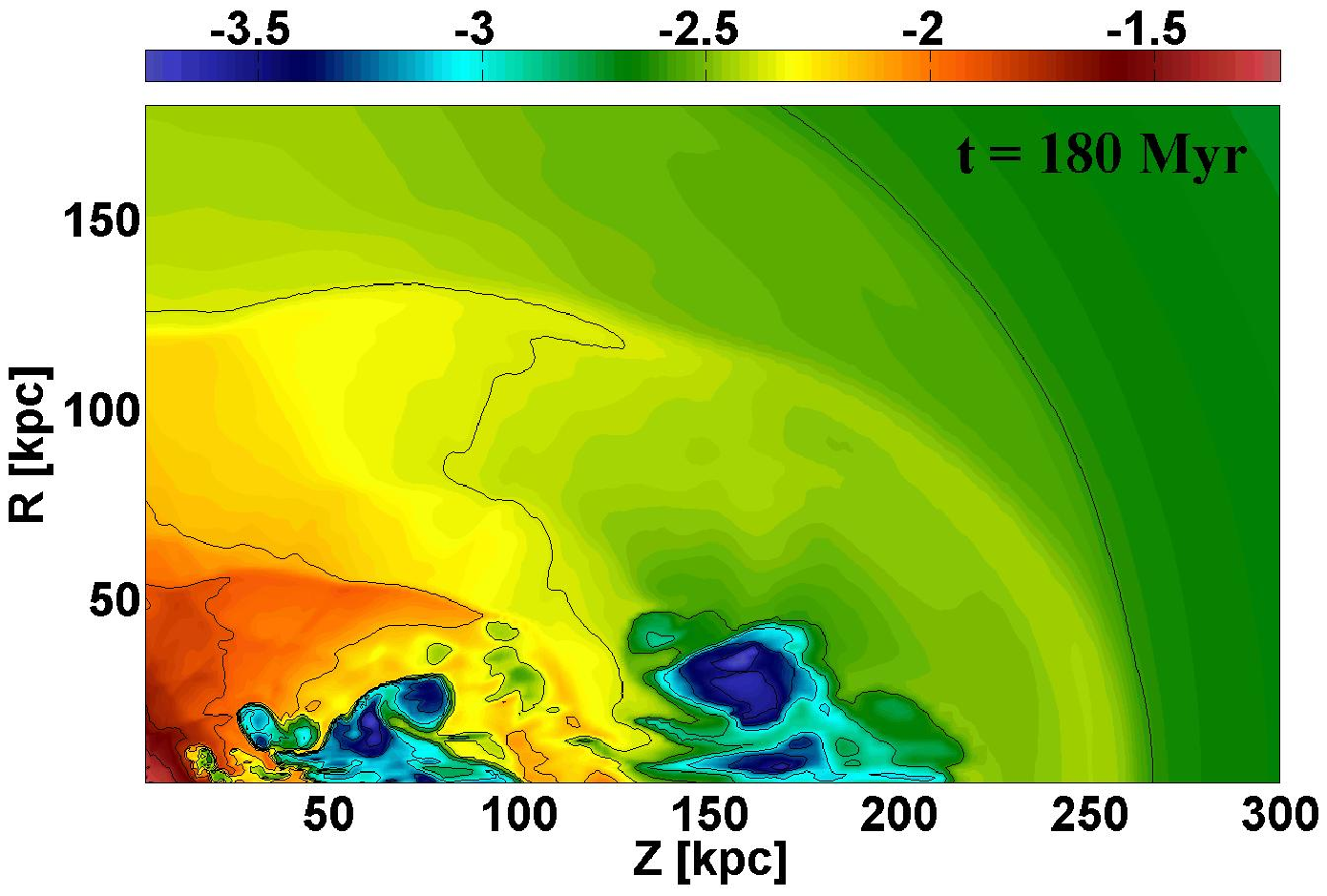}}
 \subfigure[][]{\label{subfigure:d_std_100}\includegraphics*[scale=0.5,clip=true,trim=80 0 80 0]{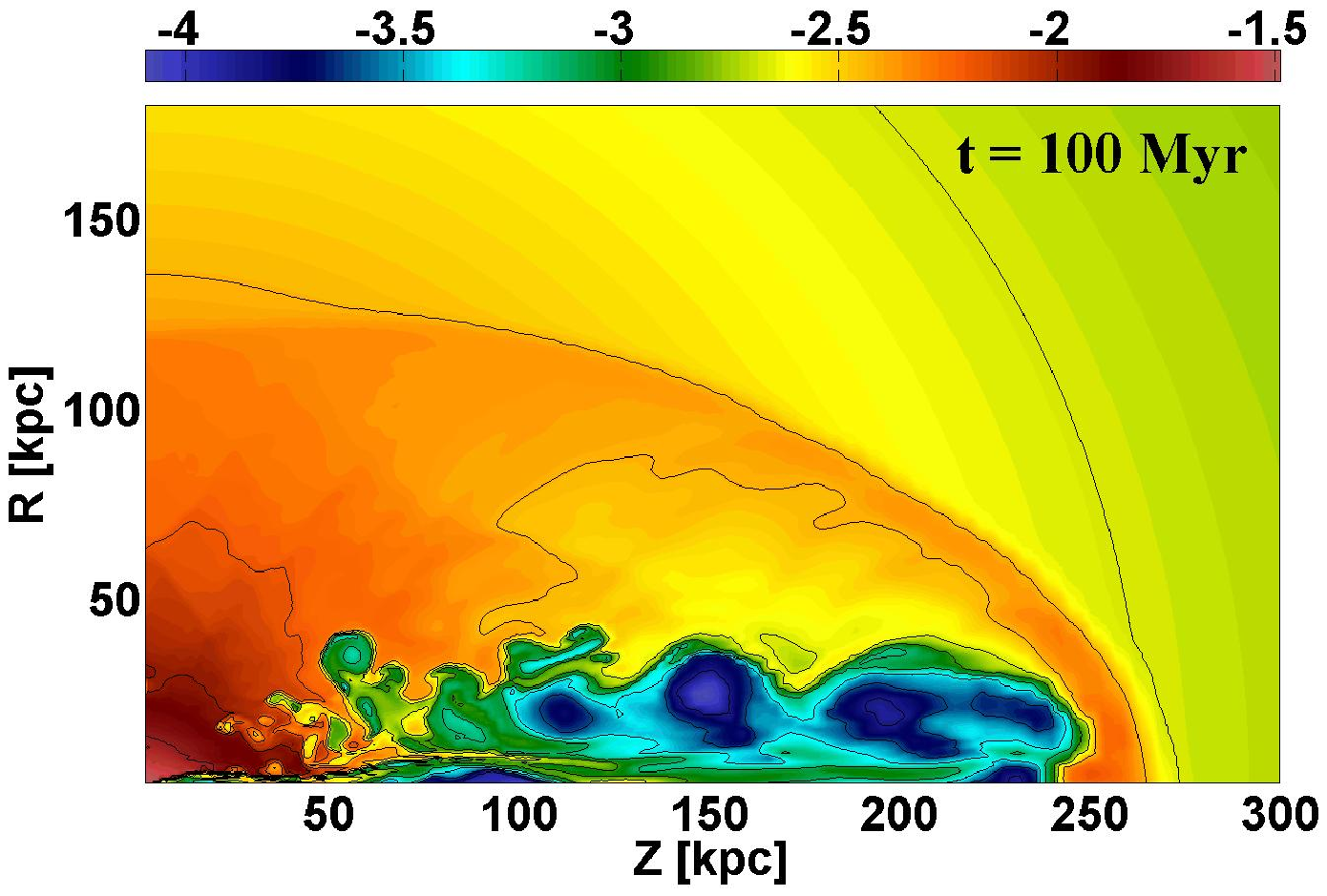}}

\caption{Colormaps of electron density in logarithmic scale and units of $\cm^{-3}$.
Results of the continuous (\subref{subfigure:d_std_20},\subref{subfigure:d_std_69},
\subref{subfigure:d_std_100})  and switched (\subref{subfigure:d_epi_20},\subref{subfigure:d_epi_69}, \subref{subfigure:d_epi_180}) runs at different times.
 Axes are in $\kpc$.  The horizontal boundary (labelled $Z$) is the symmetry
  axis of the two jets. The left vertical boundary of the figure
  (labelled $R$) is taken along the mirror-symmetry plane of the flow $z=0$; it is termed the equatorial plane.
  The jet is injected at $r=1 \kpc$, a region that is not well resolved in the
 figure. In panel \subref{subfigure:d_std_20} there are velocity arrows depicting
 the vortices present inside the bubbles.
 There are 3 velocity bins: $500-1000 \km \s^{-1}$, $1000-5000 \km \s^{-1}$ and $5000-10^4 \km \s^{-1}$ -  from shortest to longest.
 Velocities below $500 \km \s^{-1}$ and above $10^4 \km \s^{-1}$ do not appear in the figure for clarity.
}
  \label{fig:dens_epi_std}
\end{figure*}

\begin{figure*}
  \centering
 \subfigure[][]{\label{subfigure:x_epi_20}\includegraphics*[scale=0.5,clip=true,trim=50 0 70 0]{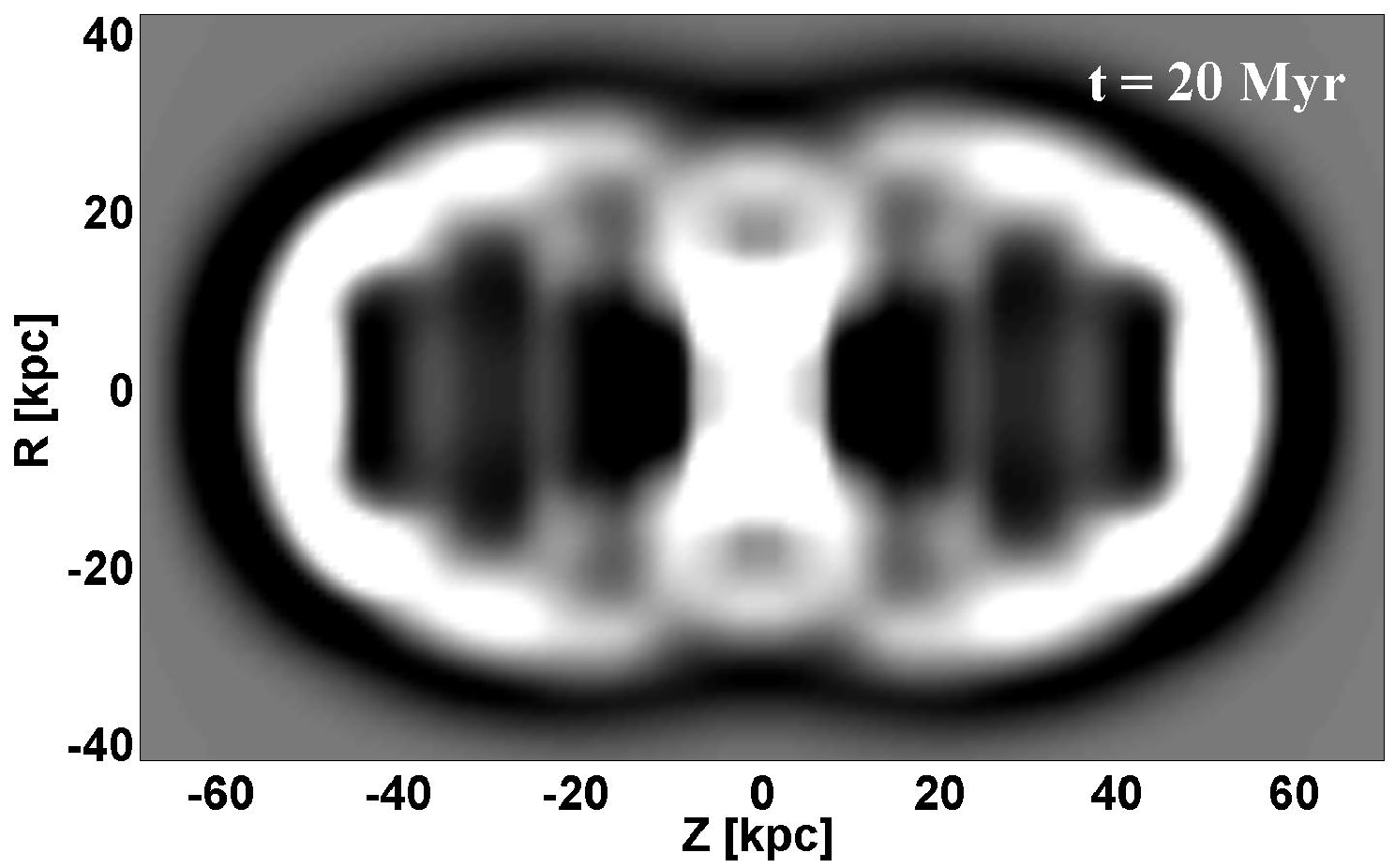}}
 \subfigure[][]{\label{subfigure:x_std_20}\includegraphics[scale=0.5,clip=true,trim=50 0 70 0]{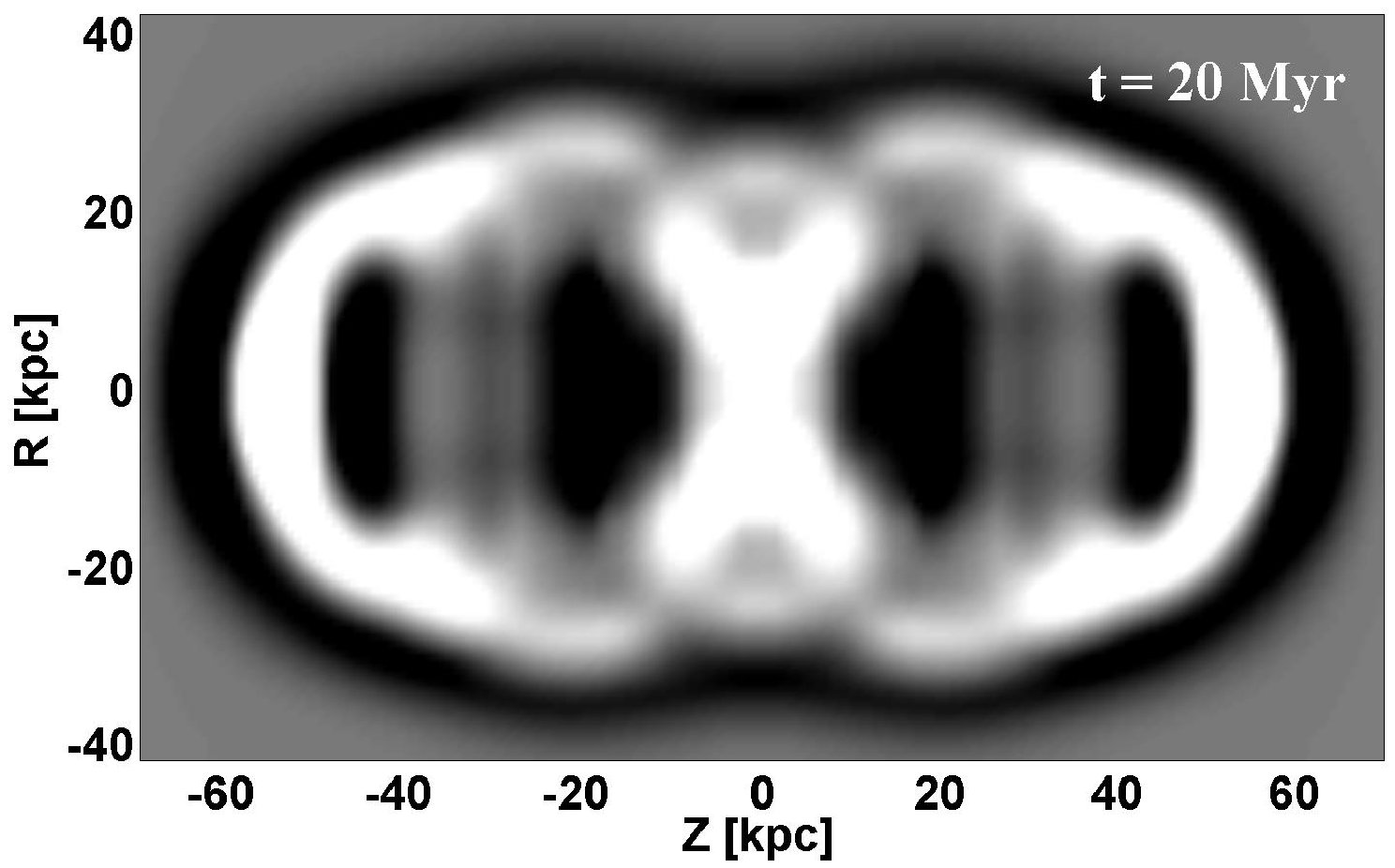}}\\
 \subfigure[][]{\label{subfigure:x_epi_69}\includegraphics*[scale=0.5,clip=true,trim=50 0 70 0]{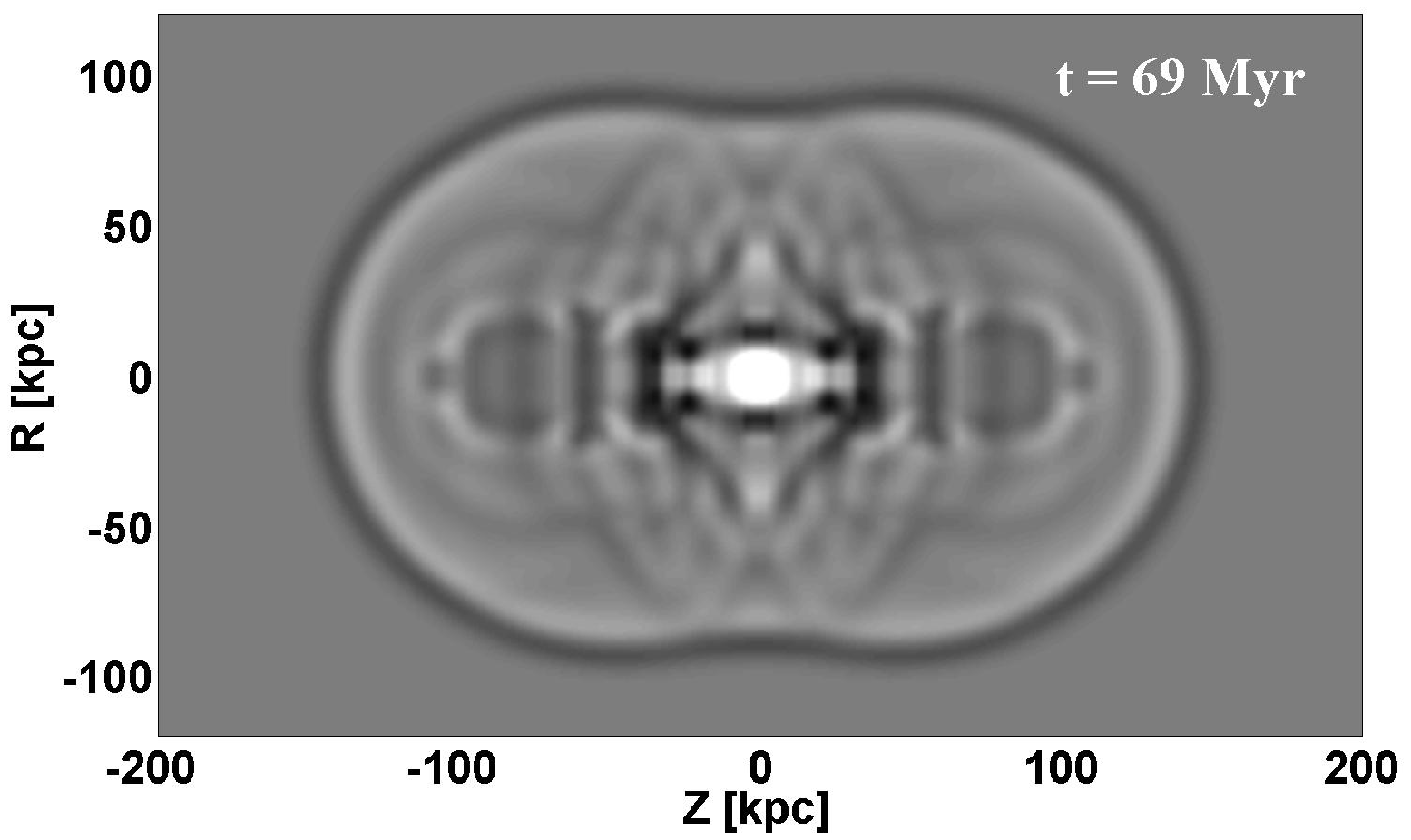}}
 \subfigure[][]{\label{subfigure:x_std_69}\includegraphics*[scale=0.5,clip=true,trim=50 0 70 0]{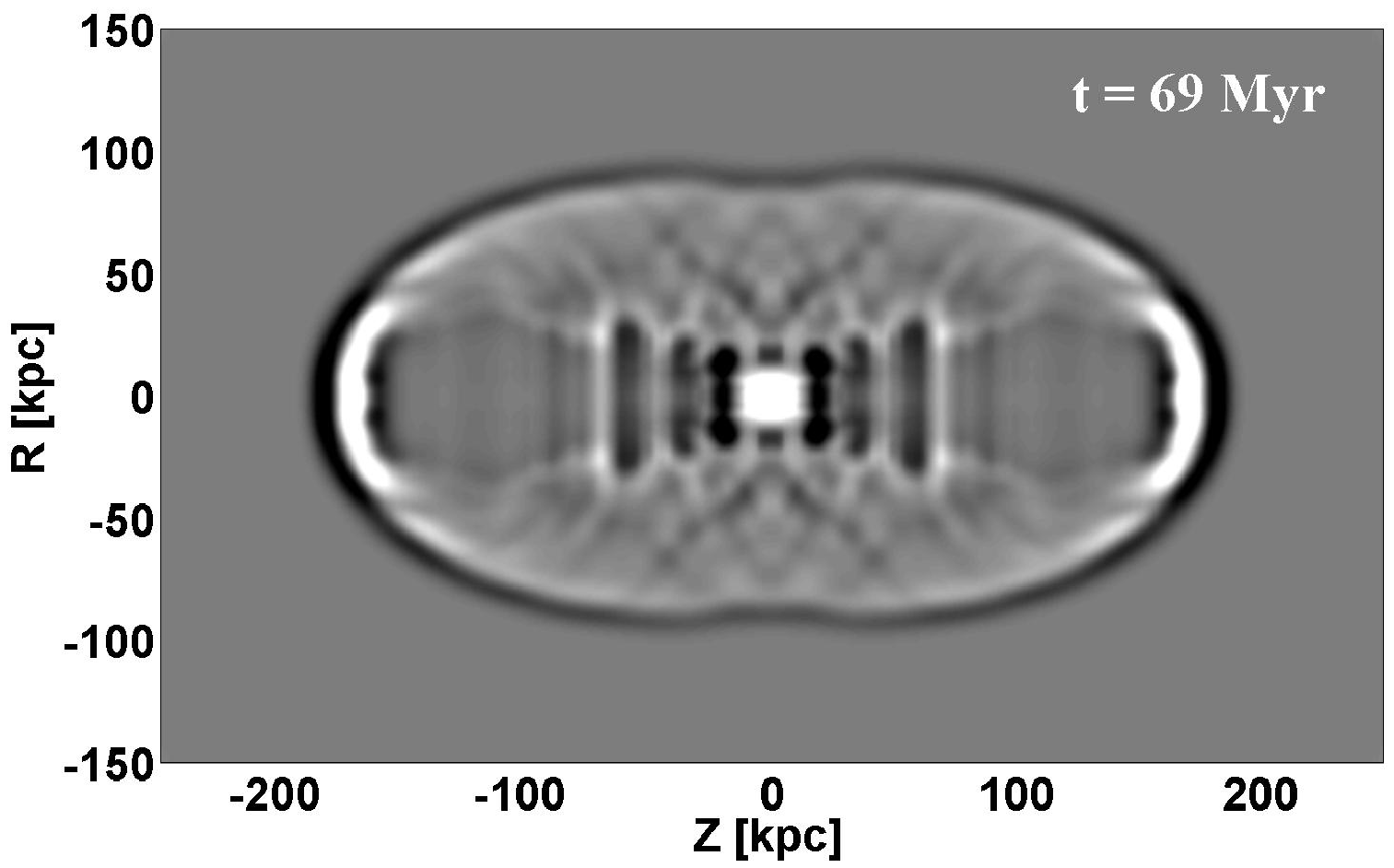}}\\
 \subfigure[][]{\label{subfigure:x_epi_180}\includegraphics*[scale=0.5,clip=true,trim=50 0 70 0]{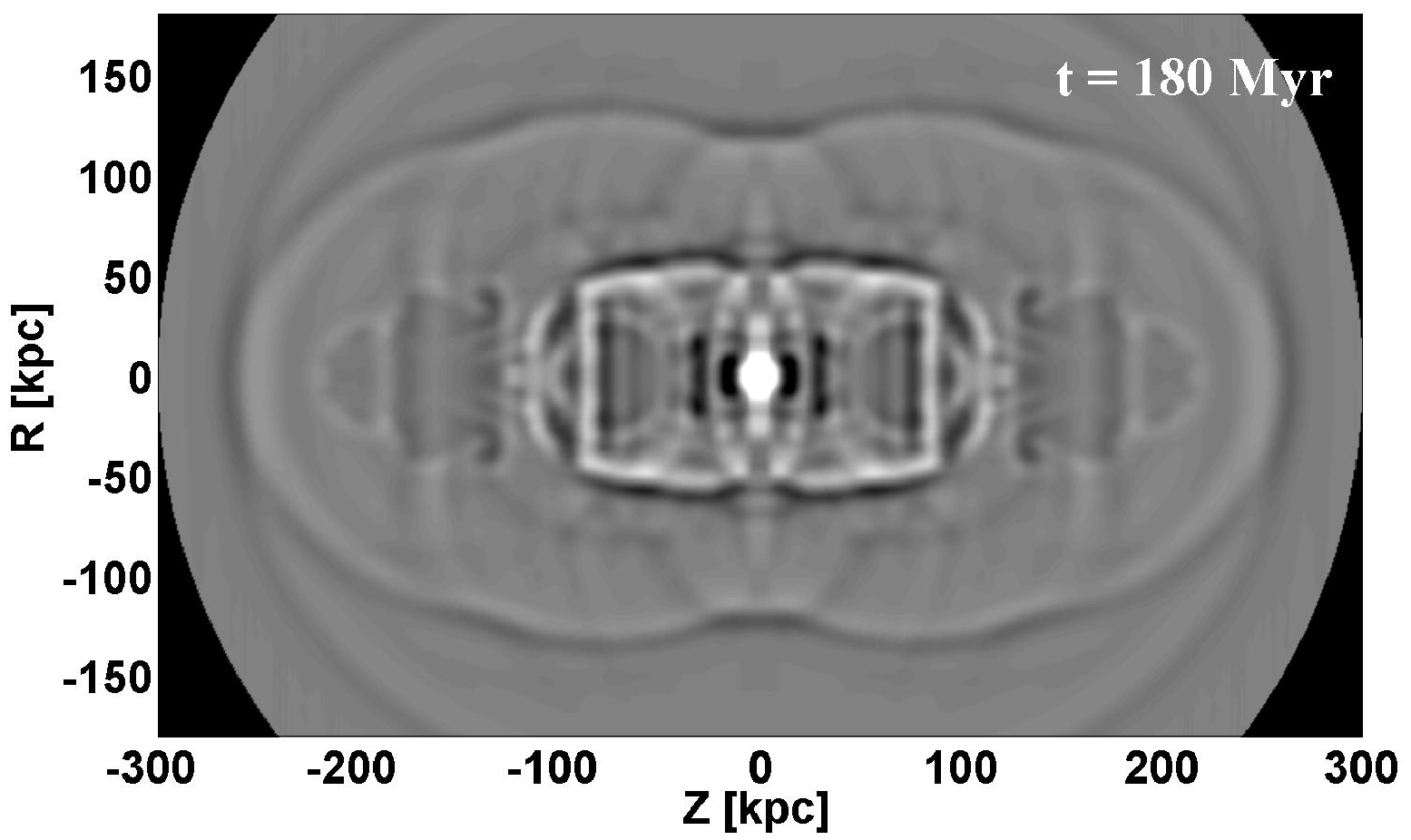}}
 \subfigure[][]{\label{subfigure:x_std_100}\includegraphics*[scale=0.5,clip=true,trim=50 0 70 0]{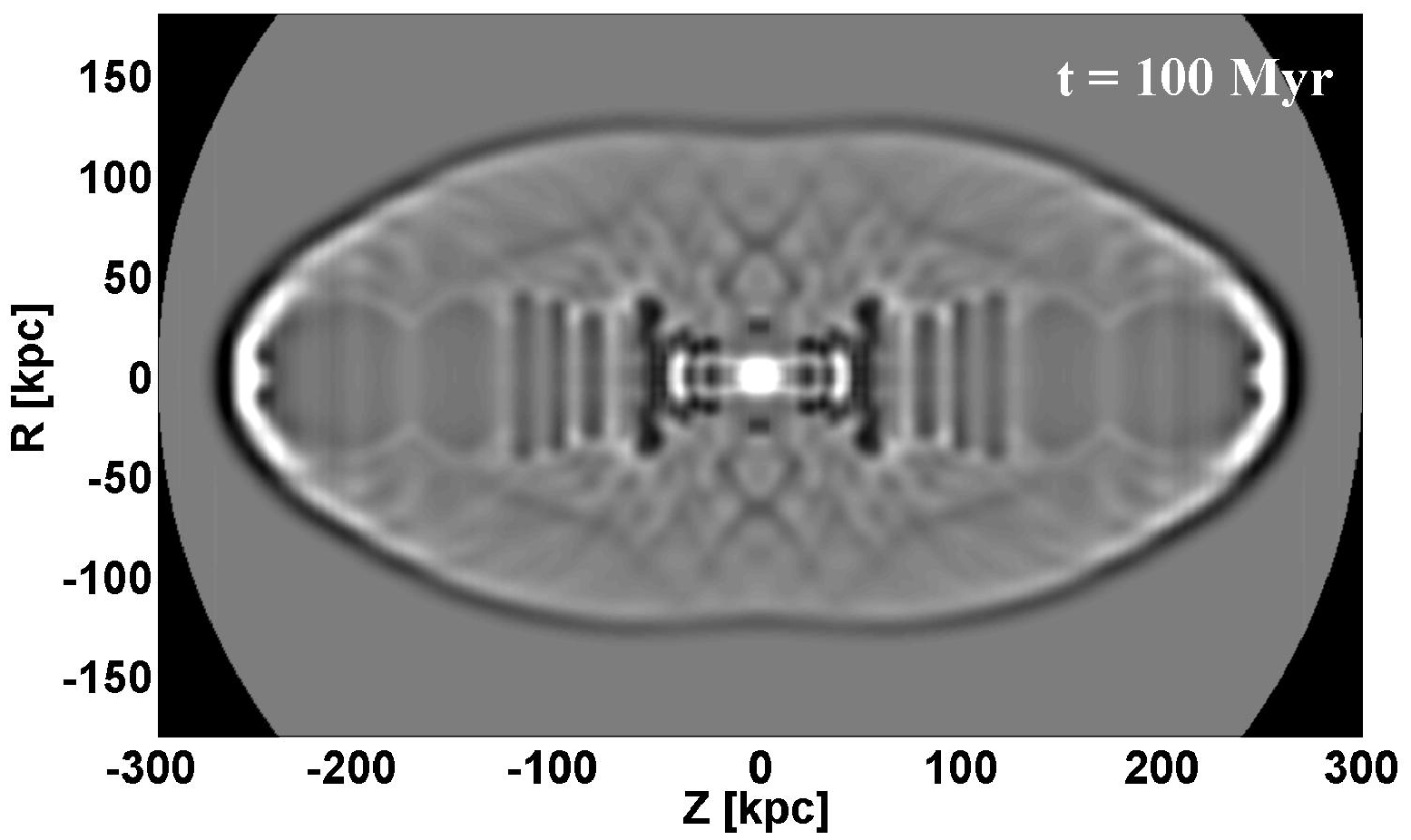}}

\caption{ Synthetic X-ray image of the simulations viewed from the equatorial plane direction (inclination of $90^\circ$) performed by integrating $n_e^2$ along the line of sight. Results of the continuous (\subref{subfigure:x_std_20},\subref{subfigure:x_std_69},
\subref{subfigure:x_std_100})  and switched (\subref{subfigure:x_epi_20},\subref{subfigure:x_epi_69}, \subref{subfigure:x_epi_180}) runs at different times.}
  \label{fig:xray_epi_std}
\end{figure*}

In Figures \ref{fig:dens_epi_std} and \ref{fig:xray_epi_std} we present the
results of the two simulations mentioned above. One run with the continuous jet
and another with the jet switched on and off. Figure \ref{fig:dens_epi_std} shows density color-maps and
Figure \ref{fig:xray_epi_std} shows their simulated X-ray map  counterparts after applying an unsharp filter as described in \cite{Refaelovich2012}.
The figures compare the two runs on  three different stages.
 \begin{enumerate}
\item $t=20\Myr$ - Just before the end of the first activity episode in the switched run.
  The positions of the shocks are identical, the exact cavity structure may be somewhat different due to convergence issue \citep{Refaelovich2012}.
\item $t=69\Myr$ - The end of the first pause episode, i.e., just before the next activity starts.
\item $t=180\Myr$ in the switched run Vs. $t=100\Myr$ in the continuous run. In the switched run it is $20\Myr$
after the end of the 3rd jet launching episode.
 \end{enumerate}

\begin{figure*}
\subfigure{
\includegraphics[scale=0.5,clip=true,trim=80 0 80 0]{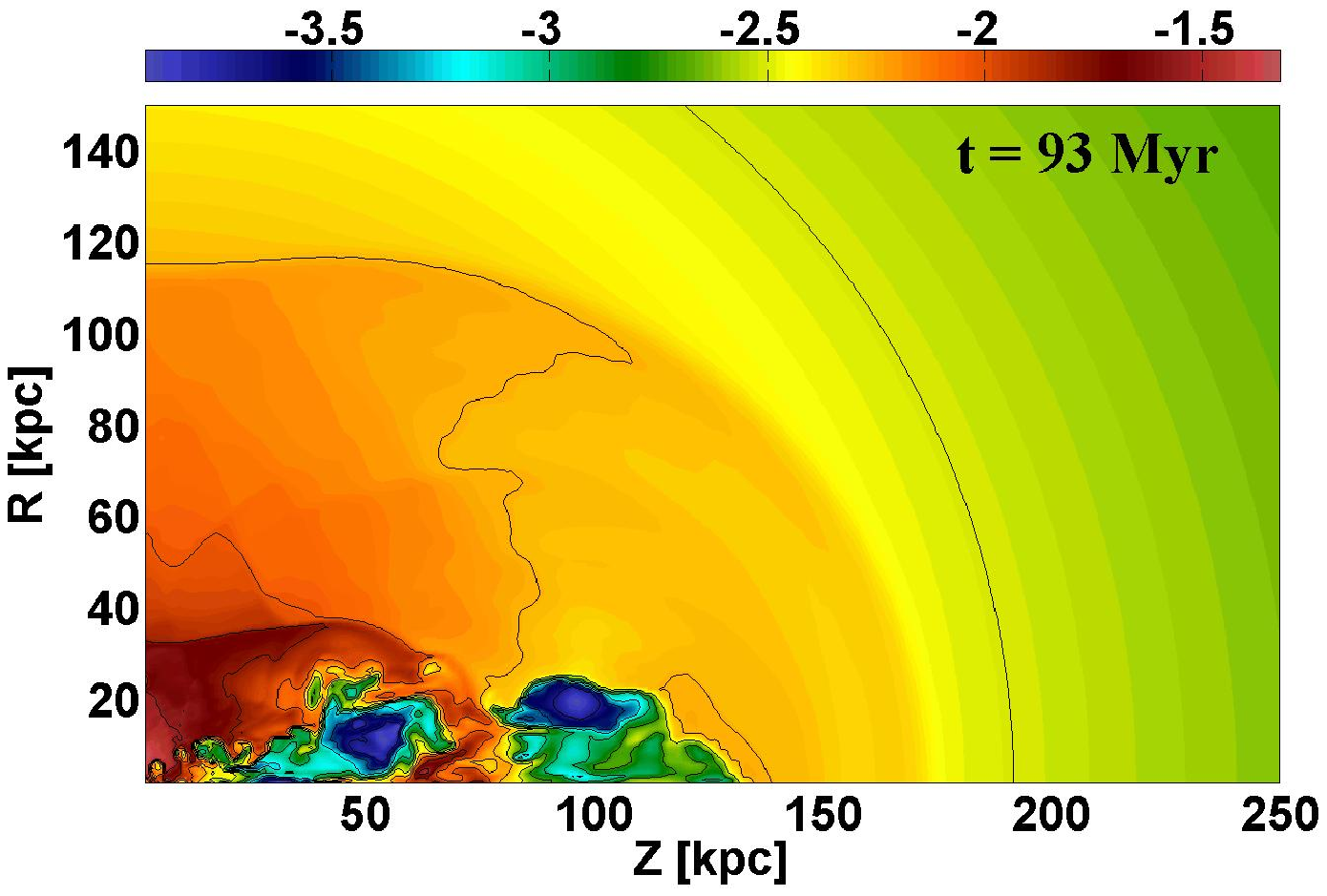}}
\subfigure{
\includegraphics[scale=0.5,clip=true,trim=50 0 70 0]{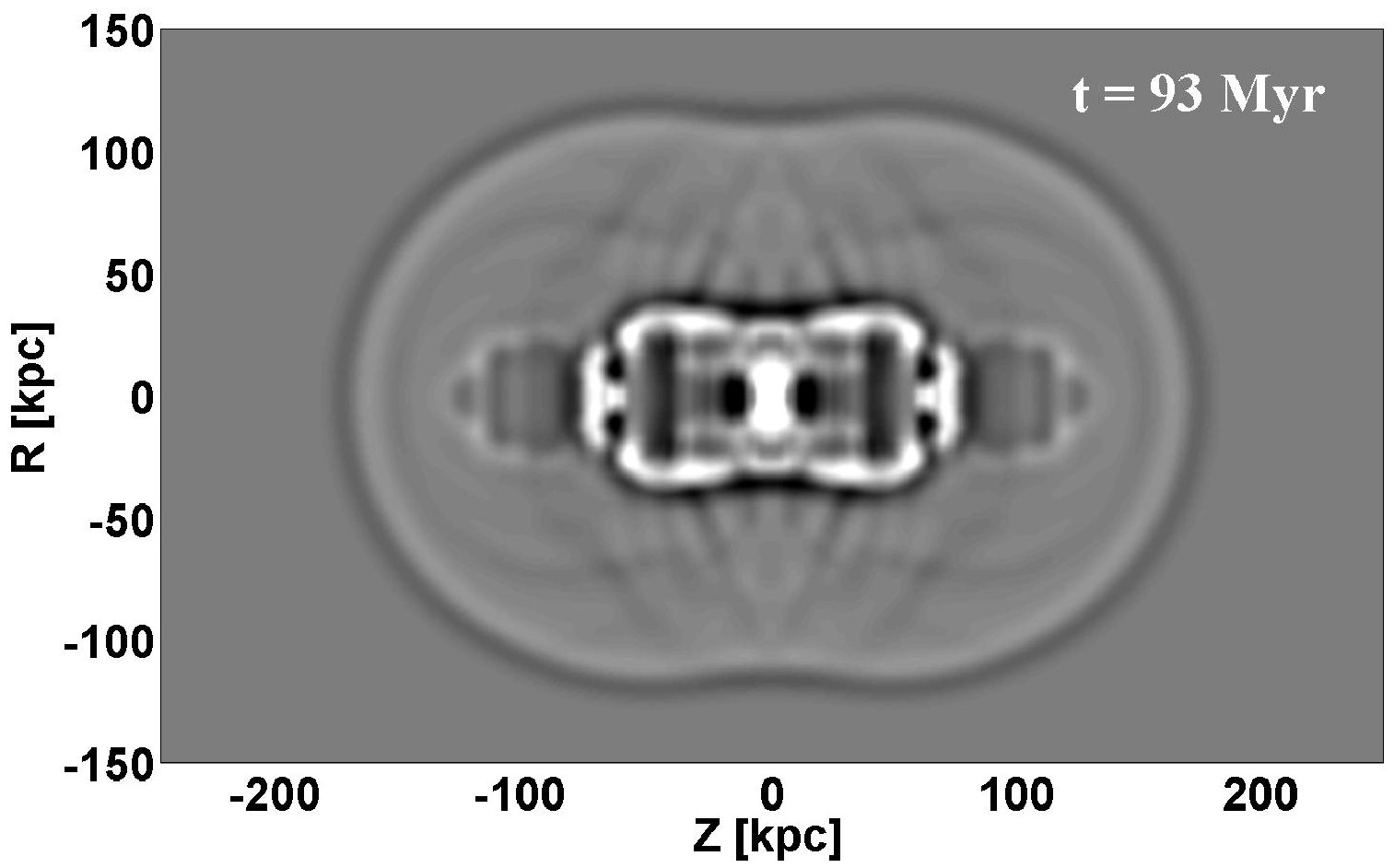}}
\caption{Electron density color-map and its X-ray map counterpart from the switched run (as in Figures \ref{fig:dens_epi_std} and \ref{fig:xray_epi_std}).
This figure shows the run $3\Myr$ after the second jet activity ceased. The first shock is at $z=175\kpc$ and the second shock is at $z \simeq 75\kpc$.
The two bubbles are clearly seen. This figure depicts the strong cavity separation which occurs when the shock from a secondary jet activity episode
is located between the newly inflated bubble and its predecessor from a previous activity episode.}
\label{fig:strongseparation}
\end{figure*}

There are several conclusions following these simulations.
\begin{itemize}
\item \textit{Cavity position relative to the shock front.} During the active phase the primary bubble (cavity) which is created
near the head of the jet is located just behind the shock. The jets' momentum transferred to the bubble keeps the bubble behind the shock.
During the idle phase, the jet stops driving the bubble and the shock separates from the cavity and runs ahead (figs. \ref{subfigure:d_epi_69} and \ref{subfigure:x_epi_69}).
This emphasizes the fact that the outward bubbles' motion is driven  mainly by the momentum of the jet, rather than by buoyancy.
It should be noted that the same process of jet driving and separation will occur with each next jet activity episode.
\item \textit{Positions of secondary shocks.}  Secondary shocks from successive activity episodes catch-up with the previous shocks very fast.
However, they catch-up only in the direction of the jet. In the perpendicular direction the shocks stay separated for much longer times
(they actually do not catch up during the entire simulation time).
This point is illustrated in Figure \ref{subfigure:d_epi_180} and its X-ray counterpart in Figure \ref{subfigure:x_epi_180}.
There are two shocks which catch-up at $z=250\kpc$ on the symmetry axis, followed by a third shock around $ z=125\kpc$.
It may be seen that all the three shocks are well separated along the $R$ direction (note that the outermost shock is partially outside the frame).

Thus, while it is probably impossible to count the number of episodes using the shocks on the chain axis,
it should be possible, at least in principle, to discriminate between them at the perpendicular direction.
\citet{Randall2011} find three such shocks in the galaxy group NGC 5813,
and argue for multiple jet-launching episodes to account for the chain of bubbles there.
\item \textit{Inter-cavity separation.}  When considering the separation between X-ray deficient cavities we should consider two different cases.
In the first case the recent shock didn't catch up yet with the bubble inflated by the previous activity episode.
At this stage we expect to see very sharp separation between the cavities because of the shock front that runs between them
as in Figure \ref{fig:strongseparation}.
Note that the outermost bubble is detached from the shock and therefore not sharply separated from the ICM in the X-ray map.
The second case occurs when a newer shock interacts and disrupts a previously inflated bubble.
At this case the cavity (which consists of a vortex) is fragmented to smaller vortices which may appear as additional cavities.
No sharp separation between the cavities exists in this case, and the image will resemble
the multiple cavities formed by a continuous jet as described by \cite{Refaelovich2012}.
We want to emphasize that both cases can in principle coexist in the same system.
For example, in a case of three episodes when the second shock already disrupted the first
cavity but the third shock didn't reach yet the second cavity as in Figures \ref{subfigure:d_epi_180} and \ref{subfigure:x_epi_180}.
\end{itemize}

\subsection{Heating by mixing}
\label{subsec:heating}

To study the heating of the ICM by jets we performed 2.5D simulations with the code PLUTO.
We launch jets with a wide half-opening angle of $\theta=70^{\circ}$ and a velocity of $9600 \km \s^{-1}$.
To follow the lasting effects of the jets we run the simulations for a significantly longer time
beyond the jet activity phase.

\begin{figure*}
\hskip -1.0 cm
\begin{tabular}{cc}
\hskip 3.3 cm
{\includegraphics*[scale=0.25,clip = true, trim=2cm 0 2cm 0]{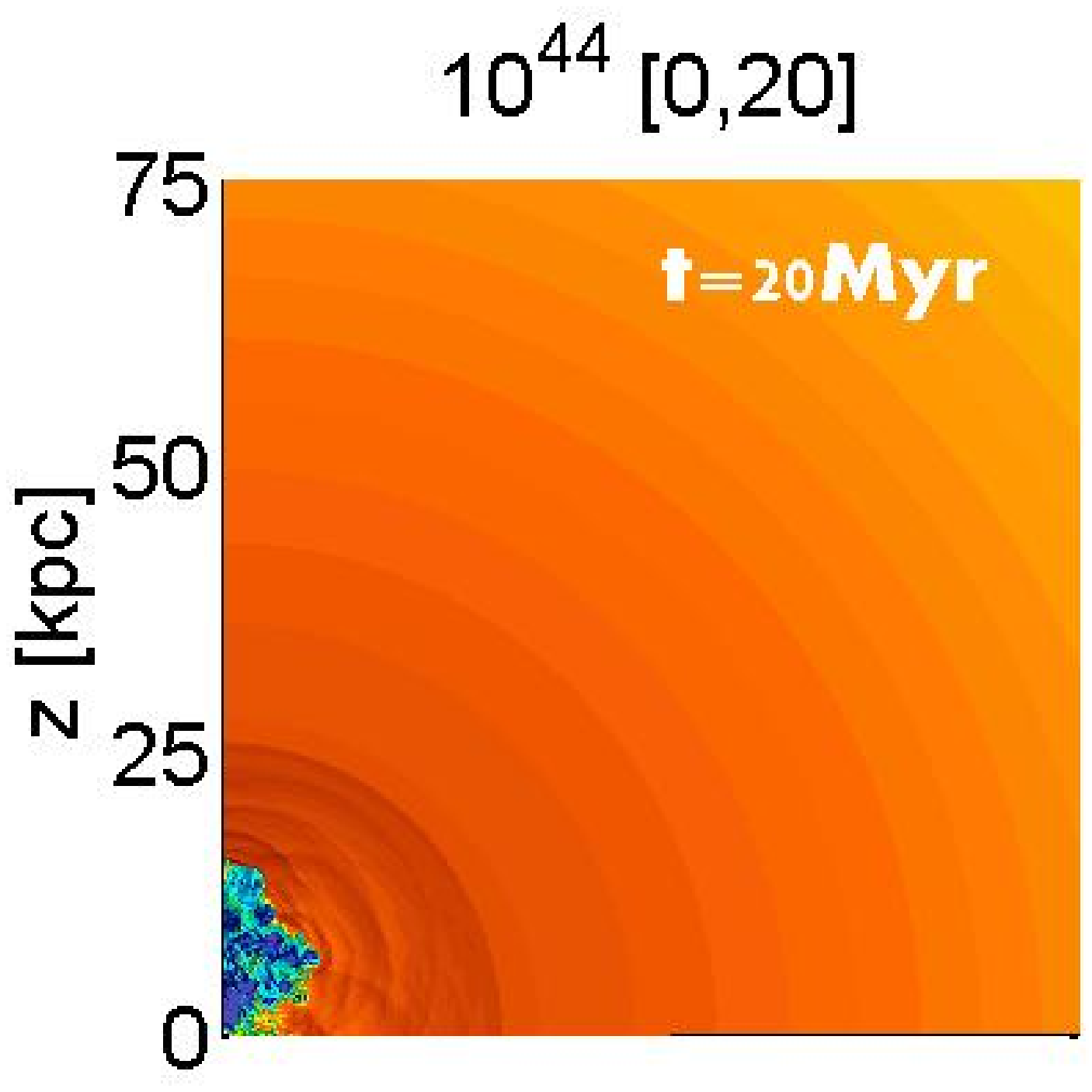}}
{\includegraphics*[scale=0.25, clip = true,trim=2cm 0 2cm 0]{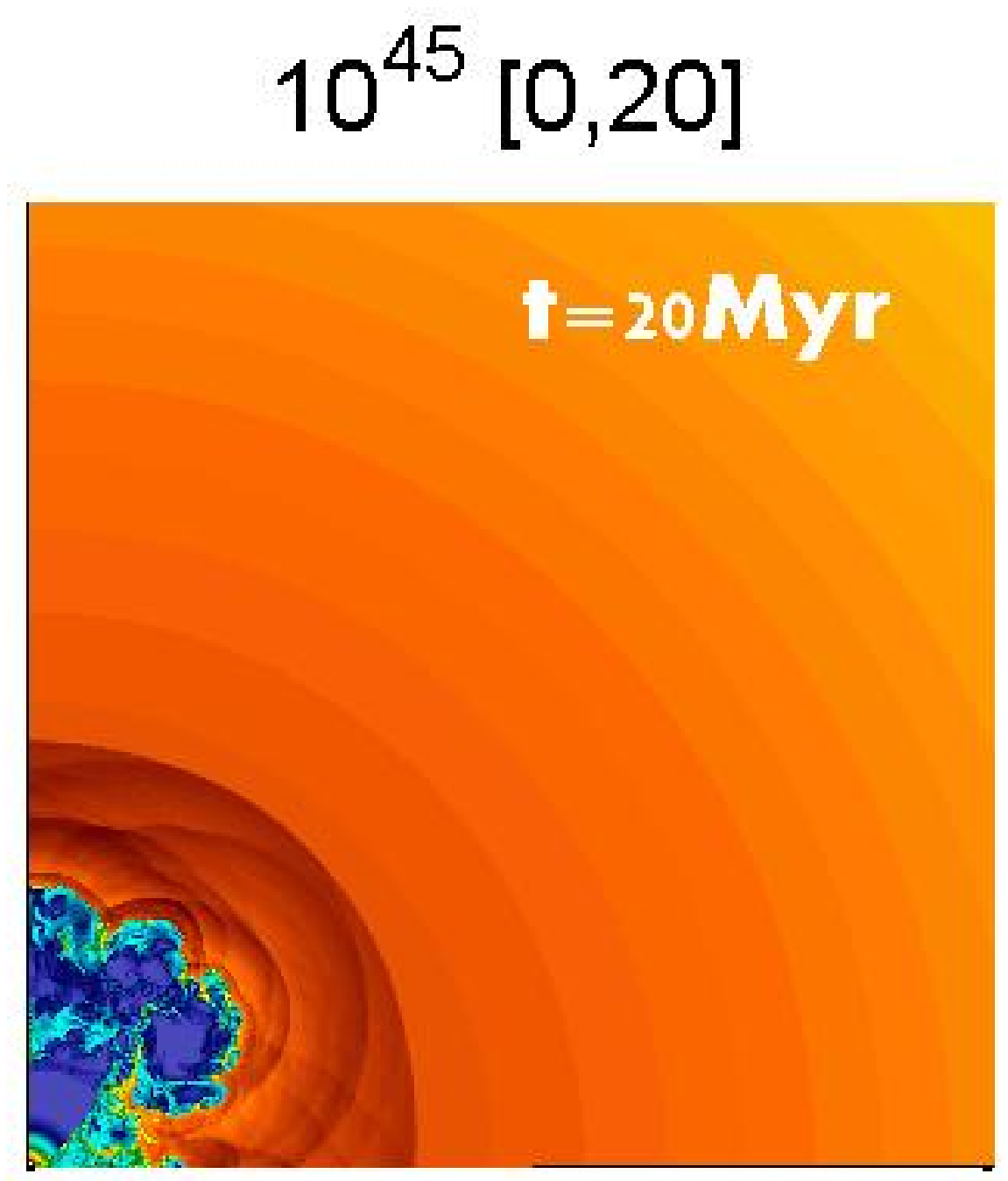}}
{\includegraphics*[scale=0.25, clip = true,trim=2cm 0 2cm 0]{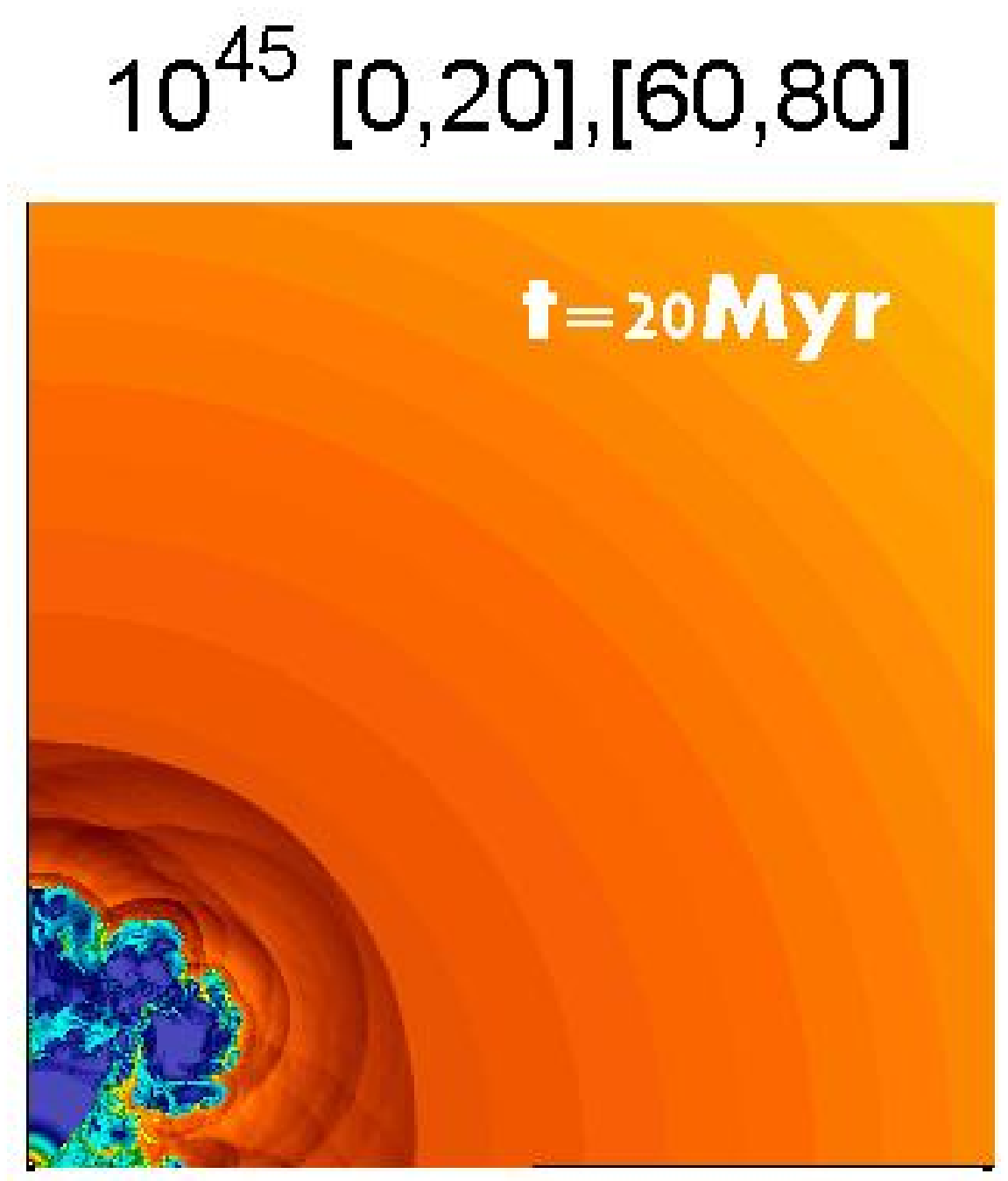}} \\
\hskip 3.3 cm
{\includegraphics*[scale=0.25, clip = true,trim=2cm 0 2cm 0]{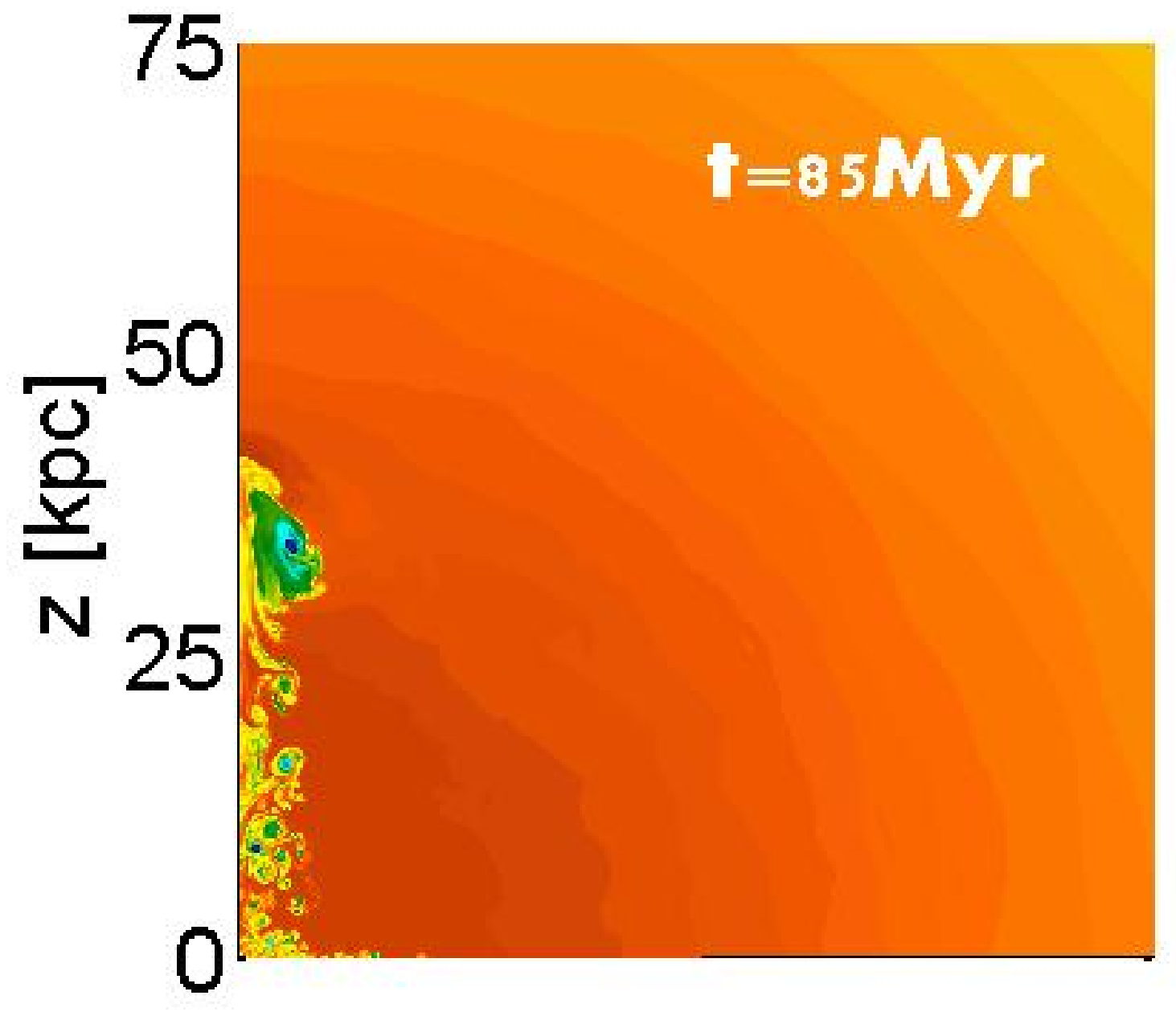}}
{\includegraphics*[scale=0.25, clip = true,trim=2cm 0 2cm 0]{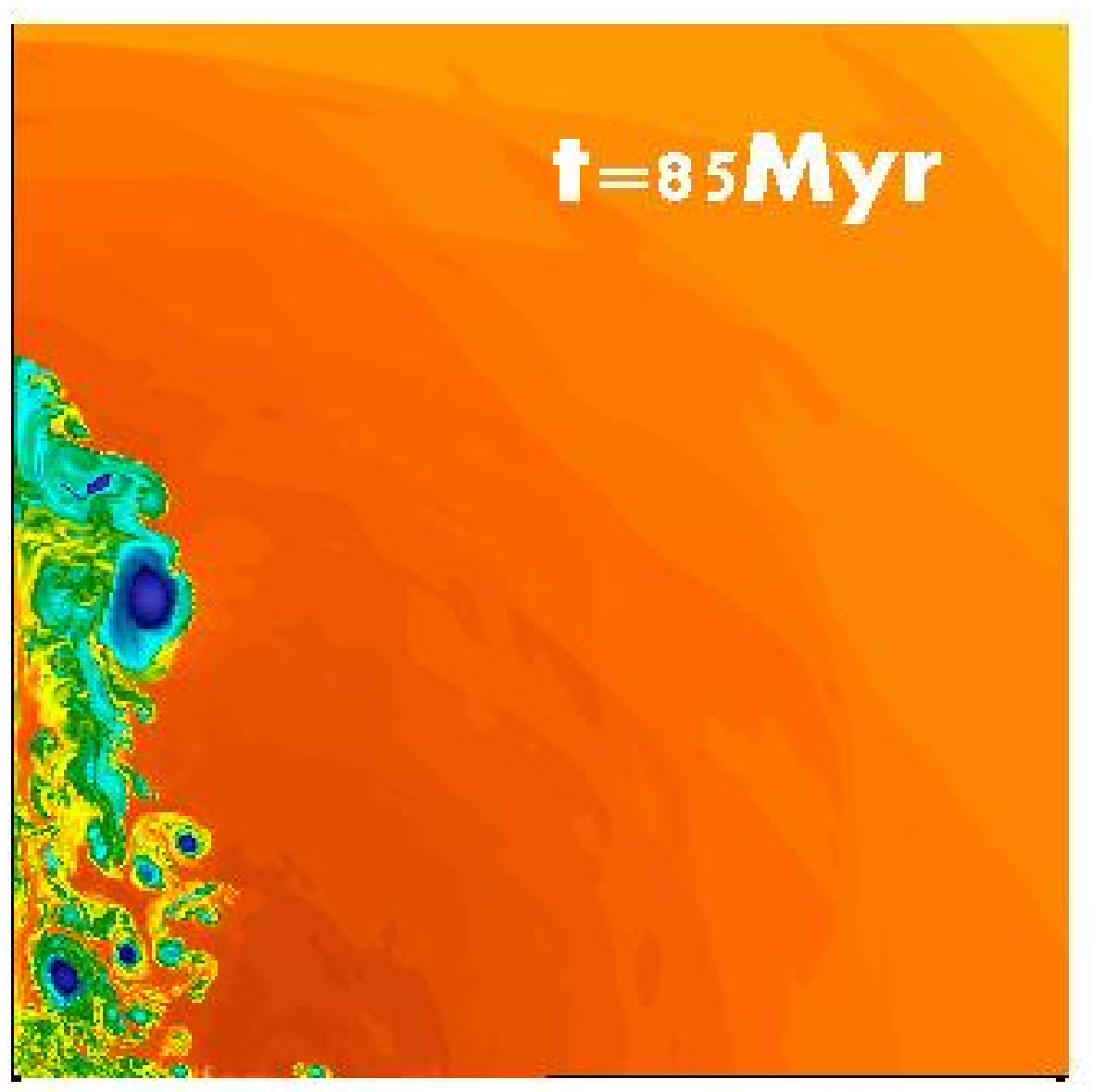}}
{\includegraphics*[scale=0.25, clip = true,trim=2cm 0 2cm 0]{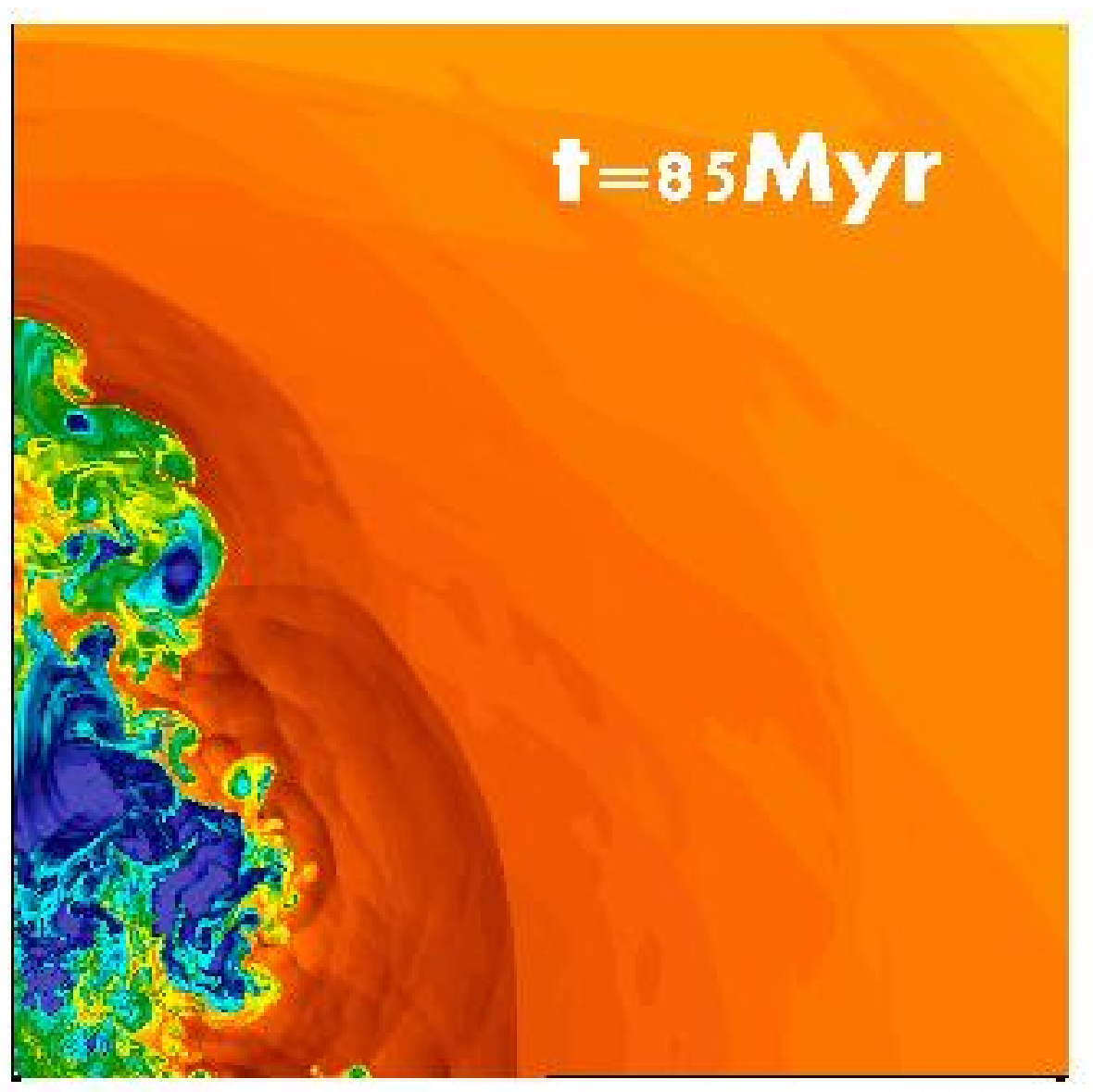}} \\
\hskip 3.3 cm
{\includegraphics*[scale=0.275, clip = true,trim=2.25cm 0 2.75cm 0]{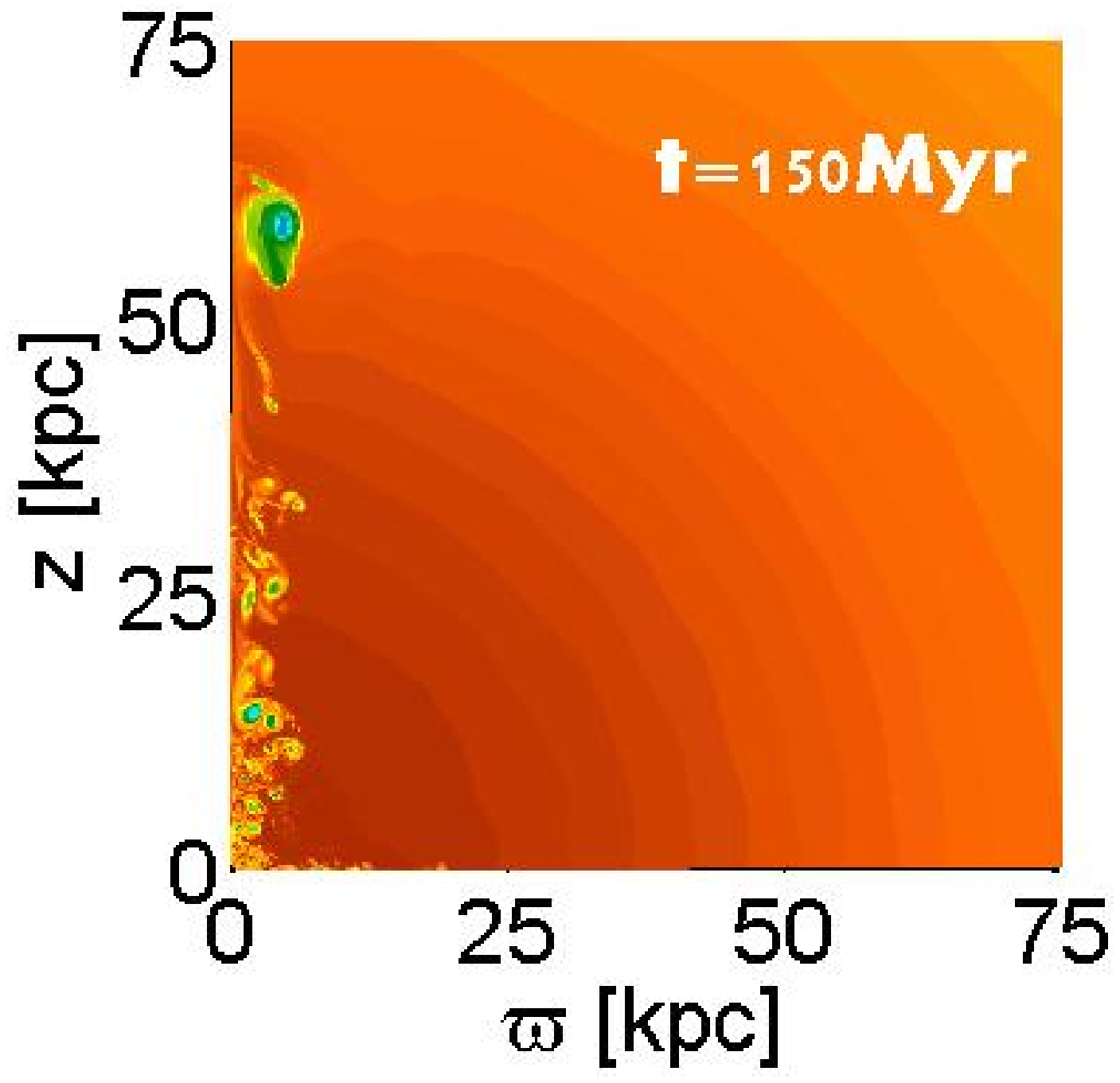}}
{\includegraphics*[scale=0.275, clip = true,trim=2.75cm 0 2.75cm 0]{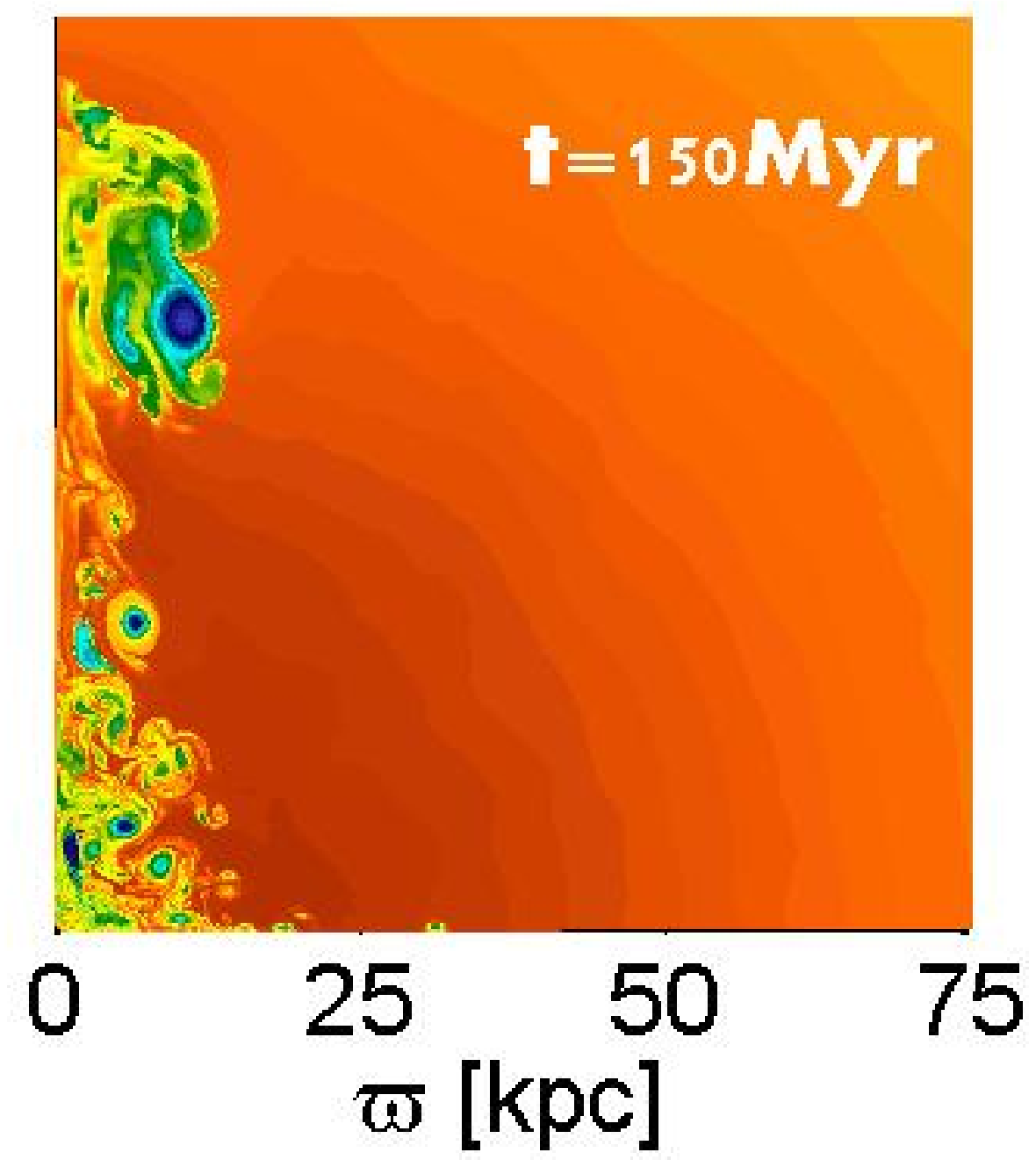}}
{\includegraphics*[scale=0.275, clip = true,trim=2.75cm 0 2.25cm 0]{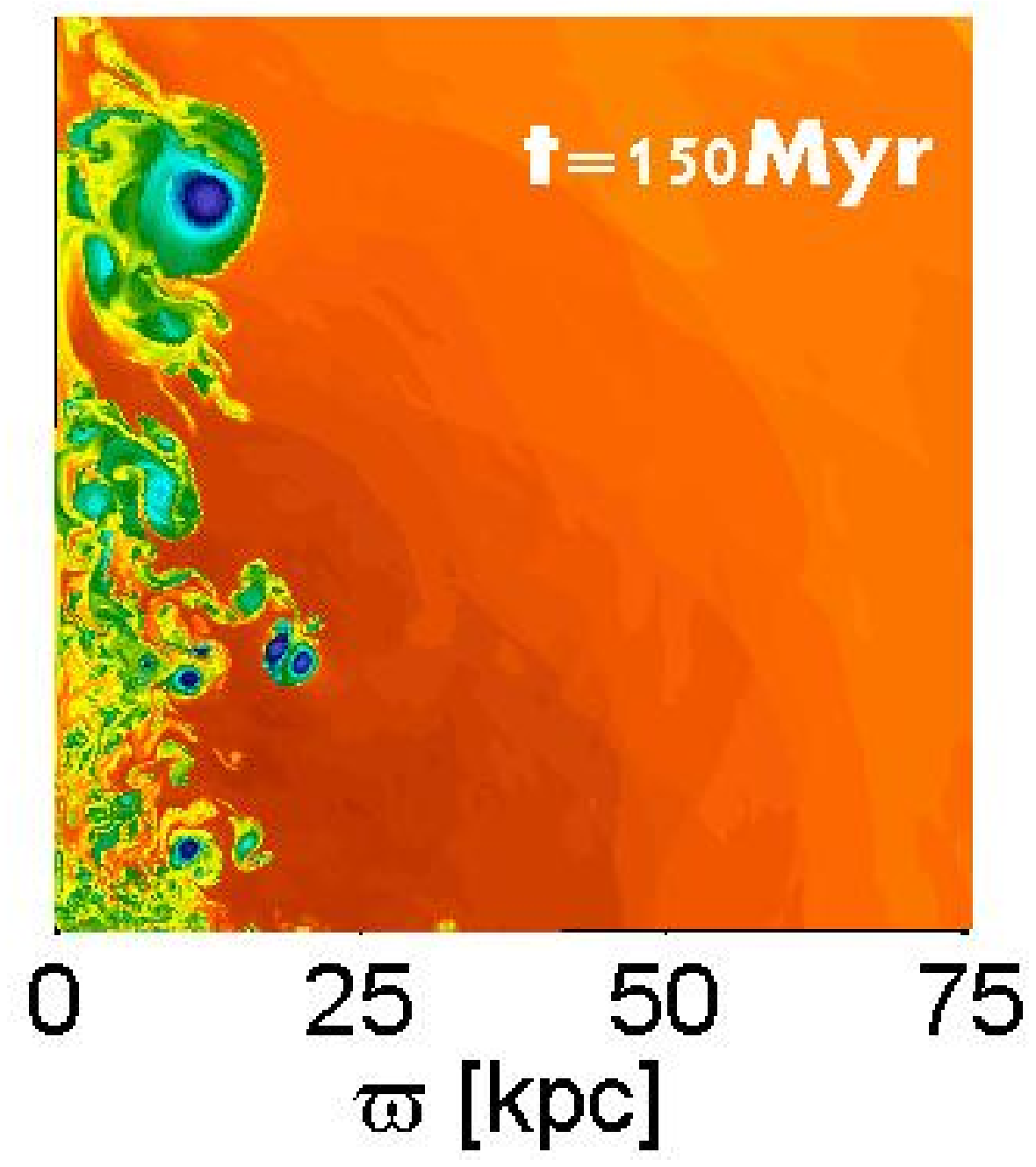}} \\
\end{tabular}
\caption{The density maps in the meridional $(\varpi,z)$ plane at three
 times (as marked on each panel),
 and for three cases (as marked above each column). The first number is
 the power of one jet in $\erg \s^{-1}$, followed by the active jet period(s)
 in $\Myr$: given inside each square parenthesis is the
 time the jet starts and ends. The density scale is logarithmic,
 from $10^{-26.5} \g \cc$ in blue to $10^{-24.5} \g \cc$ in red.
 The symmetry axis of the jets is along the vertical ($z$) axis.
}
\label{fig:mixrho}
\end{figure*}

Figure \ref{fig:mixrho} shows the evolution of the density in the meridional plane $(\varpi,z)$
at three times and for three models.
The left and middle columns show results for a single jet-launching episode lasting $20 \Myr$, and with a kinetic two-jets power
of $2\times 10^{44} \erg \s^{-1}$ and $2 \times 10^{45} \erg \s^{-1}$, respectively.
The right column shows results for two subsequent jet-launching episodes, both lasting $20 \Myr$ and with
a kinetic two-jets power of $2 \times 10^{45} \erg \s^{-1}$; the first episode starts at $t=0$ and ceases at $t=20 \Myr$,
while the second one starts at $t=60 \Myr$ and ceases at $80 \Myr$.
In each panel the $z$ (vertical) axis is the symmetry axis (the jet axis).
Only one half of the volume is simulated.

The first row shows the flow structure at the end of the (first) jet injection episode.
The inflated bubbles are clearly seen. At early times they are attached to the center and form `fat bubbles'.
 It is evident that a lower power jet (left panel) inflates a smaller bubble.
By $t=85 \Myr$ in the left and middle columns the low-density bubble has risen, leaving behind a trail of vortices.
In the right column $t=85 \Myr$ corresponds to $5 \Myr$ after the second jet episode has ceased.
The shock front excited by this second jet-launching episode is not spherical, contrary to the case with the first jet episode.
The third row shows the flow structure at the end of our simulations,
long after the end of any jet activity.

\begin{figure}
\centering
\includegraphics[width=80mm]{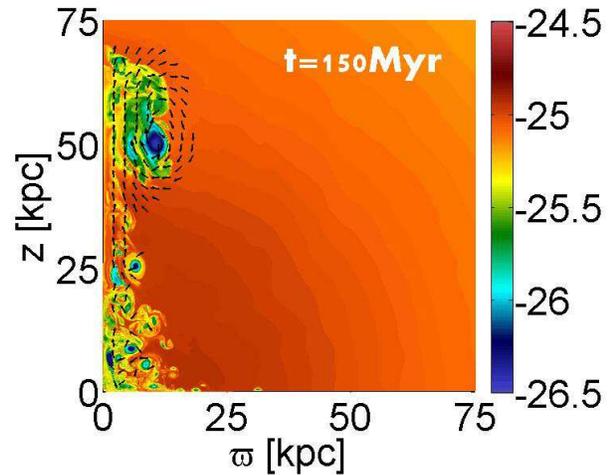} 
\caption{
The density map in the $(\varpi,z)$ plane for the simulation with one jet
episode and a one-jet kinetic power of $\dot{E}=10^{45} \erg \s^{-1}$. Also shown
are velocity vectors, for $v > 120 \kms$.
}
\label{fig:mixrhov}
\end{figure}

In Figure \ref{fig:mixrhov} we plot velocity arrows that emphasize the turbulent
nature of the bubbles and their surroundings.
Vortices are clearly seen, and can be compared to those studied in section
\ref{subsec:bubbles}.
This turbulent flow is responsible for the mixing of very hot shocked jet
material with the ICM.

\begin{figure}
\centering
\includegraphics[width=80mm]{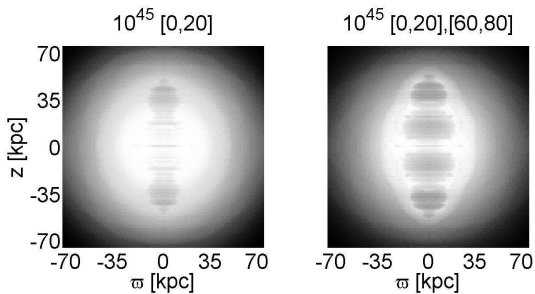} 
\caption{
The projected X-ray map in the full $(\varpi,z)$ plane at $t=85 \Myr$ for two
cases (as marked above each panel). The first number is the jet power in
$\erg \s^{-1}$, followed by the active jet period(s) in $\Myr$.
The X-ray image is obtained by integrating over the density squared.
}
\label{fig:mixxray}
\end{figure}

Figure \ref{fig:mixxray} shows the synthetic projected X-ray maps (obtained by the same procedure as described in section \ref{subsec:bubbles})
of the high-power runs at $t=85 \Myr$ (corresponding to the middle row of the middle and right columns of Figure \ref{fig:mixrho}).
We mirrored the space simulated twice (about the symmetry axis and about the equatorial plane) to obtain the full image.
One pair of bubbles away from the center is clearly seen in the left panel.
The right panel shows two pairs of bubbles that were generated by the two jet-launching episodes.
These synthetic maps can be compared with those in Figure \ref{fig:xray_epi_std} that were obtained for different physical parameters, e.g.,
narrower jets.
Rich variety of bubble morphologies (X-ray deficient bubbles) similar to those observed can be formed with different parameters.

\begin{figure}
\centering
\includegraphics[width=80mm]{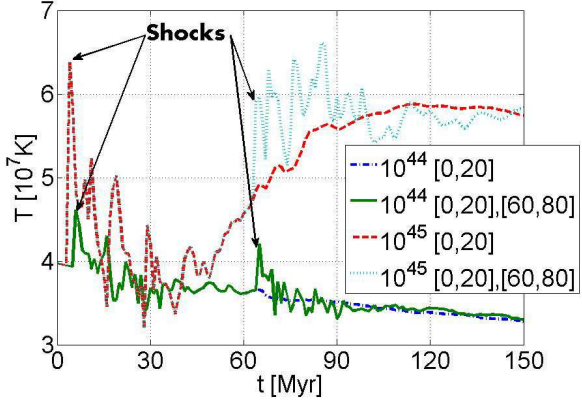} 
\caption{Average temperature of the TR91 tracer gas for four cases as marked
in the inset: the first number is the jet power in $\erg \s^{-1}$,
followed by the active jet period(s) in $\Myr$.
The TR91 tracer gas is the gas that was located initially at $(\varpi,z)=(9,1) \kpc$.
Note that the low-power jets do not manage to heat TR91 (see \citealt{GilkisSoker2012}).
}
\label{fig:tr91}
\end{figure}

\begin{figure}
\centering
\includegraphics[width=80mm]{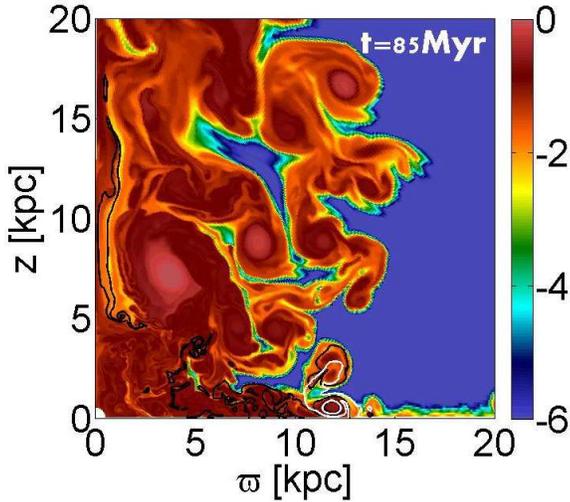} 
\caption{The concentration of jet material $\xi_{jet}$ (fraction of jet
material at each point) is shown by the color coded map in logarithmic scale,
for the simulation with one jet episode and a one-jet kinetic power of $\dot{E}=10^{45} \erg \s^{-1}$.
It is clearly seen that the jet material mixes with the ICM. We also follow
the material of two tracers, TR61 and TR91. Here TR91 is the gas that started
inside the circular region centered on $(\varpi,z)=(9,1) \kpc$ and having a
radius of $0.25 \kpc$, with a similar definition for TR61. The black (white)
contours show where the concentration of TR61 (TR91) is one percent. Both
tracers are well mixed with their surroundings.
}
\label{fig:mixing1}
\end{figure}

Figure \ref{fig:tr91} shows the evolution of the average temperature of gas (the ``TR91 tracer gas'') that was initially
in a circular region in the $(\varpi,z)$ plane (torus in 3D), centered around $(\varpi,z)=(9,1) \kpc$, and having a radius of
$0.25 \kpc$.
We follow this gas using an artificial flow variable $\xi$,
which is frozen-in to the flow. We denote this gas TR91 (for more details see \citealt{GilkisSoker2012}).
Four cases are shown, for two values of jet power, and for one or two jet episodes,
as marked in the inset of the figure.
Clearly seen are sharp rises in temperature that occur when the forward shocks hit the gas in this tracer
(marked on the figure).
Following these shocks are adiabatic oscillations of the temperature caused by sound waves.
Eventually, the temperature returns to a value very close to its starting value; the heating
by the shock has no lasting effect.
The only significant and lasting heating happens when the tracer material mixes with hot low-density shocked jet-material.
This happens for the cases of the high-power jet(s)
($\dot{E}_{1j}=10^{45} \erg \s^{-1}$) at $t \sim 60-90 \Myr$.
The mixing of the TR91 gas is shown in Figure \ref{fig:mixing1}.

 The main conclusion from these simulations is that the turbulent motion excited by the inflation process of the bubbles
 mixes the ICM with the shocked jets' material, leading to the heating of the ICM.
 In the next subsection we further show that this process is more important than shock waves in heating the ICM.

\subsection{Shock heating}
\label{subsec:shocks}

Consider a short duration explosion, namely, a short-duration bubble-inflation episode, with a total energy $E_b$ inside a constant density $\rho_0$ of the ICM.
The constant energy medium holds only in the inner few kpc, but the assumption is adequate enough for the present goal.
The shock radius and velocity for an ICM with $\gamma=5/3$ reads (e.g., \citealt{Taylor1950}) of this Sedov solution are
\begin{subequations}
\begin{align}
R_s(t)=1.15 \left( \frac{E_b t^2}{\rho_0}  \right)^{1/5}, \\
 v_s=0.46 \left( \frac{E_b}{\rho_0 t^3}  \right)^{1/5} = 1.16 \left[ \frac{E_b}{M(R_s)} \right]^{1/2}
\label{eq:vs1}
\end{align}
\end{subequations}
where $M(R_s) = \rho_0 V$ is the ICM mass inside radius $R_s$, and $V=4 \pi R_s^3 /3$ is the volume.
The shock Mach number $M_s=\frac{v_s}{C_s}$, where $C_s = (\gamma P/\rho_0)^{1/2}$ is the sound speed of the ICM, can be evaluated as follows.
For the constant density and temperature assumed here the pressure is constant as well. This is of course not accurate, but we can think
of a mass-weighted averaged pressure.
The total thermal energy of the gas within radius $r$ is $E_{\rm th} = (3/2)P V$, and the Mach number can be written as
\begin{equation}
M_s=1.1 \left( \frac{E_b}{E_{\rm th}} \right)^{1/2}.
\label{eq:mach}
\end{equation}

 The energy equation reads
\begin{equation}
\frac{d}{dt} \ln P \rho^{-5/3} = \frac{L}{e_{\rm th}} \equiv \frac{1}{t_{\rm cool}},
\label{eq:energ1}
\end{equation}
where $e_{\rm th}$ is the thermal energy per unit volume, $L$ is the cooling rate per unit volume, and $t_{\rm cool}$ is defined here as the cooling time in constant volume
 (at a constant pressure the cooling time is longer by a factor of $5/3$).
 For a change $\Delta_s \equiv \Delta \ln P \rho^{-5/3}$ in each shock, the number of shocks required within a cooling time to maintain the gas at the same entropy is
 $N_s(t_{\rm cool}) =\Delta_s ^{-1}$.
The total energy required by the AGN activity during the average cooling time within radius $r$ is given by
\begin{equation}
E_{AGN} (t_{\rm cool} ) = \frac{M_s^2}{1.2} \frac {E_{\rm th}}{\Delta_s}.
\label{eq:eagn}
\end{equation}
This gives $E_{AGN} (t_{\rm cool} ) = 112 E_{\rm th}$ for $M_s=1.3$, and $E_{AGN} (t_{\rm cool} ) = 27E_{\rm th}$ for $M_s=1.7$.
This is a very inefficient heating mechanism.
The power of bubbles in clusters of galaxies is in the order of magnitude of what is required to maintain the hot ICM (e.g., \citealt{Ma2012} for a recent reference),
namely, $E_{AGN} (t_{\rm cool} ) \sim E_{\rm th}$, and cannot account for heating by shocks.

In the group of galaxies NGC 5813 \cite{Randall2011, Randall2012} detected 3 shocks perpendicular to the axis of the three bubble pairs. The shocks' parameters are
$[R_s , M_s, \Delta_s^{-1}, t_{\rm cool}(R_s)]=(1.2 \kpc, 1.71, 10, 2\times10^8 \yr)$, $(10 \kpc, 1.52, 20, 9\times10^8 \yr)$, and $(25 \kpc, 1.3, 77, 2\times10^9 \yr)$.
Namely, to heat the gas at $10 \kpc$ over a time of $\sim 10^9 \yr$ the AGN energy release in jet activity should be $\sim 20$ times the total energy within radius of $10 \kpc$,
or $77$ times within a radius of $25 \kpc$.
This is problematic according to our findings for the following reasons.
(1) The heating along the jets direction is very efficient as mixing takes place there very efficiently.
This implies that the activity that required to keep the gas residing perpendicular to the jets' axis hot by shock heating,
will lead to very high temperatures along the jets direction.
(2) The shock-heating feedback requires many bubble-inflation episodes.
The bubbles remove gas from the center. This will cause gas to flow inward, hence being more prone to heating by mixing.
One cannot treat the gas as sitting in one place. A huge circularization of gas will star.

Our preferred explanation is that (a) mixing occurs perpendicular to the jets axis close to the center. (b) At larger distances the cooling time is longer.
Over such a time the bubbles direction is very likely to change, e.g., the Perseus ghost bubbles \citep{Fabianetal2000}.
That varying jets' axis lead to a much more efficient interaction with the ambient gas is discussed in the next section where the explosion of CCSN is discussed.

\section{Exploding core collapse supernovae (CCSN)}
\label{sec:CCSN}
In this section we describe numerical results for the jittering-jet model for CCSNe.

\subsection{Jet-driven explosion}

Jet-driven supernova explosion models have a long history
(e.g. \citealt{LeBlanc1970,  Meier1976, Bisnovatyi1976, Khokhlov1999, MacFadyen2001, Hoflich2001,
Woosley2005, Couch2009, Couch2011,Lazzati2011}).
Our \textit{jittering-jet model} for explosion \citep{papish2011, papish2012} is based on the
following points, that differ in several ingredients from the models cited above (for more detail see \citealt{papish2011}).
\newline
(1) We don't try to revive the stalled shock. To the contrary, our model requires the material near the stalled-shock
to fall inward and form an accretion disk around the newly born NS or black hole (BH). In the present simulations we do
not include the stalled shock and don't resolve the accretion disk, but we do show the accretion inflow through the equatorial plane.
\newline
(2) We conjecture that due to stochastic processes and the stationary
accretion shock instability (SASI; e.g. \citealt{Blonding2003, Blondin2007}) segments of the post-shock accreted gas
(inward to the stalled shock wave) possess local angular momentum.
\citet{Foglizzo2012} found in their lab experiment that although the inflow contains no net angular momentum,
the non-linear interactions in the inflow ultimately favor a single spiral mode to be formed.
This leads the accreted mass to temporarily posses angular momentum. This accretion of angular momentum is crucial to our model,
as it probably leads to the formation of an accretion disk with rapidly varying axis direction.
We note that in order to revive the stalled shock, as is required in neutrino-driven models, the SASI seems to be very important (e.g., \citealt{Mueller2012a}).
The SASI is obtained in new high-resolution simulations (e.g., \citealt{Mueller2012a}), and seems to be ubiquitous in CCSNe.
\newline
(3) We assume that the accretion disk launches two opposite jets. Due to the rapid change in the disk's axis, the
jets can be intermittent and their direction rapidly varying. These are termed \emph{jittering jets}.
\newline
(4) We show in \cite{papish2011} that the jets penetrate the infalling gas up to a distance of few$\times 1000 \km$, i.e.,
beyond the stalled-shock. Beyond few$\times 1000 \km$ the jittering jets
cannot penetrate the gas any more.
The jittering jets don't have the time to drill a hole through the ambient gas before their direction changes;
they are shocked before penetrating through the ambient gas.
This condition can be met if the jets' axis rapidly changes its direction.
This process of depositing jets' energy into the ambient medium to prevent further accretion is termed the {\it penetrating jet feedback mechanism.}
The jets deposit their energy inside the star via shock waves, and form \emph{hot bubbles}.
These bubbles and their similarity to hot bubbles in other astrophysical objects is the focus of this section.
\newline
(5) The jets are launched only in the last phase of accretion onto the NS.
For the required energy the jets must be launched from the very inner region of the accretion disk.
In our jittering-jets explosion model the jets are launched close to the NS where the
gas is neutron-rich (e.g., \citealt{kohri2005}).
As long as the jets and the bubbles they form don't explode the core, accretion continues.
When the jets manage to explode the core of the star, accretions stops. This is the feedback component in the model.

\subsection{Numerical Setup}
We use the FLASH code to solve the non-relativistic hydrodynamic equations on an adaptive-mesh refinement with 12 levels.
The simulation is done in a 2D cylindrical coordinates on a grid of size $1.5 \times 10^9 \cm$ in each direction.
The equation of state (EOS) is the Helmholtz equation of state \citep{timmes2000} used in FLASH.
This EOS includes contributions from partial degenerates electrons and positrons,
radiation and non degenerate ions.
For the initial conditions of the ambient gas in the core we used the results of \cite{Liebend2005} who made a 1D simulations of the core bounce.
We map their results from about $0.2 \s$ after bounce for our simulations.
For gravity we use FLASH's multipole solver with only a monopole term ($l=0$).
In addition we take the newly born-NS to be a point mass $M = 1.4 M_\odot$ in the center.
The cooling function is taken to be
\begin{equation}
Q_{\nu}^- = 5 \times 10^{30} \frac{T_{10}^{9}}{\rho} + 9\times 10^{23} X_{nuc} T_{10}^6 \erg \g^{-1} \s^{-1},
\end{equation}
here $T_{10}$ is the temperature in $10^{10} \K$, $\rho$ is the density in grams, and $X_{nuc}$ is the mass fraction of free nucleons.
The first term is the contribution of electron positron pair annihilation \citep{Itoh1996}.
The second is due to electron/positron capture on free nucleons \citep{Qian1996}.

The jets are injected in pairs, alternating between the two sides.
For each $0.1 \s$ there is a $0.05 \s$ jet-launching episode in the upper half plan, followed by a $0.05 \s$ jet-launching episode in the lower half plan.
The jets are injected in directions of $20^\circ - 50^\circ$ relative to the symmetry axes, and with a velocity of $v_j = 10^{10} \cm \s^{-1}$.

\subsection{Results}
The first jet is able to penetrate through the collapsing envelope.
Two shocks are created: a reverse shock where the jet is shocked, and a bow shock expanding outward shocking the core material.
The bow shock accelerates the gas outwards, while material more or less perpendicular to the jet initial direction continues to flow
inwards.
The two shocks heat the jet and the core material, respectively, to temperatures of $\sim 2-7 \times 10^9 \K$ creating hot bubbles.
In this range of temperatures the neutrino cooling is not significant and most of the energy is deposit inside the surrounding gas.
Nuclear reactions should occur inside the hot bubbles but this is not taken into account in the simulation
(see \citealt{papish2012} for discussion of r-process in these hot bubbles).

The distance the first jet can penetrate depends on its properties, as well as on how fast it jitters.
In our simulations this distance is about few$\times 1000 \km$.
In the next episodes the jets are injected into a more dilute environment left by the expanding gas from previous episodes.
This allows the jets to penetrate faster and catch up with previous shocks.

\begin{figure}
\centering
\includegraphics[width=80mm,clip = true, trim=13mm 10mm 30mm 20mm]{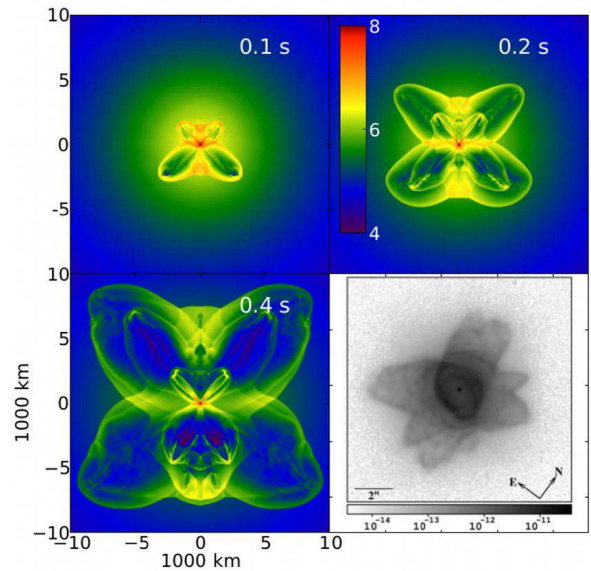} 
\caption{Comparing results from 2.5D simulations of jittering jets in core-collapse supernovae (Papish \& Soker 2012, in prep.)
to an H$\alpha$ image of the PN He 2-47, a low-excitation PN (image from \citealt{Sahai2000}).
Panels 1-3 show the log density (in cgs) of the inner most part of a $15 M_\odot$ star after bounce. The times are after the first jet is injected.
The similarity between the CCSN simulation and the PN image is in the structure where the bubbles protrude in different directions.
}
\label{fig:supervova}
\end{figure}

This mechanism, were each time a jet in a different direction is injected, cause the hot bubbles
to coalesce and form a more spherically shaped outflow; see lower left panel of Figure \ref{fig:supervova}.
As the angles of the jets' jittering are changing within only $30^\circ$, the material in other directions keeps
flowing inward to the NS.
Accretion is reversed in all directions at $\sim 3000 \km$, namely, where the hot bubbles coalesce into one large bubble that accelerates the
core material outward. The evolution is seen in the first 3 panels of Figure \ref{fig:supervova}.

There are some differences between the behavior of the jets in the supernova case and in the cluster case.
These differences are mainly because the jets in supernovae are much denser than their surrounding material,
while in the clusters the jets are less dense than their environment.

\begin{figure}
\centering
\includegraphics[width=80mm,clip = true, trim=7mm 23mm 20mm 33mm]{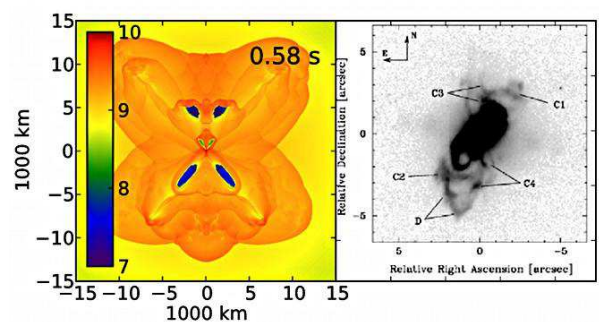} 
\caption{Left panel: The temperature in the hot bubble at $0.58 \s$
after the start of the jets injection (Papish \& Soker 2012,  in prep.);
there are 6 episodes of jets in the upper side and 6 episodes of jets in the lower side.
Right panel: HST image of the planetary nebula Hu 2-1 obtained in the [N II]6583 line (image from \citet{Miranda2001}).
The similarity between the CCSN simulation and the PN image is in the structure where bubbles are seen to catch up with previously inflated bubbles.  }
\label{fig:supervova2}
\end{figure}

\begin{figure*}
\centering
\subfigure[][]{\includegraphics[width=150mm,clip = true, trim=0mm 0mm 0mm 0mm]{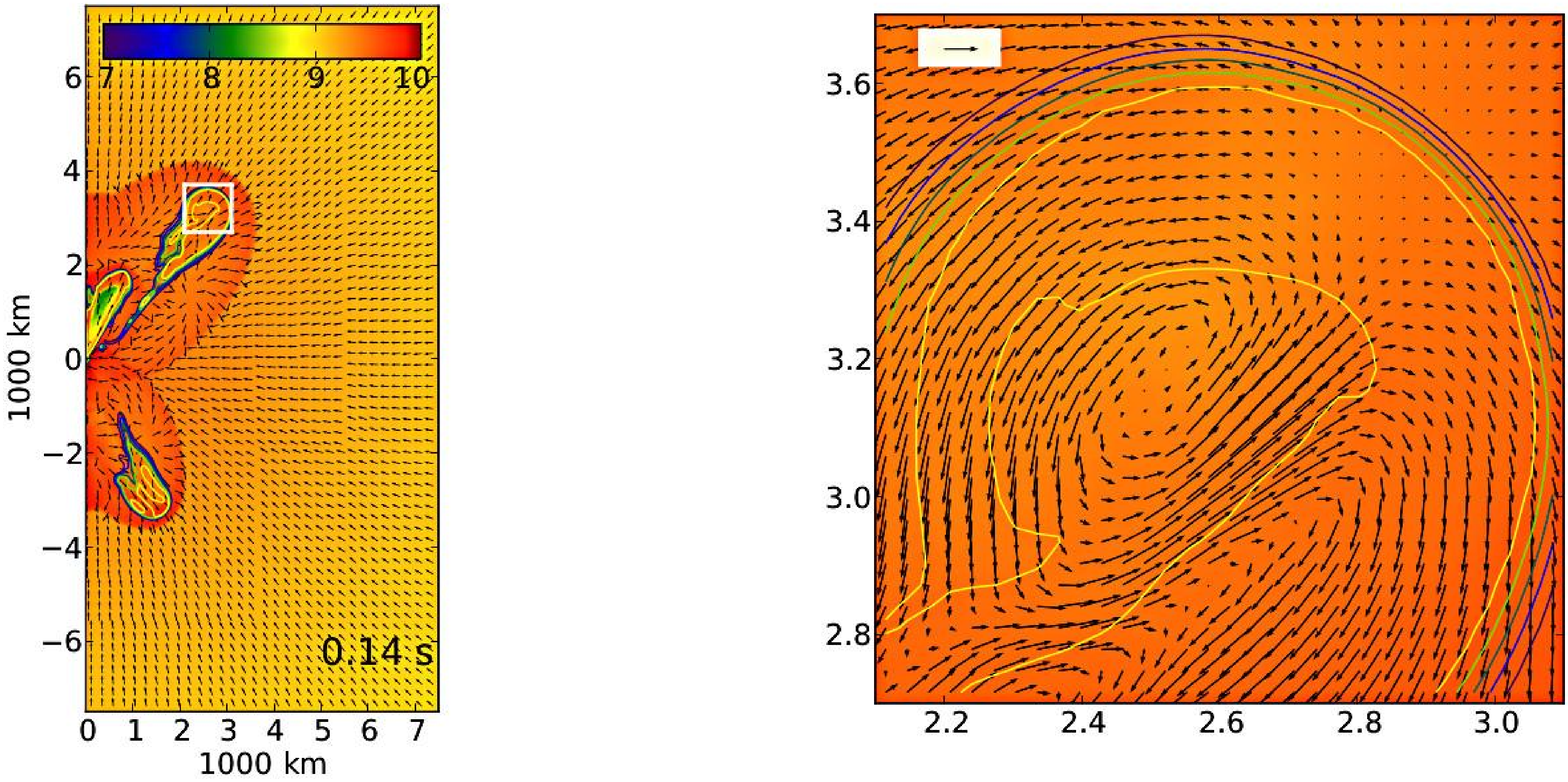}}\\ 
\subfigure[][]{\label{subfigure:d_std_20b}\includegraphics[scale=0.5,clip=true,trim=80 0 80 0]{dens_std_20_arrows}}
\subfigure[][]{\includegraphics[width=80mm,clip = true, trim=0mm 0mm 0mm 0mm]{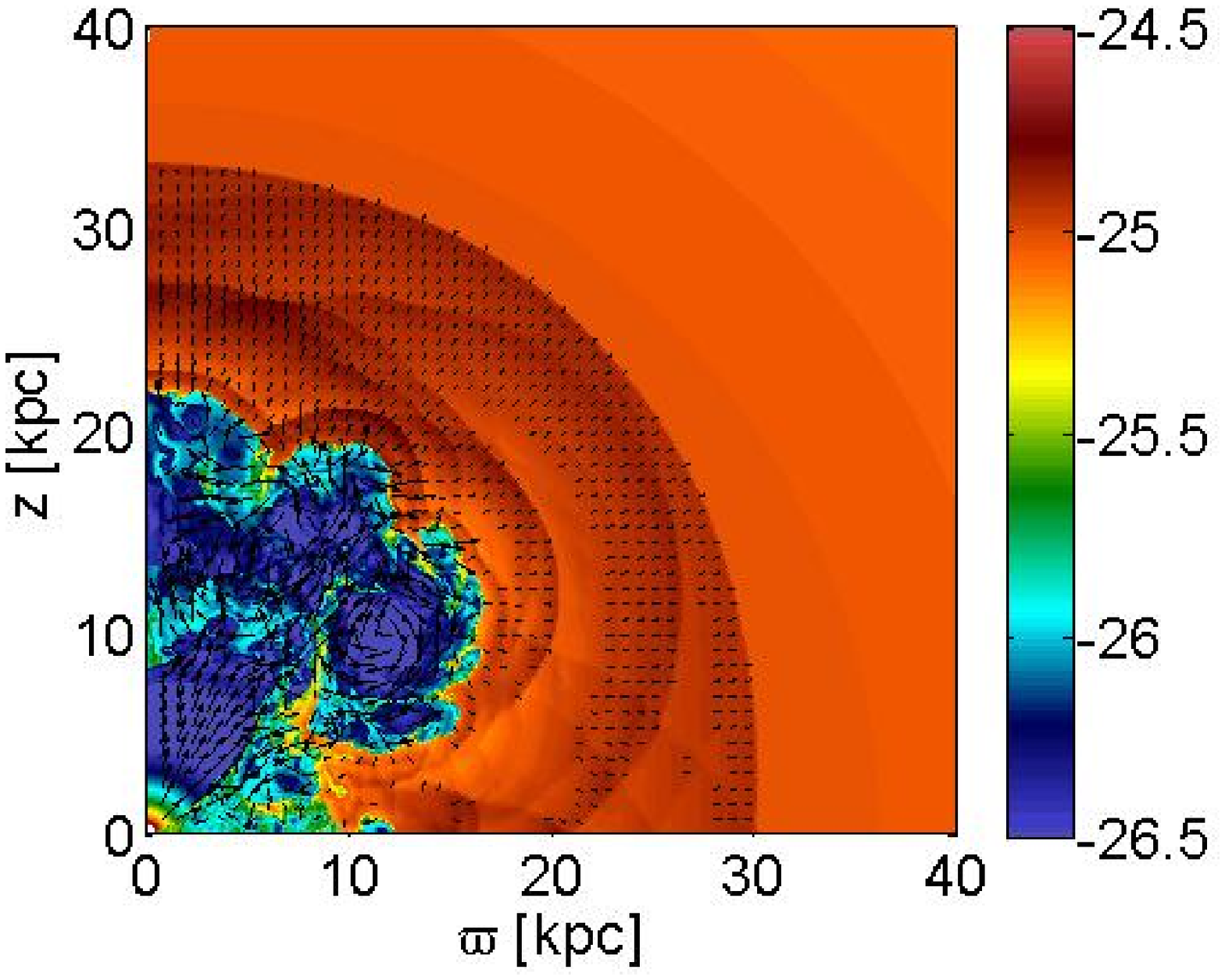}} 
\caption{Upper panel: The temperature in the hot bubbles at $0.14 \s$ after the start of the jets injection after
one episode of jets in the lower side and two episodes of jets in the upper side. The contours represent the mass fraction of the jets material.
Right upper panel: Zoom-in of the white square in the left panel. The velocity is shifted to the reference frame of the reverse shock.
The velocity arrow in the white box is scaled to $10,000 \km \s^{-1}$. Lower panels: the results from Figure \ref{fig:mixrho} panel 2,
and Figure \ref{fig:dens_epi_std} panel (b) shown here again for comparison. }
\label{fig:supervova3}
\end{figure*}

The highly dense jets in our simulations of CCSN cause the outer froward shocks running to the core to be very fast.
Hence, the forward shocks and the reverse shocks (where the jets are shocked) have similar velocities, implying similar post-shock temperatures
as seen in Figures \ref{fig:supervova2} and \ref{fig:supervova3}.
In clusters of galaxies and in planetary nebulae the jets are shocked to a temperature of two-three orders of magnitude higher than the temperature of the surrounding gas.

Another difference is in the structure of the cocoon.
A high density jet in CCSN flow creates a weak cocoon compared to the cocoon seen around jets in our simulations of clusters of galaxies.
However, a large vortex exists inside the bubble, similar to those in cluster simulations.
To present the vortex, in Figure \ref{fig:supervova3} we draw the velocity relative to the rest frame of the reverse shock.
A clear counter-clock wise vortex is seen in the inset.
The two lower panels show vortices from cluster simulations. The lower-left panel is the vortex of the flow shown in
panel 2 in Figure \ref{fig:mixrho}, while the lower-right panel is the vortex of the flow shown in panel (b) in Figure \ref{fig:dens_epi_std}.

Finally we note the very interesting similarity in morphologies of our simulated CCSN bubbles to those in some PNs, as shown in Figures \ref{fig:supervova} and \ref{fig:supervova2}.
The structure in our simulations of CCSNe is changing from a bipolar structure into a more spherical structure in about $\sim 0.5 \s$.
The PNs are relatively young. In old PNs such delicate bubble structures are smeared out.

\section{JFM inside a common envelope}
\label{sec:common}

Multidimensional numerical simulations of the common envelope (CE) process
usually include the gravity of the star and the companion (e.g.,  \citealt{Passy2011, Lombardi2006, SandquistTaam1998, RickerTaam2008, RickerTaam2012}).
One of the main goals of these simulations is to find the manner by which the envelope is ejected.
\cite{Soker2004b} suggested that in some cases accretion disks are formed around WD and NS companions spiralling-in inside a CE.
In these cases the accretion disk might launch two jets. Due to the very high velocity of jets from WDs and NS,
the temperature of the post-shock jet material is much larger than that of the surrounding envelope. Hot bubbles are formed.
These bubbles, \cite{Soker2004b} argued, can facilitate the ejection of the AGB envelope.
We here conduct a preliminary study of this process for a WD orbiting inside the envelope of a massive AGB star.
More realistic would be to study the jets blown by a NS, as it is not clear an accretion disks can be formed around s WD in a CE.
Simulating jets from NS requires more sophisticated ingredients in the numerical code, and is left for a future study.

The AGB star has a total mass of $M_{\rm AGB} = 5 M_{\odot}$, a core mass of$M_{\rm core} = 0.77 M_{\odot}$, and a radius  of $R_{\rm AGB} = 310 R_\odot$.
The density profile in the envelope is taken to be $\rho \propto r^{-2}$.
The relative velocity of the WD to the envelope is taken to be $50 \km\s^{-1}$, somewhat lower the the Keplerian velocity due to envelope rotation.
In this preliminary study we neglect the gravity of the WD, so that accretion is not treated here.
This will have to be added in future simulations.
The orbital separation is taken to be $a=150 R_\odot$.

\begin{figure}[h!]
\begin{center}
\includegraphics[width=0.45\textwidth]{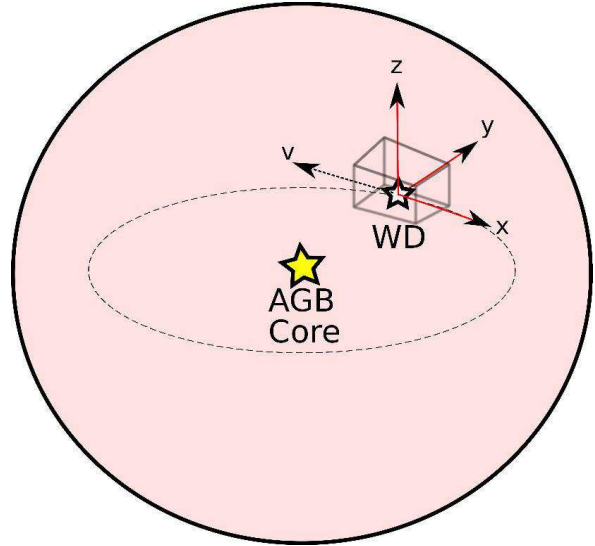}
\caption{A not-to-scale drawing of the WD motion within the AGB envelope.
Our computational domain is marked by the box.
The $x$ direction is opposite to the velocity vector of the WD, $v$.
The $y$ direction is along the radial direction from the center of the AGB star.
The fast jet is injected in the $z$ direction.}
\label{fig:cewd_setup}
\end{center}
\end{figure}

The simulations are performed using the high-resolution multidimensional hydrodynamics code FLASH 4.0b, using a 3D uniform grid with Cartesian $(x,y,z)$ geometry.
We treat a small volume of the AGB envelope, and simulate the flow in a box with dimensions of $60 R_\odot \times 30 R_\odot \times 30 R_\odot$
in the $x,y,z$ directions, with $512 \times 256 \times 256$ cells along each axis (single cell size is$\simeq 0.1 R_\odot$).
We set the $x$ axis along the relative velocity of the WD and the AGB envelope, and inject the AGB envelope gas with velocity $v_x=50 \km \s^{-1}$
from the minimum $x$ boundary ($x=0$) through the $y-z$ plane.
We impose outflow boundary conditions on all the other sides of the simulation box.
The $y$ axis is taken along the radial direction, such that the gravity is taken along the $y$ axis and the density is $\rho(y) = 3.1\times10^{19}(y+120 R_\odot)^{-2} \g\cm^{-3}$.
The temperature is taken to be polytropic with  $T \propto \rho^{2/3}$, and with a value that ensures hydrostatic equilibrium along the $y$ axis.
There is a mirror symmetry about the $z=0$ plane, and we apply a reflective boundary condition there.
A schematic drawing of the computational grid is presented in Figure \ref{fig:cewd_setup}.

\begin{figure*}[htp!]
\begin{center}
\subfigure[][]{\label{subfigure:cewd_dens1_xz}\includegraphics*[scale=0.23,clip=false,trim=0 0 0 0]{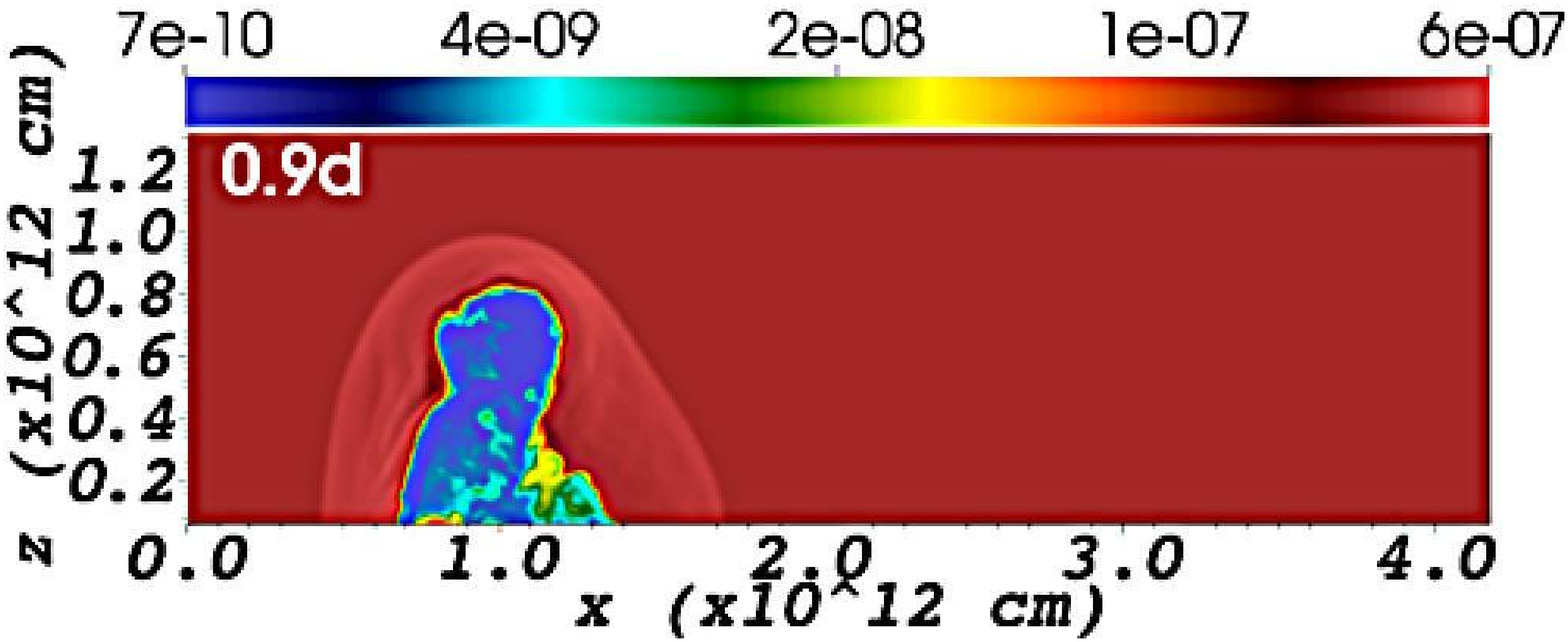}}
\subfigure[][]{\label{subfigure:cewd_dens2_xz}\includegraphics[scale=0.23,clip=false,trim=0 0 0 0]{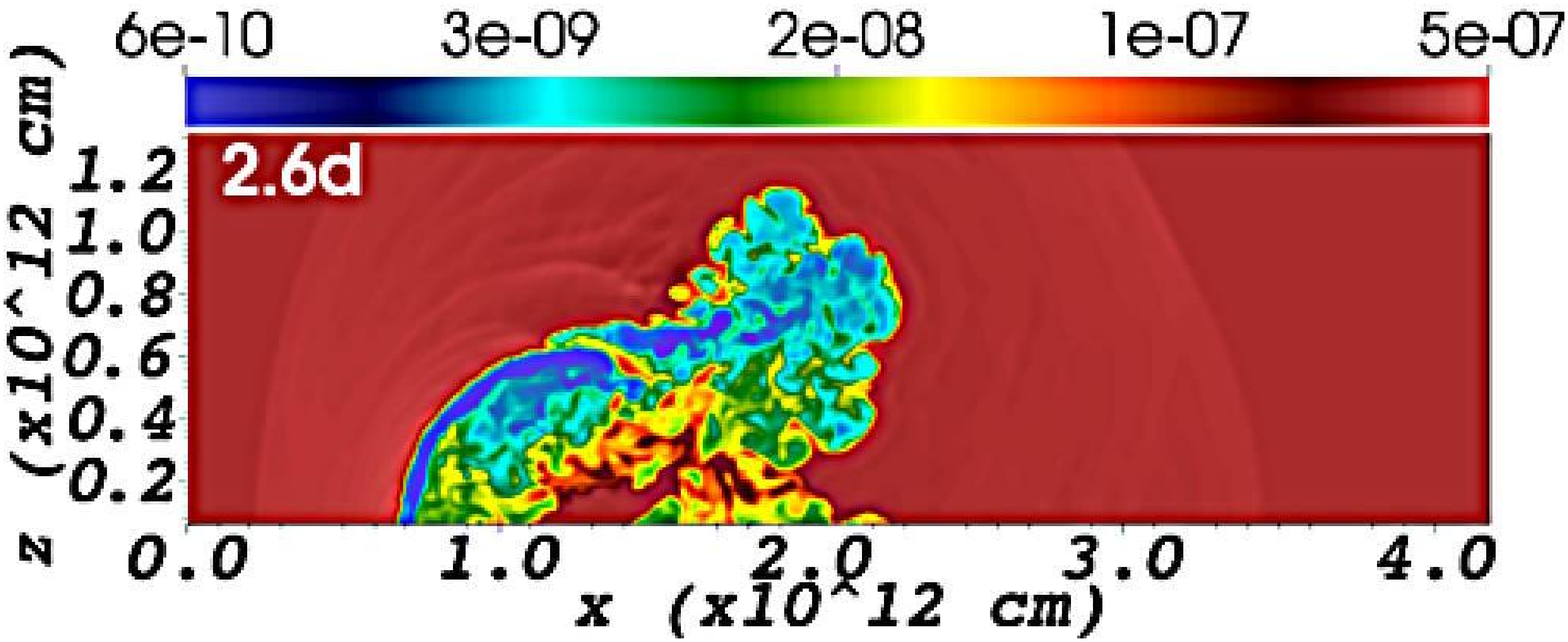}}\\
\subfigure[][]{\label{subfigure:cewd_dens3_xz}\includegraphics*[scale=0.23,clip=false,trim=0 0 0 0]{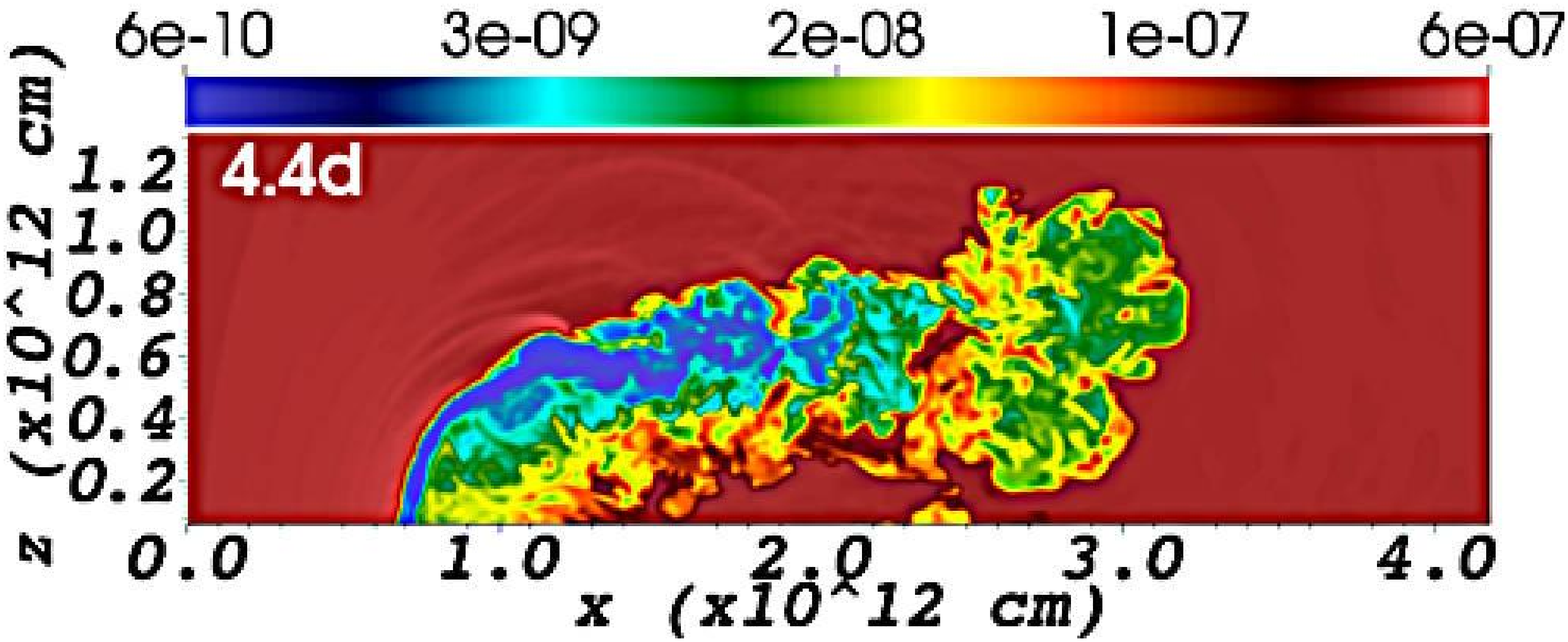}}
\subfigure[][]{\label{subfigure:cewd_temp3_xz}\includegraphics*[scale=0.23,clip=true,trim=0 0 0 0]{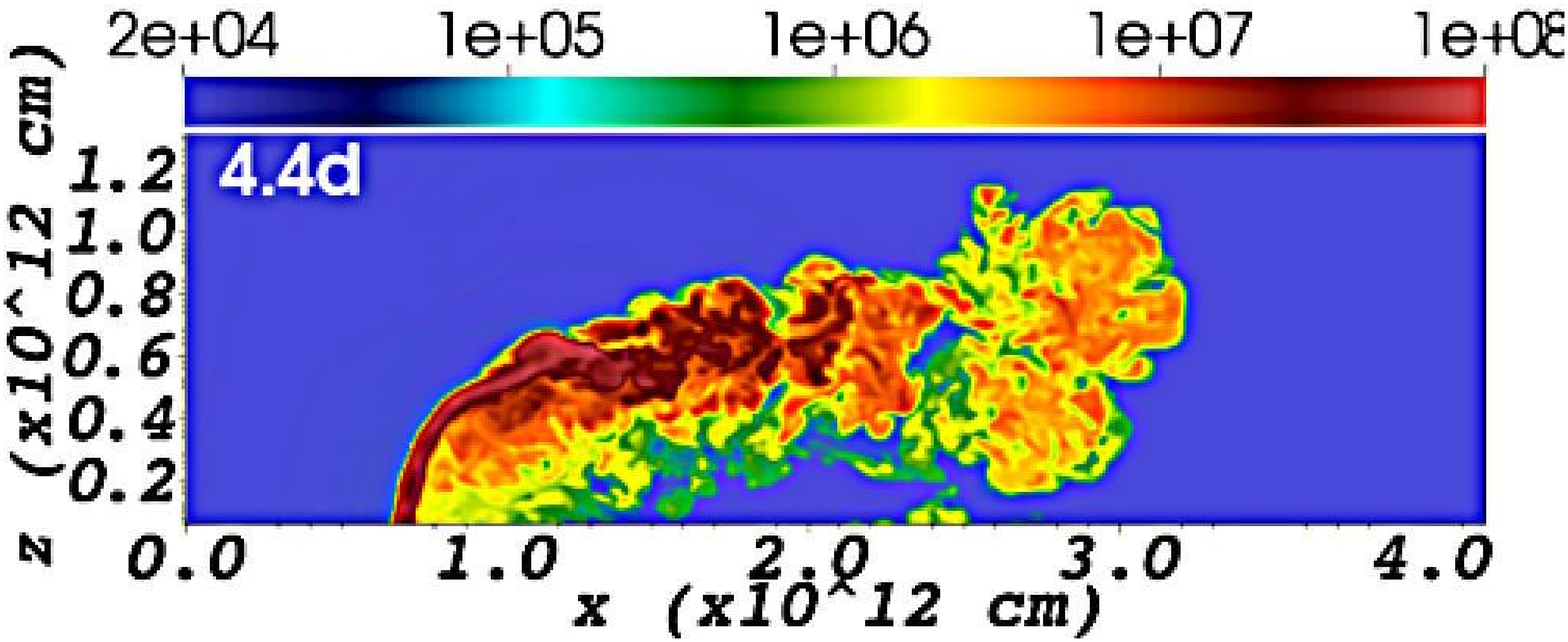}}\\
\subfigure[][]{\label{subfigure:cewd_vel3_xz}\includegraphics[scale=0.23,clip=false,trim=0 0 0 0]{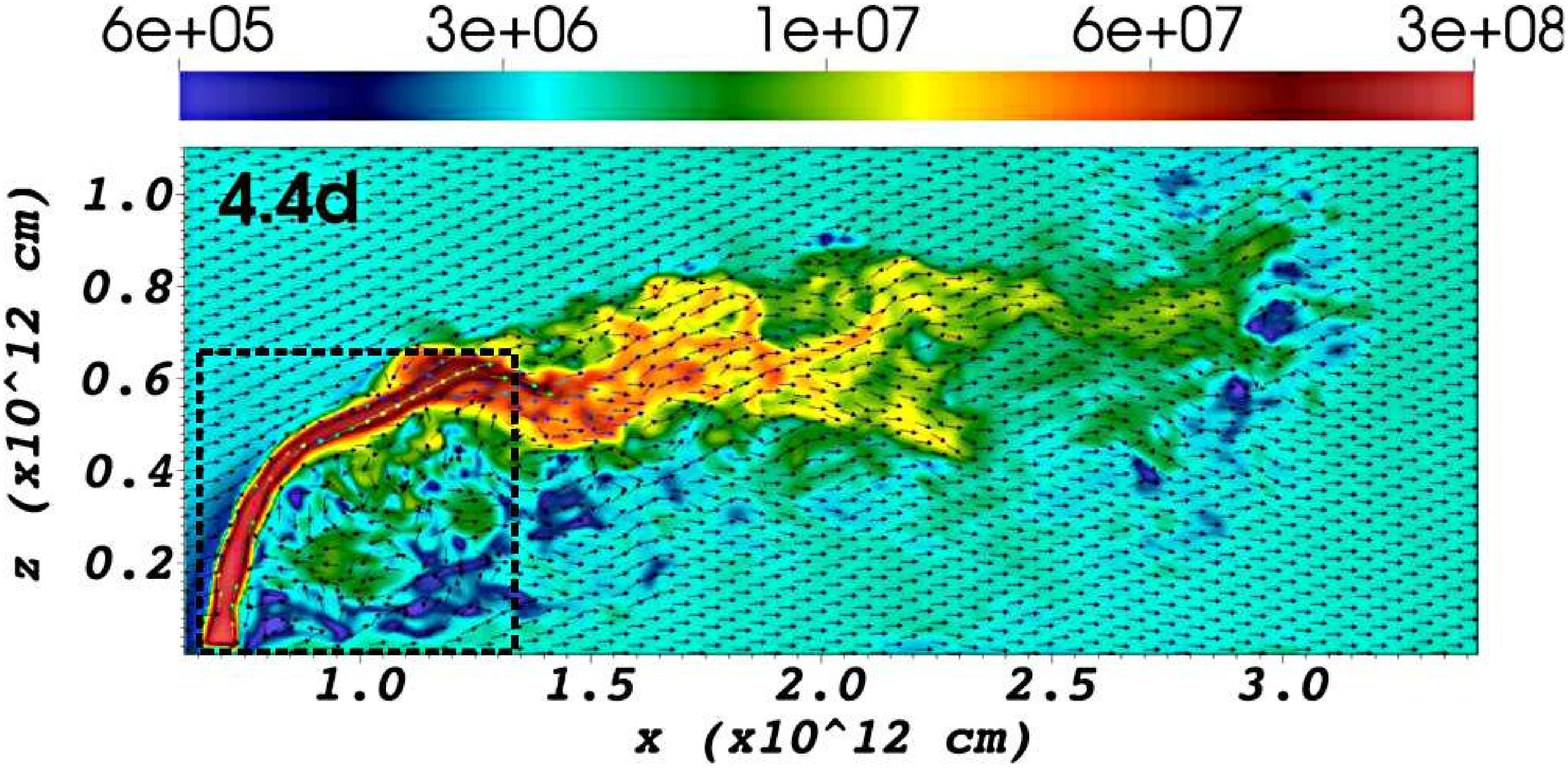}}
\subfigure[][]{\label{subfigure:cewd_dens3_xy}\includegraphics[scale=0.23,clip=false,trim=0 0 0 0]{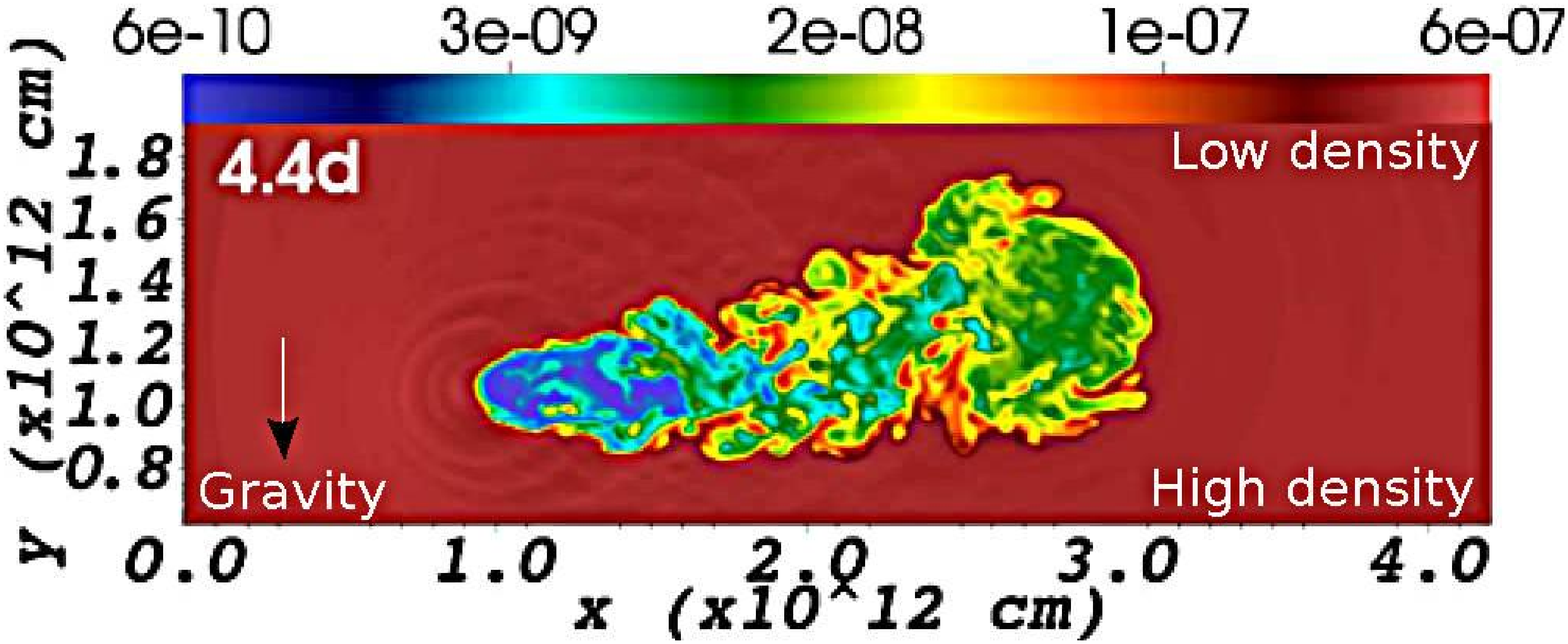}}\\
\subfigure[][]{\label{subfigure:cewd_velzoom_xz}\includegraphics[scale=0.3,clip=false,trim=0 0 0 0]{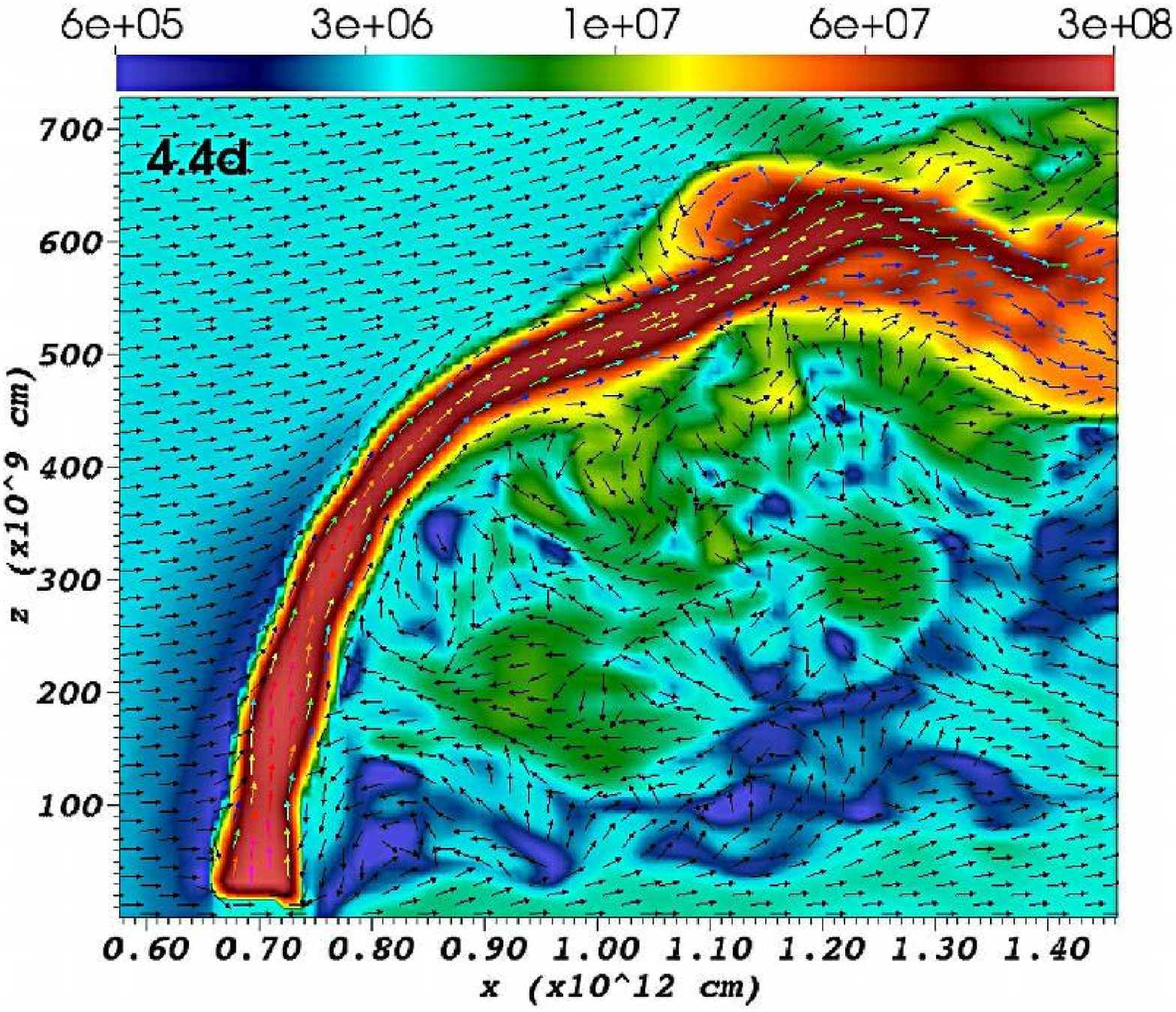}}
\caption{The first three images (a,b,c) with the red background show the density maps at three different times given in days in the $z-x$ plane passing through the WD source of the jet.
The jet is injected from $(x,z)=(10^{12} \cm , 0)$ in the $+z$ direction. The envelope is injected from the left, in the $+x$ direction. The density scale is in units $\g \cm^{-3}$.
Panel (d) shows the temperature in the same plane in units of $\K$. Panel (e) presents the velocity magnitude map in the same plane. Enlargement of the marked part is shown in panel (g).
Panel (f) shows the density in the $z=0.5 \times 10^{12} \cm$ plane.}
\label{fig:cewd_dens}
\end{center}
\end{figure*}

We model the half space $z>0$, and inject the jet in that side with a velocity of $v_z=3000 \km \s^{-1}$.
The jet is launched from a square of 9 cells ($3 \times 3$) in the plane $z=0$.
The mass loss rate of one jet is $\dot{M}_{1}=2.5\times10^{-5} M_\odot \yr^{-1}$, such that the kinetic power of the two-jets is about the
Eddington luminosity of a $1 M_\odot$ WD.
Our main results are presented in Figure \ref{fig:cewd_dens}.

In the present study we take the source of the jets to be a WD, but the results are relevant to accreting NS as well.
As with the rest of the present study, we focus on morphologies of bubbles and the velocity structure in the bubbles and their surroundings.
\begin{enumerate}
\item \emph{Morphology.} As expected, the AGB envelope bends the jet. This morphology is very similar to the one observed when
radio jets are bent when a radio galaxy is moving through the intra-galactic medium, e.g., \cite{Odea1986}.
These are called narrow-angle-tail radio sources.
The bending of radio jets was simulated in the past, e.g., \cite{BalsaraNorman1992}.
We basically see a morphology similar to that of smoke flowing out of a chimney in a wind.
However, we neglected convection in our simulation, something not completely justified as
AGB envelopes possess a vigorous convection.
The typical size of the convective cells in AGB stars is the pressure scale height of $l_p(r) \simeq 0.3 r$,
which in our case is more or less the size of our grid. The convection velocity is close to the sound speed.
With these typical values, we expect the convection to substantially influence the morphology.
\item \emph{The flow near the jets' source.} In addition to the shocked jets' material that is dragged by the envelope to large distances,
some jets' material is flowing just behind the source. This flow can be seen in the panel (g) of Figure \ref{fig:cewd_dens},
and was studied in some detail by \cite{BalsaraNorman1992}.
The hot and low-density region formed by this flow can substantially affect the accretion process onto the WD.
\item \emph{Bubble fragmentation.} As best seen in the two density panels at $t=4.4$~days, a separated bubble of hot shocked jet's gas is formed. This fragmentation is caused by
instabilities, mainly the Kelvin-Helmholtz instability, as we already discussed for bubbles in clusters in section \ref{subsec:bubbles} and can be seen in
Figure \ref{fig:dens_epi_std}.
\item \emph{Buoyancy.} Hot bubbles are known to buoy outward in clusters of galaxies.
The buoyancy of the hot bubble is seen in panel (f) of Figure \ref{fig:cewd_dens}, as a motion in the $+y$ direction.
Gravity goes down in this panel, and the bubble buoys upward.
\item \emph{Feedback: the AGB envelope}. \cite{Soker2004b} discussed the influence of the hot bubbles formed by the jets on the AGB envelope.
Although there is some energy input from the bubbles, jets from WDs are not expected to influence much the AGB envelope
(maybe only in the very outer envelope regions). This is also seen in our simulation by the small volume of the hot bubbles.
The volume of the bubble is smaller than the typical volume of a convective cell.
However, NS can accrete at a very high rate due to neutrino cooling
\citep{Chevalier1993}, and if they blow jets they can substantially disturb the envelope.
\item \emph{Feedback: the accretion process.} The flow behind the jets' source, as discussed in point (2) above,
forms a hot region close to the source. This might substantially reduce the
accretion rate, hence forming a negative feedback of the accretion process. This reminds us of the feedback in cooling flows.
This feedback will be studied in the future when gravity is added to the jets' source.
\end{enumerate}

\section{Summary}
\label{sec:summary}

A generic feature of many gravitationally bound systems is the tendency of compact regions to become smaller, and of the more extended regions to expand.
The fundamental reason is that there is no real equilibrium state, and entropy (or another similar function, sometimes called H-function) can
in principle increase unlimited.
The gravitational energy released by a contracting small volume can substantially, and even catastrophically, affect the evolution
of extended regions.

There is the question of how the energy is transferred from the compact region to the extended regions.
In stellar systems the energy transfer is done by gravity itself.
Examples include a triple system where the orbit of two stars shrinks and the tertiary body is ejected, and globular clusters where the core shrinks and
the outer region expand.

In most systems where the extended regions are filled with gas, radiation is inefficient in depositing energy to the gas.
In cluster of galaxies and in galaxies, the hot gas is transparent to the radiation emitted by the AGN.
As well, PNe absorbs a small fraction of the central stellar luminosity.
In CCSNe at most $\sim 1 \%$ of the energy carried by neutrinos is absorbed.

More efficient in influencing the ambient gas are jets launched by the compact object.
The systems studied here, or to whom the results are relevant, and their order of magnitude properties are listed in Table \ref{table:compare}.
In some of these the interaction of the jets with the ambient gas is regulated by a feedback mechanisms (JFM).
The emphasis of this paper is that despite the large differences between the systems listed in Table \ref{table:compare},
there are intriguing similarity in the interaction of jets with the ambient gas.
These similarities can be used to learn from one system on another.
For example, bubbles are observed directly only in CF clusters and in PNe.
The direct observations can be used to compare the numerical simulations with observations.

Our results strengthen the jet feedback mechanism (JFM) as a common process in many astrophysical objects.

Part of the simulations were performed on the TAMNUN HPC cluster at the Technion.
The software used in this work was in part developed by the DOE NNSA-ASC OASCR Flash Center at the University of Chicago.
This research was supported by the Asher Fund for Space Research and the E. and J. Bishop Research Fund at the Technion, and the Israel Science foundation.


\begin{thebibliography}{}

\bibitem[Akashi \& Soker(2008)]{Akashi2008} Akashi, M., \& Soker, N.\ 2008, New Astronomy, 13, 157

\bibitem[Allen et al.(2006)]{Allen2006} Allen, S.~W., Dunn, R.~J.~H., Fabian, A.~C., Taylor, G.~B., \& Reynolds, C.~S.\ 2006, \mnras, 372, 21

\bibitem[Alouani Bibi et al.(2007)]{AlouaniBibi2007} Alouani Bibi, F., Binney, J., Blundell, K., \& Omma, H.\ 2007, \apss, 311, 317

\bibitem[Balsara \& Norman(1992)]{BalsaraNorman1992} {Balsara, D.~S. \& Norman, M.~L.},\ 1992, \apj, 393, 628

\bibitem[Basson \& Alexander(2003)]{Basson2003} Basson, J.~F., \& Alexander, P.\ 2003, \mnras, 339, 353

\bibitem[Bisnovatyi-Kogan et al.(1976)]{Bisnovatyi1976} Bisnovatyi-Kogan, G.~S., Popov, I.~P., \& Samokhin, A.~A.\ 1976, \apss, 41, 287

\bibitem[Blondin et al.(2003)]{Blonding2003} Blondin, J.~M., Mezzacappa, A., \& DeMarino, C.\ 2003, \apj, 584, 971

\bibitem[Blondin \& Mezzacappa(2007)]{Blondin2007} Blondin, J.~M., \& Mezzacappa, A.\ 2007, Nature, 445, 58

\bibitem[Br{\"u}ggen et al.(2007)]{Bruggen2007} Br{\"u}ggen, M., Heinz, S., Roediger, E., Ruszkowski, M., \& Simionescu, A.\ 2007, \mnras, 380, L67

\bibitem[Cavagnolo et al.(2011)]{Cavagnolo2011} Cavagnolo, K.~W., McNamara, B.~R., Wise, M.~W., Nulsen,  P.~E.~J., Br{\"u}ggen, M., Gitti, M., \& Rafferty, D.~A.\ 2011, \apj, 732, 71

\bibitem[Chevalier(1993)]{Chevalier1993} Chevalier, R.~A.\ 1993, \apjl, 411, L33

\bibitem[Couch et al.(2009)]{Couch2009} Couch, S.~M., Wheeler, J.~C., \& Milosavljevi{\'c}, M.\ 2009, \apj, 696, 953

\bibitem[Couch et al.(2011)]{Couch2011} Couch, S.~M., Pooley, D., Wheeler, J.~C., \& Milosavljevi{\'c}, M.\ 2011, \apj, 727, 104

\bibitem[Cuadra et al.(2006)]{Cuadra2006} Cuadra, J., Nayakshin, S., Springel, V., \& Di Matteo, T.\ 2006, \mnras, 366, 358

\bibitem[Dong et al.(2010)]{Dong2010}Dong, R., Rasmussen, J., \& Mulchaey, J.~S.\ 2010, \apj, 712, 883

\bibitem[Fabian(2012)]{Fabian2012} Fabian, A.~C.\ 2012, ARAA, (arXiv:1204.4114)

\bibitem[Fabian et al.(2000)]{Fabianetal2000} Fabian, A. C., et al.\ 2000, \mnras, 318, L65     

\bibitem[Falceta-Goncalves et al.(2010)]{Falceta-Goncalves2010} Falceta-Goncalves, D., Caproni, A., Abraham, Z., Teixeira, D. M., \& de
  Gouveia Dal Pino, E. M. 2010, ApJ, 713, L74

\bibitem[Farage et al.(2012)]{Farage2012} Farage, C.~L., McGregor, P.~J., \& Dopita, M.~A.\ 2012, \apj, 747, 28

\bibitem[Foglizzo et al.(2012)]{Foglizzo2012} Foglizzo, T., Masset,
F., Guilet, J., \& Durand, G.\ 2012, Physical Review Letters, 108, 051103

\bibitem[Freytag \& H\"ofner(2008)]{Freytag2008} Freytag, B., H\"ofner, S.\ 2008, \aap, 483, 571

\bibitem[Fryxell et al.(2000)]{Fryxell2000} Fryxell, B., Olson, K., Ricker, P., et al.\ 2000, \apjs, 131, 273

\bibitem[Gaspari et al.(2012a)]{Gaspari2012a} Gaspari, M., Brighenti, F., \& Temi, P.\ 2012a, arXiv:1202.6054

\bibitem[Gaspari et al.(2012b)]{Gaspari2012b} Gaspari, M., Ruszkowski, M., \& Sharma, P.\ 2012b, \apj, 746, 94

\bibitem[Gilkis \& Soker(2012)]{GilkisSoker2012} Gilkis, A., \& Soker, N.\ 2012,  (arXiv1205.3571)

\bibitem[Guerrero et al.(2003)]{Guerreroetal2003}  Guerrero, M. A., Chu, Y.-H., Manchado, A., Kwitter, K. B. 2003, \aj, 125, 3213

\bibitem[Harpaz(1984)]{Harpaz1984} Harpaz, A.\ 1984, \mnras, 210, 633

\bibitem[Heinz et al.(2006)]{Heinz2006} Heinz, S., Br{\"u}ggen, M., Young, A., \& Levesque, E.\ 2006, \mnras, 373, L65

\bibitem[Hillel \& Soker(2012)]{HillelSoker2012} Hillel, S., \& Soker, N.\ 2012,  (arXiv1206.6029)

\bibitem[H{\"o}flich et al.(2001)]{Hoflich2001} H{\"o}flich, P., Khokhlov, A., \& Wang, L.\ 2001, 20th Texas Symposium on relativistic astrophysics, 586, 459

\bibitem[Huarte-Espinosa et al.(2012)]{Huarte-Espinosa2012} Huarte-Espinosa, M., Frank, A., Balick, B., Blackman, E. G., De Marco, O., Kastner, J. H., \& Sahai, R.\ 2012, \mnras, 424, 2055

\bibitem[Itoh et al.(1996)]{Itoh1996} Itoh, N., Hayashi, H., Nishikawa, A., \& Kohyama, Y.\ 1996, \apjs, 102, 411

\bibitem[Kashi et al.(2012)]{Kashi2012} Kashi, A., Nagamine, K., Proga, D., Ostriker, J. P., \& Varghese, S. 2012, preprint

\bibitem[Khokhlov et al.(1999)]{Khokhlov1999} Khokhlov, A.~M., H{\"o}flich, P.~A., Oran, E.~S., et al.\ 1999, \apjl, 524, L107

\bibitem[Kohri et al.(2005)]{kohri2005} Kohri, K., Narayan, R.,
     \& Piran, T.\ 2005, \apj, 629, 341

\bibitem[Kormendy et al.(1996)]{Kormendy1996} Kormendy, J., et al.\ 1996, \apjl, 459, L57

\bibitem[Lazzati et al.(2011)]{Lazzati2011} Lazzati, D., Morsony, B.~J., Blackwell, C.~H., \& Begelman, M.~C.\ 2011, arXiv:1111.0970

\bibitem[LeBlanc \& Wilson(1970)]{LeBlanc1970} LeBlanc, J.~M., \& Wilson, J.~R.\ 1970, \apj, 161, 541

\bibitem[Liebend{\"o}rfer et al.(2005)]{Liebend2005} Liebend{\"o}rfer, M., Rampp, M., Janka, H.-T., \& Mezzacappa, A.\ 2005, \apj, 620, 840

\bibitem[Lombardi et al.(2006)]{Lombardi2006} Lombardi, J.~C., Jr., Proulx, Z.~F., Dooley, K.~L., Theriault, E.~M., Ivanova, N., \& Rasio, F.~A.\ 2006, \apj, 640, 441

\bibitem[Ma et al.(2012)]{Ma2012}Ma, C.-J., McNamara, B.~R., \& Nulsen, P. E. J. \ 2012,  arXiv:1206.7071

\bibitem[MacFadyen et al.(2001)]{MacFadyen2001} MacFadyen, A.~I., Woosley, S.~E., \& Heger, A.\ 2001, \apj, 550, 410

\bibitem[McCourt et al.(2012)]{McCourt2012} McCourt, M., Sharma, P., Quataert, E., \& Parrish, I.~J.\ 2012, \mnras, 419, 3319

\bibitem[McNamara \& Nulsen(2012)]{McNamara2012} McNamara, B.~R., \& Nulsen, P.~E.~J.\ 2012, New Journal of Physics Focus Issue on Clusters of Galaxies
   (arXiv:1204.0006)

\bibitem[McNamara et al.(2005)]{McNamaraetal2005} McNamara, B. R., Nulsen, P. E. J., Wise, M. W., Rafferty, D. A.,
      Carilli, C., Sarazin, C. L. \& Blanton, E. L. 2005, Nature, 433 ,45

\bibitem[McNamara et al.(2011)]{McNamara2011} McNamara, B.~R., Rohanizadegan, M., \& Nulsen, P.~E.~J.\ 2011, \apj, 727, 39

\bibitem[Meier et al.(1976)]{Meier1976} Meier, D.~L., Epstein, R.~I., Arnett, W.~D., \& Schramm, D.~N.\ 1976, \apj, 204, 869

\bibitem[Mendygral et al.(2011)]{Mendygral2011} Mendygral, P.~J., O'Neill, S.~M., \& Jones, T.~W.\ 2011, \apj, 730, 100

\bibitem[Mendygral et al.(2012)]{Mendygral2012} Mendygral, P., Jones, T., \& Dolag, K.\ 2012, \apj, 750, 166

\bibitem[Mignone et al.(2007)]{Mignone2007}Mignone, A., Bodo, G., Massaglia,
S., Matsakos, T., Tesileanu, O., Zanni, C., Ferrari, A., 2007, \apjs, 170, 228

\bibitem[Miranda et al.(2001)]{Miranda2001} Miranda, L.~F.,
Torrelles, J.~M., Guerrero, M.~A., V{\'a}zquez, R.,
\& G{\'o}mez, Y.\ 2001, \mnras, 321, 487

\bibitem[Morsony et al.(2010)]{Morsony2010} Morsony, B.~J., Heinz, S., Br{\"u}ggen, M., \& Ruszkowski, M.\ 2010, \mnras, 407, 1277

\bibitem[Mueller et al.(2012a)]{Mueller2012a} Mueller, B., Janka, H.-T., \& Marek, A.\ 2012, arXiv:1202.0815

\bibitem[Mueller et al.(2012b)]{Mueller2012b} Mueller, B., Janka, H.-T., \& Heger, A.\ 2012, arXiv:1205.7078

\bibitem[Narayan \& Fabian(2011)]{Narayan2011} Narayan, R., \& Fabian, A.~C.\ 2011, \mnras, 926

\bibitem[Nesvadba et al.(2011)]{Nesvadba2011} Nesvadba, N.~P.~H., Boulanger, F., Lehnert, M.~D., Guillard, P., \& Salome, P.\ 2011, \aap, 536, L5

\bibitem[Nordhaus et al.(2010)]{Nordhausetal2010}Nordhaus, J., Burrows, A., Almgren, A., \& Bell, J.\ 2010, \apj, 720, 694

\bibitem[O'Dea \& Owen(1986)]{Odea1986} O'Dea, C.P., Owen, F.N\ 1986, \apj, 301, 84

\bibitem[Omma et al.(2004)]{Omma2004} Omma, H., Binney, J., Bryan, G., \& Slyz, A.\ 2004, \mnras, 348, 1105

\bibitem[O'Neill \& Jones(2010)]{ONeill2010} O'Neill, S.~M., \& Jones, T.~W.\ 2010, \apj, 710, 180

\bibitem[Ostriker et al.(2010)]{Ostriker2010} Ostriker, J.~P., Choi, E., Ciotti, L., Novak, G.~S., \& Proga, D.\ 2010, \apj, 722, 642

\bibitem[Papish \& Soker(2011)]{papish2011} Papish, O., \& Soker, N.\ 2011, \mnras, 416, 1697

\bibitem[Papish \& Soker(2012)]{papish2012} Papish, O., \& Soker, N.\ 2012, \mnras, 421, 2763

\bibitem[Passy et al.(2011)]{Passy2011} Passy, J.-C., De Marco, O., Fryer, C.~L., et al.\ 2012, \apj, 744, 52

\bibitem[Perets(2010)]{Perets2010} Perets, H.~B.\ 2010, arXiv:1001.0581

\bibitem[Pizzolato \& Soker(2005)]{Pizzolato2005} Pizzolato, F., \& Soker, N.\ 2005, Advances in Space Research, 36, 762

\bibitem[Pope(2009)]{Pope2009} Pope, E.~C.~D.\ 2009, \mnras, 395, 2317

\bibitem[Qian \& Woosley(1996)]{Qian1996} Qian, Y.-Z., \& Woosley, S.~E.\ 1996, \apj, 471, 331

\bibitem[Randall et al.(2011)]{Randall2011}Randall, S.~W., et al. \ 2011, \apj, 726, 86

\bibitem[Randall et al.(2012)]{Randall2012}Randall, S.~W., et al. \ 2012, Proceedings of this meeting.

\bibitem[Refaelovich \& Soker(2012)]{Refaelovich2012}Refaelovich, M., \& Soker, N.\ 2012, \apjl, 755, L3

\bibitem[Revaz et al.(2008)]{Revaz2008} Revaz, Y., Combes, F., \& Salom{\'e}, P.\ 2008, \aap, 477, L33

\bibitem[Ricker \& Taam(2008)]{RickerTaam2008} Ricker, P.~M., \& Taam, R.~E.\ 2008, \apjl, 672, L41

\bibitem[Ricker \& Taam(2012)]{RickerTaam2012} Ricker, P.~M., \& Taam, R.~E.\ 2012, \apj, 746, 74

\bibitem[Russell et al.(2010)]{Russell2010} Russell, H.~R., Fabian, A.~C., Sanders, J.~S., Johnstone, R.~M.,  Blundell, K.~M., Brandt, W.~N., \& Crawford, C.~S.\ 2010, \mnras, 402, 1561

\bibitem[Sahai(2000)]{Sahai2000} Sahai, R.\ 2000, \apjl, 537, L43

\bibitem[Sandquist et al.(1998)]{SandquistTaam1998} Sandquist, E.~L., Taam, R.~E., Chen, X., Bodenheimer, P., \& Burkert, A.\ 1998, \apj, 500, 909

\bibitem[Schwarz et al.(1992)]{Schwarzetal1992} Schwarz, H. E., Corradi, R. L. M., \& Melnick, J. 1992, A\&A Suppl. Ser., 96, 23.

\bibitem[Sharma et al.(2012)]{Sharma2012} Sharma, P., McCourt, M., Quataert, E., \& Parrish, I.~J.\ 2012, \mnras, 420, 3174

\bibitem[Soker(2004a)]{Soker2004a} Soker, N.\ 2004a, \apj, 612, 1060

\bibitem[Soker(2004b)]{Soker2004b} Soker, N.\ 2004a, New Astronomy, 9, 399

\bibitem[Soker(2009)]{Soker2009} Soker, N.\ 2009, \mnras, 398, L41

\bibitem[Soker \& Bisker(2006)]{SokerBisker2006} Soker, N., \& Bisker, G.\ 2006, \mnras, 369, 1115

\bibitem[Soker et al.(2001)]{Sokeretal2001} Soker, N., White, R.~E., III, David, L.~P., \& McNamara, B.~R.\ 2001, \apj, 549, 832

\bibitem[Sternberg et al. (2007)]{Sternberg2007} Sternberg, A., Pizzolato, F. \& Soker N. 2007, \apj, 656, L5

\bibitem[Sternberg \& Soker(2008a)]{Sternberg2008a} Sternberg, A., \& Soker N. 2008a, \mnras, 384, 1327 

\bibitem[Sutherland \& Dopita (1993)]{SutherlandDopita1993} Sutherland R.S., Dopita M.A., \ 1993, \apjs, 88, 253

\bibitem[Terzian \& Hajian(2000)]{TerzianHajian2000} Terzian, Y., \& Hajian, A. R. 2000,   ``Asymmetrical Planetary Nebulae II: From Origins to Microstructures,'' eds. J.H. Kastner, N. Soker, \&
  S. Rappaport, ASP Conf.\ Ser.\, Vol.\ 199, p. 33

\bibitem[Timmes \& Swesty(2000)]{timmes2000} Timmes, F.~X., \& Swesty, F.~D.\ 2000, \apjs, 126, 501

\bibitem[Taylor(1950)]{Taylor1950} Taylor, G.\ 1950, Royal Society of London Proceedings Series A, 201, 159

\bibitem[Vernaleo \& Reynolds(2006)]{Vernaleo2006} Vernaleo, J.~C., \& Reynolds, C.~S.\ 2006, \apj, 645, 83

\bibitem[Wilman et al.(2009)]{Wilman2009} Wilman, R.~J., Edge, A.~C., \& Swinbank, A.~M.\ 2009, \mnras, 395, 1355

\bibitem[Wilman et al.(2011)]{Wilman2011} Wilman, R.~J., Edge, A.~C., McGregor, P.~J., \& McNamara, B.~R.\ 2011, \mnras, 416, 2060

\bibitem[Wise et al.(2007)]{Wise2007}Wise, M.~W., McNamara, B.~R., Nulsen,
P.~E.~J., Houck, J.~C., \& David, L.~P.\ 2007, \apj, 659, 1153

\bibitem[Wong et al.(2011)]{Wong2011} Wong, K.-W., Irwin, J.~A., Yukita, M., Million, E.~T.,   Mathews, W.~G., \& Bregman, J.~N.\ 2011, \apjl, 736, L23

\bibitem[Woosley \& Janka(2005)]{Woosley2005} Woosley, S., \& Janka, T.\ 2005, Nature Physics, 1, 147

\end{thebibliography}
\end{document}